\documentclass[twocolumn]{aastex631}
\usepackage{import}
\usepackage{xcolor}
\usepackage{comment}
\usepackage{ulem}
\usepackage{booktabs}
\usepackage{xspace}

\newcommand{\mdot}{$\dot{M}$\xspace}

\newcommand{\lsun}{L$_{\odot}$\xspace}

\newcommand{\rstar}{R$_{\star}$\xspace}
\newcommand{\lstar}{L$_{\star}$\xspace}
\newcommand{\lline}{L$_{Line}$\xspace}
\newcommand{\lacc}{L$_{acc}$\xspace}
\newcommand{\rin}{R$_{in}$\xspace}

\newcommand{\accrateunits}{$\times$10$^{-8}$ M$_{\odot}$ yr$^{-1}$\xspace}
\newcommand{\accrate}{{M$_{\odot}$} yr$^{-1}$\xspace}
\newcommand{\Ha}{H$\alpha$\xspace}
\newcommand{\Hb}{H$\beta$\xspace}
\newcommand{\Hy}{H$\gamma$\xspace}
\newcommand{\hh}{H$_2$\xspace}

\defcitealias{RE19}{RE19}

\shorttitle{Accretion Variability in T Tauri Stars. I. FUV/NUV Spectra}
\shortauthors{Wendeborn et al.}

\begin{document}

\title{A Multi-wavelength, Multi-epoch Monitoring Campaign of Accretion Variability in T Tauri Stars from the ODYSSEUS Survey. I.~\emph{HST} FUV and NUV Spectra}


\author[0000-0002-6808-4066]{John Wendeborn}
\affil{Institute for Astrophysical Research, Department of Astronomy, Boston University, 725 Commonwealth Avenue, Boston, MA 02215, USA}

\author[0000-0001-9227-5949]{Catherine C. Espaillat}
\affil{Institute for Astrophysical Research, Department of Astronomy, Boston University, 725 Commonwealth Avenue, Boston, MA 02215, USA}

\author[0009-0003-8647-672X]{Sophia Lopez}
\affil{Institute for Astrophysical Research, Department of Astronomy, Boston University, 725 Commonwealth Avenue, Boston, MA 02215, USA}

\author[0000-0003-4507-1710]{Thanawuth Thanathibodee}
\affil{Institute for Astrophysical Research, Department of Astronomy, Boston University, 725 Commonwealth Avenue, Boston, MA 02215, USA}

\author[0000-0003-1639-510X]{Connor E. Robinson} 
\affil{Department of Physics \& Astronomy, Amherst College, C025 Science Center 25 East Drive, Amherst, MA 01002, USA}

\author[0000-0001-9301-6252]{Caeley V. Pittman}
\affil{Institute for Astrophysical Research, Department of Astronomy, Boston University, 725 Commonwealth Avenue, Boston, MA 02215, USA}

\author[0000-0002-3950-5386]{Nuria Calvet}
\affil{Department of Astronomy, University of Michigan, 1085 South University Avenue, Ann Arbor, MI 48109, USA}

\author[0009-0002-0185-9098]{Nicole Flors}
\affil{Institute for Astrophysical Research, Department of Astronomy, Boston University, 725 Commonwealth Avenue, Boston, MA 02215, USA}

\author[0000-0001-7796-1756]{Fredrick M. Walter}
\affiliation{Department of Physics and Astronomy, Stony Brook University, Stony Brook NY 11794-3800, USA}

\author[0000-0001-7157-6275]{\'{A}gnes K\'{o}sp\'{a}l}
\affil{Konkoly Observatory, HUN-REN Research Centre for Astronomy and Earth Sciences, CSFK, MTA Centre of Excellence, Konkoly-Thege Mikl\'os \'ut 15-17}
\affil{Institute of Physics and Astronomy, ELTE E\"otv\"os Lor\'and University, P\'azm\'any P\'eter s\'et\'any 1/A, 1117 Budapest, Hungary}
\affil{Max Planck Institute for Astronomy, K\"onigstuhl 17, 69117 Heidelberg, Germany}

\author[0000-0001-5707-8448]{Konstantin N. Grankin}
\affil{Crimean Astrophysical Observatory, 298409 Nauchny, Republic of Crimea}

\author[0000-0002-0233-5328]{Ignacio Mendigut\'ia}
\affil{Centro de Asrobiolog\'ia, (CSIC-INTA), Departamento de Astrof\'isica, ESA-ESAC Campus,E-28691 Madrid, Spain}

\author[0000-0003-4243-2840]{Hans Moritz G\"unther}
\affil{Kavli Institute for Astrophysics and Space Research, Massachusetts Institute of Technology, 77 Massachusetts Ave, Cambridge MA, 02139}

\author[0000-0001-6496-0252]{Jochen Eisl\"offel}
\affil{Th\"uringer Landessternwarte, Sternwarte5, D-07778 Tautenburg, Germany}

\author[0000-0003-0292-4832]{Zhen Guo}
\affil{Instituto de F{\'i}sica y Astronom{\'i}a, Universidad de Valpara{\'i}so, ave. Gran Breta{\~n}a, 1111, Casilla 5030, Valpara{\'i}so, Chile}

\author[0000-0002-1002-3674]{Kevin France}
\affil{Laboratory for Atmospheric and Space Physics, University of Colorado Boulder, Boulder, CO 80303,USA}

\author[0000-0002-5261-6216]{Eleonora Fiorellino}
\affil{INAF-Osservatorio Astronomico di Capodimonte, via Moiariello 16, 80131 Napoli, Italy}

\author[0000-0002-3747-2496]{William J. Fischer}
\affil{Space Telescope Science Institute, 3700 San Martin Drive, Baltimore, MD 21218, USA}

\author[0000-0001-6015-646X]{P\'eter \'Abrah\'am}
\affil{Konkoly Observatory, HUN-REN Research Centre for Astronomy and Earth Sciences, CSFK, MTA Centre of Excellence, Konkoly-Thege Mikl\'os \'ut 15-17}
\affil{ELTE E\"otv\"os Lor\'and University, Institute of Physics, P\'azm\'any P\'eter s\'et\'any 1/A, 1117 Budapest, Hungary}
\affil{University of Vienna, Dept. of Astrophysics, T\"urkenschanzstr. 17, 1180, Vienna, Austria}

\author[0000-0002-7154-6065]{Gregory J. Herczeg}
\affil{Kavli Institute for Astronomy and Astrophysics, Peking University, Beijing 100871, People's Republic of China}
\affil{Department of Astronomy, Peking University, Beijing 100871, People's Republic of China}

\begin{abstract}

The Classical T Tauri Star (CTTS) stage is a critical phase of the star and planet formation process. In an effort to better understand the mass accretion processes, which can dictate future stellar evolution and planet formation, a multi-epoch, multi-wavelength photometric and spectroscopic monitoring campaign of four CTTSs (TW~Hya, RU~Lup, BP~Tau, and GM~Aur) was carried out in 2021 and 2022/2023 as part of the Outflows and Disks Around Young Stars: Synergies for the Exploration of ULLYSES Spectra (ODYSSEUS) program. Here we focus on the $HST$ UV spectra obtained by the $HST$ Director's Discretionary Time UV Legacy Library of Young Stars as Essential Standards (ULLYSES) program. Using accretion shock modeling, we find that all targets exhibit accretion variability, varying from short increases in accretion rate by up to a factor of 3 within 48 hours, to longer decreases in accretion rate by a factor of 2.5 over the course of 1 year. This is despite the generally consistent accretion morphology within each target. Additionally, we test empirical relationships between accretion rate and UV luminosity and find stark differences, showing that these relationships should not be used to estimate the accretion rate for an individual target. Our work reinforces that future multi-epoch and simultaneous multi-wavelength studies are critical in our understanding of the accretion process in low-mass star formation. 

\end{abstract}


\section{Introduction}
\label{sec: Introduction}

Classical T Tauri Stars (CTTSs) are young ($<$10 Myr), low-mass ($<$2 M$_{\odot}$) stars that are actively accreting from a surrounding protoplanetary disk. This disk is the reservoir from which planets will form \citep[or in some cases have already formed, e.g.,][]{Keppler2018, Haffert2019, Perez2019, Zhou2022}, and dictates much of the future evolution of the system. The process of disk-to-star mass accretion, which takes place at the star-disk boundary, typically within $\sim$10 R$_*$, is highly energetic, emitting X-rays and UV radiation \citep{Kastner2002, Herbst1994, Gunther2007} which can have a profound impact on the disk. These effects include modifying disk chemistry, modulating the presence and distance of snowlines, inhibiting dust grain growth, annealing solid dust grains, and depleting/dispersing disk material \citep[see reviews by][]{PPVII10, PPVII12, PPVII14}. As such, it is important to fully understand the accretion process in order to paint a complete picture of the star/planet formation processes.

Due to the strong magnetic fields in CTTSs \citep[1--3 kG;][]{Johns-Krull2007, Donati2020}, accretion occurs via magnetospheric accretion \citep[see reviews by][]{Bouvier2007, Hartmann2016}. The magnetic field truncates the disk at some radius (the inner truncation radius, \rin, typically assumed to be 5 \rstar), where the hot plasma is bound to and travels along the magnetic fields lines. Along its path, it eventually reaches super-sonic, free-fall velocities (v$_{ff}$) until it reaches the photosphere, producing highly energetic shocks and accretion hotspots. 

Simulations show the accretion flow can vary in density \citep{Romanova2008, Kulkarni2008, Romanova2012, Zhu2023a} which affects the temperature of the resulting hotspot. Denser flows create hotter hotspots which radiate more strongly at UV wavelengths, while less dense flows (i.e. cooler hotspots) peak in the optical \citep{Calvet1998, Ingleby2013, Pittman2022}. This excess UV/optical emission can be modeled, assuming it originates from accretion columns of various energy densities \citep{Calvet1998}. Such modeling (often using $HST$ UV-optical spectra) has revealed that the accretion structure is not only time-variable, but also varies greatly from object to object. Some stars are dominated by large, cooler hotspots driven by low density accretion columns, while in others the accretion is dominated by high density columns producing small but high temperature hotspots \citep{Ingleby2013, Ingleby2015, RE19, Pittman2022}. 

The accretion onto CTTSs is known to be intrinsically variable on timescales from minutes--hours (e.g. inhomogenous accretion flows, hotspot plasma oscillations), days--weeks (e.g. inner disk inhomogeneity, inner disk thermal instabilities), to months--years (e.g. MRI, gravitational instability); see review by \citet{PPVII12}. Observationally, the variable mass accretion process is often associated with other sources of variability, including stellar rotation, chromospheric activity, disk occultation, variable extinction, winds/outflows, dark spots, and flares.

Accretion has been correlated with UV \citep{Calvet2004, Ardila2013, RE19} and optical \citep{Alcala2014, Alcala2017} line luminosity, though there is typically notable scatter/outliers within these empirical relationships, implying that they are an imperfect analogue for the true accretion rate. One explanation for this scatter is intrinsic variability, though until now no extensive multi-epoch FUV/NUV campaign has been carried out to gauge how these relationships vary in individual stars as the accretion varies over the span of several rotation periods.

In an effort to better understand accretion variability in CTTSs, and to connect it to other observables in the UV-IR, a multi-wavelength, multi-epoch observing campaign of the CTTSs TW~Hya, RU~Lup, BP~Tau, and GM~Aur was carried out in 2021 and 2022. These observations include multi-epoch $HST$ Cosmic Origins Spectrograph (COS) UV spectra taken as part of the ULLYSES (\textit{UV Legacy Library of Young Stars as Essential Standards}) $HST$ Director's Discretionary Time Program, moderate cadence multi-band $uBgVriz$ and {\it Transiting Exoplanet Survey Satellite} \citep[$TESS$,][]{TESS} photometry, and multi-epoch high resolution optical spectra. In this paper (Paper I) we present the $HST$ COS FUV and NUV results of our multi-epoch, multi-wavelength study of these four CTTSs. The photometry and optical spectra, partly obtained as part of the \textit{Outflows and Disks Around Young Stars: Synergies for the Exploration of ULLYSES Spectra} (ODYSSEUS) collaboration and PENELLOPE Very Large Telescope (VLT) program \citep{Espaillat2022, Manara2021}, will be presented in follow-up works, Paper II and Paper III, respectively.

Here in Paper I, we describe our sample in Section \ref{sec: Sample} and then our observations and data in Section \ref{sec: Observations}. Section \ref{sec: Analysis and Results} describes our accretion shock model and the results of our analyses. In Section \ref{sec: Discussion} we discuss these results in more detail and speculate on their implications. We summarize our final conclusions in Section \ref{sec: Conclusion}.

\section{Sample} \label{sec: Sample}

Our sample consists of TW~Hya, RU~Lup, BP~Tau, and GM~Aur, chosen by the ULLYSES implementation team. This sample was chosen as they are well-studied, prototypical CTTSs with well-known magnetic configurations and reasonably constrained rotation periods. Below we provide more background information on each target, with some properties summarized in Table~\ref{tab: Targets}.

\subsection{TW~Hya}

TW~Hya is a nearby \citep[$d\sim60$ pc][]{GAIADR3} CTTS of spectral type K7 \citep{Ingleby2013}. Its surrounding disk has many concentric gaps and rings \citep{Andrews2012, Huang2018b}, which may have been carved out by an embedded planet \citep{Nayakshin2020, Zhu2023b}. The star is viewed nearly pole-on \citep[18\textdegree, from $v$sin$i$][]{Alencar2002} and the disk is viewed nearly face-on \citep[5.5--7\textdegree, from scattered light and (sub)millimeter continuum][]{Debes2023, Qi2004, Huang2018b}. We adopt an inclination of 7{\textdegree} from millimeter observations here and in Paper II and Paper III. \citet{Donati2011} estimate the star's magnetic obliquity, the angle between its rotation and dipolar magnetic axis, to be low, at about 10\textdegree.

Despite its nearly pole-on geometry, TW~Hya has shown photometric periodicity, though at a wide and inconsistent range of periods \citep[0.16-4.7 days;][]{Siwak2011, Siwak2014, Siwak2018}. Radial velocity studies \citep{Huelamo2008, SiciliaAguilar2023} reveal a strong, stable, sinusoidal rotational modulation at about 3.57-days.

TW~Hya is old \citep[8--10 Myr,][]{Sokal2018}, yet is still actively accreting from a surrounding accretion disk beyond the assumed $\lesssim$5 Myr lifespan of protoplanetary disks \citep{Evans2009, Williams2011} and therefore challenges our understanding of disk dispersal mechanisms. The range of measured accretion rates in TW~Hya is 0.05--0.6\accrateunits \citep{Ingleby2013, Manara2014, RE19, Sousa2023,Herczeg2023}. This accretion is thought to primarily occur through a single, large, pole-ward flow \citep{Donati2011, SiciliaAguilar2023} along with many smaller, more sporadic accretion tongues near the equator \citep{Siwak2018}.

\subsection{RU~Lup}

Previous studies of RU~Lup have shown it is a highly veiled \citep[r$_K$=6.3, r$_V$=2--9;][]{Stempels2002, Stempels2007, Herczeg2008, Sousa2023}, strongly accreting \citep[\mdot$\sim$2--7\accrateunits;][]{Herczeg2005, Herczeg2008, Alcala2017, Stock2022} K7 \citep{Sousa2023} CTTS with a powerful accretion-driven outflow \citep{Takami2001, Whelan2021}. Its distance is about 156 pc \citep{GAIADR3}. The rotation period of RU~Lup is generally assumed to be 3.71 days \citep{Stempels2007}, detected via radial velocity measurements. However, this period has not been recovered in other time-dependant photometric studies of RU~Lup, likely due to the combination of stochastic accretion activity \citep{Siwak2016, Stock2022} and its low inclination angle.

RU~Lup possesses a 63 au radius dust disk (with a gas disk extending to 120 au) with concentric gaps and rings \citep{Andrews2018, Huang2018} and unique gas structure. The gas structure, primarily traced by $^{12}$CO, is a large, clumpy 5-armed spiral that extends for $>$1000 au \citep{Huang2020a}. Millimeter continuum \citep{Huang2018} and NIR (K-band) interferometric \citep{GRAVITY2021} observations find disk inclinations of 18.8\textdegree\ and 16\textdegree, respectively. Other observations of RU~Lup have placed the stellar inclination between 23--24\textdegree\ using line broadening \citep{Herczeg2005} and radial velocity variations \citep{Stempels2007}, slightly higher than the millimeter observations, but still in line with a close-to face-on disk. Here  we assume an inclination of 16\textdegree\ from millimeter observations.

\subsection{BP~Tau}

BP~Tau is a $\sim$3 Myr \citep{Grankin2016} K7 \citep{Ingleby2013} system in the Taurus-Auriga complex at a distance of about 128 pc \citep{GAIADR3}. Previous studies have estimated a wide range of accretion rates for BP~Tau, between 0.09--3 \accrateunits \citep{Gullbring1998, Schmitt2005, Long2011, Ingleby2013}. Its rotation period has been estimated to be between 6.1--11 days \citep{Vrba1986, Simon1990, Osterloh1996, Gomez1997, Percy2006a, Percy2006b}, but it is generally considered to be either 7.6 or 8.2 days from photometric variability, assumed to originate from accretion signatures rotating with the stellar surface.

\citet{Donati2008} performed a spectropolarimetric study of BP~Tau in 2006 to map the structure of its magnetic field. Their study revealed a field dominated by a dipolar component, but still with a strong octupolar component, both tilted from the rotation axis by $\sim$10\textdegree, though in opposite directions. They predicted accretion hotspots on its surface near the octupoles that cover about 2\% of the surface. \citet{Long2011} also recovered similar accretion structure, revealing that the hotspots are mostly symmetric in longitude but elongated in latitude. 

Observations of BP~Tau's disk reveal a $\sim$60 au radius dust disk with 2--3 gap/ring pairs inclined at 37--39\textdegree\ \citep{Long2019, Zhang2023}. The stellar inclination has also been estimated at 48--52\textdegree\ \citep{Simon1990, Johns-Krull1999} from estimates of $v$sin$i$ and an assumed rotation period of 7.6 days. We adopt an inclination of 38.2{\textdegree} from the millimeter observations of \citet{Zhang2023}.

\subsection{GM~Aur}

GM~Aur is a young \citep[2 Myr][]{Jensen1997} K5 \citep{Manara2014} system in the Taurus-Auriga complex at a distance of about 155 pc \citep{GAIADR3}. It possesses a transitional disk with an inner cavity of radius 20 au and multiple outer concentric rings \citep{Espaillat2011, Macias2018, Huang2020b}. Such studies place the disk inclination between 52.8--53.2\textdegree and we adopt 53\textdegree. 

It is a moderate accretor, with accretion rates between 0.4--2.0\accrateunits \citep{Ingleby2013, Manara2014, RE19, Nature}. The accretion occurs primarily via a single, large hotspot near the pole which, in combination with its moderate inclination and rotation, causes a strong, stable periodicity of about 6 days \citep{Percy2010, Nature, Bouvier2023}. The magnetic obliquity has been estimated at 13\textdegree\ \citep{McGinnis2020}.

\citet{Nature} showed that short-wavelength photometry ($ug$) peaked about a day prior to longer-wavelength bands ($ri/TESS$), which they attribute to a large, asymmetric, and azimuthally stratified accretion hotspot. \citet{Bouvier2023} recently conducted a multiwavelength study of GM~Aur in 2021, including optical/NIR photometry and spectra. They found a $\sim$6-day period in their photometry, as well as at low red-shifted velocities ($\sim$0--100 km/s) in \Ha, \Hb, and \Hy, and in the wings ($\pm$200-400 km/s) of \Ha and \Hb. At moderate blue-shifted velocities (-200--0 km/s), where variable absorption is strongest, they found non-periodic variability. They attribute these variability characteristics (both periodic and non-periodic) to the presence of a stable, rotating accretion hotspot along with stochastically variable outflows.

\section{Observations and Data Reduction} \label{sec: Observations}

\begin{figure*}[htp]
    \centering
    \includegraphics[width=0.98\textwidth]{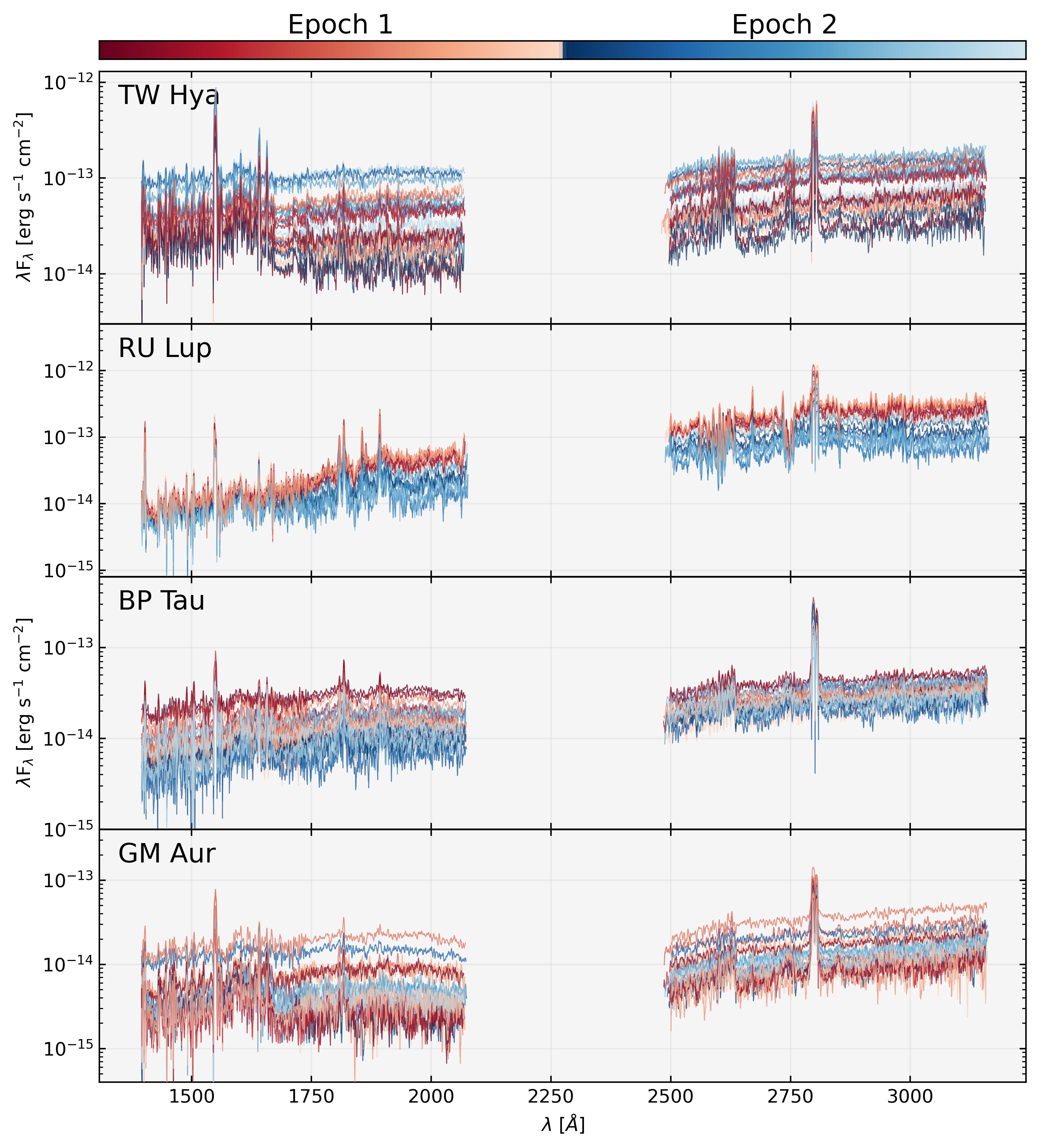}
    \caption{$HST$ COS spectra of TW~Hya, RU~Lup, BP~Tau, and GM~Aur. The red-colored spectra were taken in Epoch 1, while the blue-colored spectra were taken in Epoch 2. Darker lines denote spectra obtained earlier in their respective epoch. Spectra have not been dereddened here. For clarity, data have been smoothed using a Savitzky-Golay filter. More details on the timing of the observations are given in Table \ref{tab: HST Observations Table}.} 
    \label{fig: COS Spectra All}
\end{figure*}

$HST$ observations were carried out with the Cosmic Origins Spectrograph (COS) instrument for TW~Hya, RU~Lup, BP~Tau, and GM~Aur as part of the ULLYSES DDT program through proposals 16107--16110 and 16589--16592 (Roman-Duval et al. 2020, PI: Julia Roman-Duval). These spectra are presented in Figure \ref{fig: COS Spectra All}. The observations utilized the G160M and G230L gratings, which cover a wavelength range of about 1400--3200 {\AA}, with a gap between about 2100--2500 {\AA}. Below about 1760 {\AA} the spectral resolution, R, is $\sim$18,000 and elsewhere R$\sim$2,900. All spectra have been de-reddened according to the \citet{Whittet2004} extinction law and the extinction values from Table~\ref{tab: Targets}, assuming R$_\mathrm{V}$=3.1. The spectra presented here were obtained via ULYSSES Data Release 6 \citep{ULYSSES}. Note that a portion of the TW~Hya data used here is presented in \citep{Hinton2022} in a study of flaring in CTTSs.

Each target was observed approximately four times per rotation period for three rotation periods in two epochs separated by about a year (see Table \ref{tab: HST Observations Table}), totaling 21--24 visits per target. All $HST$ spectra used in this work can be found in MAST \dataset[10.17909/a530-qm96]{https://doi.org/10.17909/a530-qm96}. For clarity, in this work ``epoch" will refer to the different halves of the monitoring (Epoch 1/E1, corresponding to 2021 and Epoch 2/E2, corresponding to 2022/2023), while ``visit" will refer to individual $HST$ observations.

There were some changes to the observing schedule of the targets, typically due to problems with guide star and/or target acquisition. Several visits of TW~Hya failed during the original scheduled window in E2 and were rescheduled to 2022 April 25--26 (MJD: 59694.5-59695.4). $HST$ went offline during a visit of GM~Aur on 2021 October 5 (MJD: 59492) in E1. The remaining observations were carried out from 2021 December 6--12 (MJD: 59554.3-59560.2). In other cases, individual observations failed and were not re-observed, resulting in less than 24 total observations for each target, except BP~Tau for which all 24 total observations were carried out.  

\begin{deluxetable*}{c c c c c c c c c c c}[htp]
\setlength{\tabcolsep}{10pt}
\tablecaption{Target Parameters \label{tab: Targets}}
\centering
\tablehead{
\colhead{Object} & \colhead{RA/Dec} & \colhead{Distance} & \colhead{SpT} & \colhead{Period} & \colhead{M$_*$} & \colhead{T$_{eff}$} & \colhead{R$_*$} & \colhead{A$_\mathrm{V}$} & \colhead{i} \\
\colhead{} & \colhead{[J2000]} & \colhead{[pc]} & \colhead{} & \colhead{[days]} & \colhead{[M$_{\odot}$]} & \colhead{[K]} & \colhead{[R$_{\odot}$]} & \colhead{[mag]} & \colhead{[$^{\circ}$]} 
}
\startdata
TW~Hya & 11:01:51.905/ & 59.96$^{+0.37}_{-0.11}$\tablenotemark{ a} & K7\tablenotemark{b} & 3.57\tablenotemark{c} & 0.79\tablenotemark{d} & 4060\tablenotemark{d} & 0.93\tablenotemark{e} & 0.00\tablenotemark{d} & 7\tablenotemark{f} \\
 & -34:42:17.033 & & & & & & & & & \\
RU~Lup & 15:56:42.311/ & 156.0$^{+1.2}_{-1.1}$\tablenotemark{ a} & K7\tablenotemark{g} & 3.71\tablenotemark{h} & 0.65\tablenotemark{i} & 3950\tablenotemark{i} & 1.83\tablenotemark{*i} & 0.07\tablenotemark{i} & 16\tablenotemark{j} \\
 & -37:49:15.474 & & & & & & & & & \\
BP Tau & 04:19:15.834/ & 128.3$^{+0.7}_{-0.6}$\tablenotemark{ a} & K7\tablenotemark{b} & 7.6\tablenotemark{k}, 8.2\tablenotemark{l} & 0.7\tablenotemark{m} & 4055\tablenotemark{n} & 1.77\tablenotemark{*n} & 0.51\tablenotemark{m} & 38.2\tablenotemark{o} \\
 & +29:06:26.927 & & & & & & & & & \\
GM Aur & 04:55:10.982/ & 155.0$^{+1.4}_{-2.0}$\tablenotemark{ a} & K5\tablenotemark{d} & 6.0\tablenotemark{e} & 1.36\tablenotemark{d} & 4350\tablenotemark{d} & 2.00\tablenotemark{e} & 0.6\tablenotemark{b} & 53\tablenotemark{p} \\
 & +30:21:59.574 & & & & & & & & & \\
\enddata
\tablenotetext{}{a: \citet{GAIADR3}, 
b: \citet{Ingleby2013}, 
c: \citet{SiciliaAguilar2023}, 
d: \citet{Manara2014},
e: \citet{RE19} (from \citet{Manara2014} with updated Gaia distances),
f: \cite{Andrews2016},
g: \citet{Sousa2023}, 
h: \citet{Stempels2007},
i: \citet{Herczeg2005},
j: \cite{GRAVITY2021},
k: \citet{Vrba1986}, 
l: \citet{Percy2006a},
m: \citet{Gullbring1998}, 
n: \citet{Johns-Krull1999},
o: \citet{Zhang2023},
p: \citet{Macias2018},
*: Updated with current Gaia distances
}
\end{deluxetable*}\begin{deluxetable*}{c c c c c}[htp] \label{tab: HST Observations Table}
\tablecaption{$HST$ COS Observations} 
\centering
\tablehead{
\colhead{Object} & \colhead{Epoch} & \colhead{Date (UT)} & \colhead{MJD} & \colhead{\# of} \\
\colhead{} & \colhead{} & \colhead{[Begin/End]} & \colhead{[Begin-End]} & \colhead{Observations} 
}
\startdata
TW~Hya & 1 & 2021-03-29/2021-04-08 & 59302.7-59312.6 & 11 \\
TW~Hya & 2 & 2022-03-30/2022-04-27 & 59668.5-59695.4 & 10 \\
\hline
RU~Lup & 1 & 2021-08-10/2021-08-21 & 59436.9-59447.3 & 10 \\
RU~Lup & 2 & 2022-08-10/2022-08-24 & 59801.3-59815.4 & 12 \\
\hline
BP~Tau & 1 & 2021-08-20/2021-09-12 & 59446.8-59469.1 & 12 \\
BP~Tau & 2 & 2022-12-13/2023-01-06 & 59926.2-59950.5 & 12 \\
\hline
GM~Aur & 1 & 2021-10-13/2021-12-12 & 59500.5-59560.2 & 11 \\
GM~Aur & 2 & 2022-11-26/2022-12-12 & 59909.9-59925.2 & 11 \\
\enddata
\tablecomments{All $HST$ spectra used in this work, including targets and templates, can be found in MAST \dataset[10.17909/a530-qm96]{https://doi.org/10.17909/a530-qm96}.}
\end{deluxetable*}


\section{Analysis and Results} \label{sec: Analysis and Results}

In the following sections we analyze and model the UV spectra presented in Section~\ref{sec: Observations}. We characterize the accretion properties of TW~Hya, RU~Lup, BP~Tau, and GM~Aur by fitting the data with accretion shock models.  We measure the luminosities of UV lines and features and compare these to the accretion rates derived with the accretion shock models.

\subsection{Accretion Shock Modeling} \label{sec: Shock Model} \label{sec: Shock Model Description} \label{sec: Shock Model Results}

We use the accretion shock models of \citet{Calvet1998} following similar methods as \citet{RE19} \citepalias[hereafter][]{RE19}. We adopt the stellar parameters listed in Table~\ref{tab: Targets}.

To accurately estimate the excess emission caused by accretion, one must have an estimate of the underlying photospheric+chromospheric emission (F$_{\rm{Phot,\ True}}$) of a non-accreting star (see Equation \ref{eq: Fphot}, where F$_{\rm{WTTS}}$ is the raw spectrum of non-accreting Weak T Tauri Star (WTTS) and V$_{CTTS}$ and V$_{WTTS}$ are the flux of the CTTS and WTTS spectra at V-band (5500 {\AA}), $f_i$ is the surface coverage fraction of an individual accretion column, $i$, and r$_V$ is the optical veiling). To estimate this, we utilize contemporaneous optical veiling measurements (r$_V$) to scale a raw WTTS spectrum (F$_{\rm{WTTS}}$) to the appropriate $V$-band level for any visit with contemporaneous $V$ photometry (see Paper II). We calculate veiling using medium--high resolution (R$\sim$5,400--140,000) optical ($\lambda \sim4000-8000 \AA$) spectra from SMARTS/CHIRON, VLT/ESPRESSO, VLT/XSHOOTER, VLT/UVES, Haute-Provence/SOPHIE, and Tautenburg/TCES. We follow the basic procedure of \citet{Hartigan1989}. See Paper III, Section 3 for a more detailed description of our procedure and findings. For TW~Hya, RU~Lup, and BP~Tau, we use $HST$ STIS spectra of the K7 WTTS HBC 427 (PID 11616) as the basis for the photospheric template, while for GM~Aur we use the K5 WTTS RECX-1 (PID 11616). 

The result is the emission from the fraction of the stellar photosphere not covered by accretion shocks (F$_{\rm{Phot,\ Obs.}}$) that contributes towards the observed CTTS spectrum for that particular visit. We can then fit for the excess emission due to accretion (see below) and divide F$_{\rm{Phot,\ Obs.}}$ by the fraction of the stellar surface not covered by hotspots to obtain F$_{\rm{Phot,\ True}}$ that can be used for fitting. This results in multiple photospheric templates for each target (with the exception of RU~Lup), one for each visit with contemporaneous V-band observation. For TW~Hya, we obtain 21 templates, 15 for BP~Tau, and 14 for GM~Aur.

\begin{equation} \label{eq: Fphot}
    \mathrm{F}_{\rm{Phot,\ True}} = \mathrm{F}_{\rm{WTTS}}\left(\frac{V_{\rm{CTTS}}}{V_{\rm{WTTS}}}\right)\left(\frac{1}{1+r_V}\right)\left(\frac{1}{1-\sum f_i}\right)
\end{equation}

In the case of RU~Lup, we are unable to estimate r$_V$ for any visit, owing to the numerous, strong emission lines in its optical spectra \citep[see Paper III,][]{Gahm2008}. We therefore cannot estimate F$_{\rm{Phot,\ True}}$ as above. Instead, we assume F$_{\rm{Phot,\ True}}$ is the raw $HST$ spectrum of HBC~427 (F$_{\rm{WTTS}}$) scaled to the distance and radius of RU~Lup via Equation \ref{eq: RU Lup Phot}. We discuss this further in Appendix \ref{appendix: RU Lup}.

\begin{equation} \label{eq: RU Lup Phot}
    F_{\rm{Phot,\ True}}=\mathrm{F}_{\rm{WTTS}}\left(\frac{d_{\rm{WTTS}}R_{\rm{CTTS}}}{d_{\rm{CTTS}}R_{\rm{WTTS}}}\right)^2
\end{equation}.

To reproduce the excess emission, here we follow \citet{Ingleby2013} and use multiple accretion columns. We employ a four-column model with energy fluxes, ${\mathcal{F}}=1/2\rho v_s^3$, of 10$^{10}$ (low), 10$^{11}$ (medium), 10$^{12}$ (high), and 3$\times$10$^{12}$ (very~high) erg s$^{-1}$ cm$^{-2}$. \citetalias{RE19} note that the inclusion of this fourth, very high density column was necessary for several observations of DM~Tau and GM~Aur to fit the spectrum near 1800 {\AA}. Here we use four columns for all of our fitting for consistency. The energy flux (${\mathcal{F}}$) depends on $\rho$, the density of material in the accretion column, and $v_s$, the infall velocity. The infall velocity is set to be the free-fall velocity at the magnetospheric truncation radius (\rin) of 5 stellar radii. We note that \rin has been measured in TW~Hya (3.5 \rstar, \citet{Gravity2020} and 8.3--12.4 \rstar depending on assumed magnetic field strength \citealt{Gravity2023}) and RU~Lup \citep[3.3--6.6 \rstar depending on assumed magnetic field strength][]{Gravity2023}. We opt to use 5 \rstar for all objects due to the uncertainties in these estimates and for consistency; the free fall velocity changes little for any choice of \rin within the measured uncertainties.

We use filling factors ($f_{i}$) to scale the resulting emission and these are left as free parameters. These four filling factors ($f_{low}$, $f_{medium}$, $f_{high}$, $f_{very~high}$) measure the fraction of the visible stellar surface that is covered by the four accretion columns with the energy fluxes noted above. We calculate \mdot by adding the contributions of the four accretion columns with
\begin{equation}
\dot{M}=\frac{8\pi R_{\ast}^2}{v_s^2} \sum_i \mathcal{F}_i f_i =
\frac{8\pi R_{\ast}^2}{v_s^2} \mathcal{F}_{\rm{weighted}} f_{\rm{total}}~~. 
\end{equation}
One can additionally calculate the fractional accretion rate ($\dot{m}_k$) for each column as the fraction of the total mass accretion rate contributed by that individual column as
\begin{equation} 
    \dot{m}_k = \frac{f_k\mathcal{F}_k}{\sum_{i} f_i\mathcal{F}_i},
\end{equation} where $f$ is the filling factor and $\mathcal{F}$ is the associated energy flux.

We re-sample our COS spectra to a lower resolution (R$\sim$1500, similar to that of $HST$ STIS G230L grating used in \citealt{RE19}) and mask notable lines (e.g. Mg II: 2800 {\AA}). We also remove data below $\sim$1800 {\AA} due to the abundance of emission lines and added uncertainty in estimating the continuum. Including some optical emission is important to constrain the contribution from lower density columns, which can cover a large fraction of the star and contribute significantly to the total mass accretion. While \citetalias{RE19} use $HST$ STIS NUV-optical spectra, which extend continuously from $\sim$1800--5700 {\AA}, our $HST$ COS spectra extend to just 3200 {\AA} and so do not span any optical wavelengths. Thus, we combine our COS spectra with contemporaneous (within 6 hours) $BgV$ photometry (see Paper II). We do not include anything redder than $V$ (such as $ri$) since we truncate our model at 5700 {\AA}. Visits without contemporaneous photometry are still modeled in the same manner and are noted in Tables \ref{tab: Shock Model Results, TW Hya/RU Lup} and \ref{tab: Shock Model Results, BP Tau/GM Aur}, but should be considered carefully and we discuss the reliability of these results in Appendix \ref{appendix: No Photometry Fits}.

Based on modeling that has shown that accretion hotspots can cover up to $\sim$40\% of the stellar surface \citep{Romanova2003, Ingleby2013, Ingleby2015, RE19}, we allow individual $f_{i}$ to vary between 0 and 40\% and limit the total surface coverage of all columns ($\Sigma f_i$) to 50\%. The accretion column filling factors are fit using a Markov Chain Monte Carlo \citep[$emcee$,][]{emcee} technique. 

Every $HST$ visit for a given target is fit with each template for that target. For example, each of the 24 visits of BP~Tau was fit using each of the 15 templates. The final value for a given parameter in a single fit is the median of the posterior distribution, and the upper/lower uncertainties are the difference between the 84th/16th percentile and the median ($\pm$1-$\sigma$). Finally, these values and uncertainties are averaged across all templates, resulting in the values listed in Tables \ref{tab: Shock Model Results, TW Hya/RU Lup}--\ref{tab: Shock Model Results, BP Tau/GM Aur}.

\begin{figure*}
    \centering
    \includegraphics[width=0.975\textwidth]{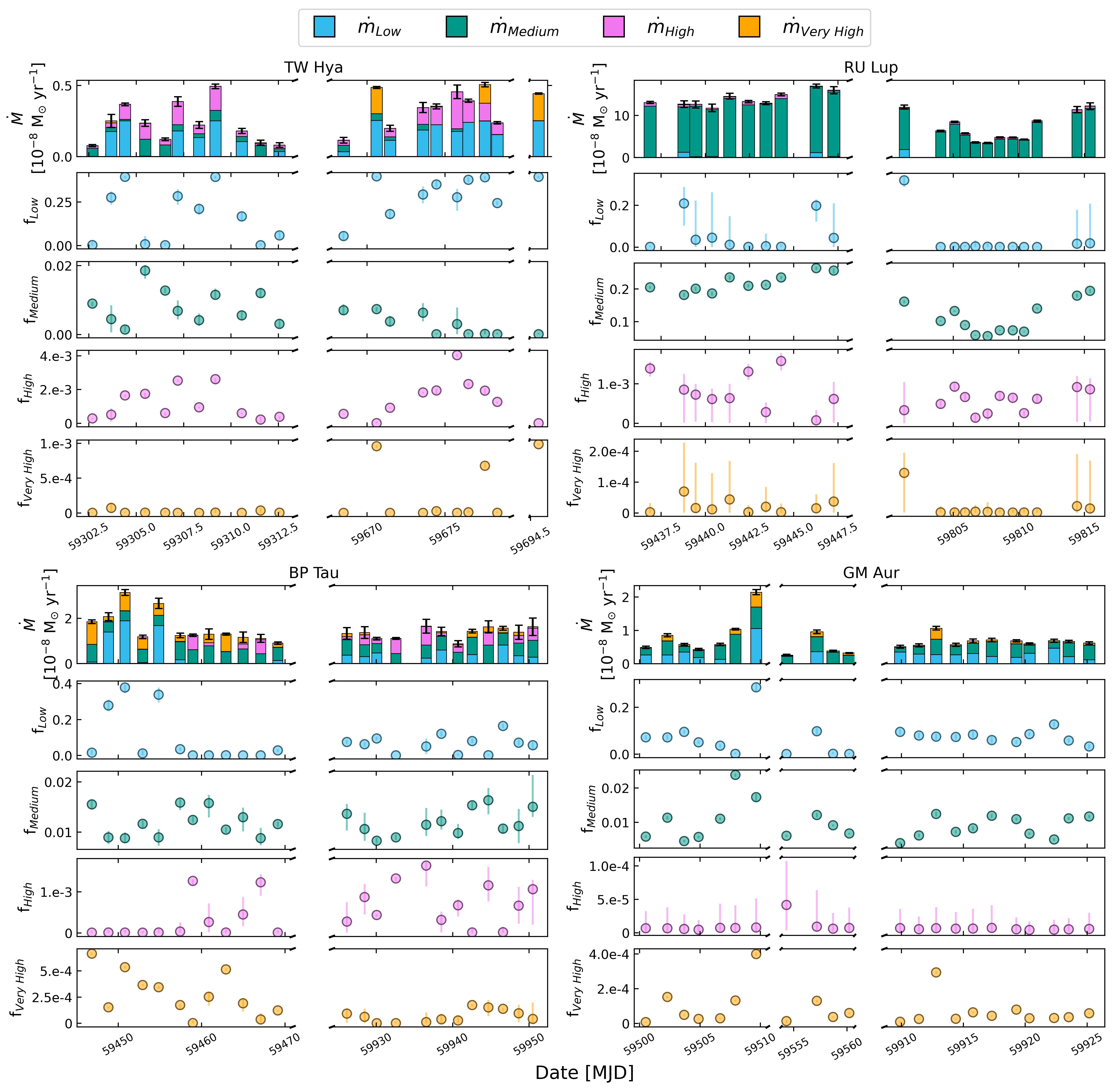}
    \caption{Results of our shock model fitting for TW~Hya (top left), RU~Lup (top right), BP~Tau (bottom left), and GM~Aur (bottom right). The top row of each panel shows the mass accretion rate: the total height of the bar represents the total accretion rate while the colored bars show the accretion contributed by each respective accretion column. The errorbar atop each bar represents the uncertainty on the total accretion rate. The subsequent rows are the filling factors of the low, medium, high, and very high density accretion columns. The left and right columns of each panel show Epochs 1 and 2. Note the additional break in the Date axes for TW~Hya E2 and GM~Aur E1 due to the uneven sampling from $HST$.}
    \label{fig: Shock Model Results All}
\end{figure*}

Figure \ref{fig: Shock Model Results All} shows the results of our shock modeling, including accretion rates, filling factors ($f$), and fractional accretion rates ($\dot{m}$) of the four accretion columns. Individual fits for each $HST$ visit can be found in Figures \ref{fig: Shock Model Fits - TW Hya - E1}--\ref{fig: Shock Model Fits - GM Aur - E2} in Appendix \ref{appendix: Shock Model}. In most cases there are anti-correlations in the posterior distributions of adjacent accretion columns (f$_{\mathrm{Low}}$/f$_{\mathrm{Medium}}$, f$_{\mathrm{Medium}}$/f$_{\mathrm{High}}$, and f$_{\mathrm{High}}$/f$_{\mathrm{Very\ High}}$) resulting from degeneracies between the models. Some degeneracy is expected, given that the emission from like-density accretion columns is similar. We discuss the details of the fits for each individual object in the sections below.

\subsubsection{Modeling Results for TW~Hya} 

TW~Hya's accretion is primarily driven by the low density accretion column, which generally covers 20--40\% of the stellar surface and contributes 40--70\% of the total mass accretion. In some cases, the medium column becomes significant (often when f$_{\mathrm{Low}}$ is very low, likely due to their degeneracy) and can contribute up to 90\% of the mass accretion, though it typically covers $\sim$1\% of the surface and contributes 0--50\% of the mass accretion. The high density column consistently contributes 10--50\% of the mass accretion, while the very high density column is generally insignificant in TW~Hya, except for two visits of high \mdot in E2.

While the total mass accretion is typically dominated by the lower density columns, the variability in the accretion appears to be driven by the high density columns. The Spearmann $r$ coefficient between $\dot{m}_{\mathrm{Low+Medium}}$ and \mdot is -0.74, while it is +0.74 between $\dot{m}_{\mathrm{High+Very\ High}}$, suggesting that changes in the accretion are driven by higher density, hotter accretion flows. For reference, $r=1$ implies a perfect positive correlation, $r=-1$ implies a perfectly negative correlation, and $r=0$ implies no correlation.

Overall, the accretion rate (top left panel, uppermost row, Figure \ref{fig: Shock Model Results All}) varies between 0.08--0.51\accrateunits peak-to-peak across both epochs, a factor of 6.5 difference (0.80 dex), with a median accretion rate of 0.25\accrateunits. The median accretion parameters and variability are similar in E1 and E2, with TW~Hya accreting more strongly in E2 (0.23 vs 0.35\accrateunits).

\subsubsection{Modeling Results for RU~Lup}

RU~Lup (top right panel of Figure \ref{fig: Shock Model Results All}) is the strongest accretor in our sample, with accretion rates between 11.8--17.0\accrateunits in E1, with a median of 13.2\accrateunits. The accretion rate drops considerably from E1 to E2, with rates between 3.5--12.2\accrateunits in E2 and a median of 6.0\accrateunits. The peak-to-peak variability in RU~Lup is a factor of 4.9, or 0.69 dex.

The accretion is dominated by the medium density column in both E1 and E2. In E1, it covers 18--26\% of the stellar surfaces and contributes 85--98\% of the total mass accretion rate. In E2, it covers 6.9--19.4\% of the surface and contributes 80--97\% to \mdot. The decrease in \mdot from E1 to E2 is driven entirely by the decrease in f$_{\mathrm{Medium}}$. No other accretion column contributes more than 16\% of the accretion in any visit. Combined, $\dot{m}_{\mathrm{Low}}$, $\dot{m}_{\mathrm{High}}$, and $\dot{m}_{\mathrm{Very\ High}}$ contribute just 4.6\% of the mass accretion on average.

\subsubsection{Modeling Results for BP~Tau}

BP~Tau is a moderate accretor, with accretion rates between 0.9--3.2\accrateunits in E1 and 0.9--1.7\accrateunits in E2, and median accretion rates of 1.3 and 1.4\accrateunits, respectively. Overall, the accretion rate varies by a factor of 3.6 (0.55 dex) peak-to-peak. 

The medium density column typically dominates the accretion in BP~Tau, covering 1--2\% of the stellar surface and accounting for 40--60\% of the mass accretion. That said, each column, in at least one visit each, contributes $>$50\% to \mdot in that visit, suggesting that all columns are important. This is reinforced by the fact that the $r$ correlation coefficient does not exceed 0.4 for the lower/higher columns in either E1 or E2, showing that no one column consistently dominates the total accretion rate. 

From E1 to E2, the accretion appears to become cooler/less dense. The low and high density columns, on average, contribute more to the accretion while the very high density column diminishes. This can be qualitatively observed in Figures \ref{fig: Shock Model Fits - BP Tau - E1} and \ref{fig: Shock Model Fits - BP Tau - E2} in Appendix \ref{appendix: Shock Model Fits}, where the slope of the BP~Tau spectra in E1 appears flatter than E2, implying a hotter/denser accretion flow in E1.

\subsubsection{Modeling Results for GM~Aur}

GM~Aur is dominated by low and medium density accretion columns which, save for the accretion burst near MJD=59509 discussed in more detail below, together fill up to 15\% of the stellar surface and typically constitute 90\% of the total mass accretion. The high density accretion is largely insignificant, generally contributing $<$1\% to the total accretion, with visit 1.7 being the only exception, when it jumps to 6.0\%. While the very high density accretion column only contributes $\sim$10\% to \mdot on average, it is highly correlated with \mdot, with $r$ values of 0.56 and 0.93 in E1 and E2, respectively. This is similar to TW~Hya, where the bulk of the accretion comes from lower density flows, but the variability is primarily driven by higher density, hotter flows.

The accretion varies between 0.32--2.15\accrateunits in E1 (which includes a burst) and 0.5--1.1\accrateunits in E2. The median accretion rates are 0.58\accrateunits and 0.67\accrateunits. Most of the variability is driven by small bursts, with up to 4 seen in our $HST$ observations at MJD$\sim$59502, 59509, 59557, and 59912. With the exception of the burst at MJD=59509, they persist across only one $HST$ visit (no more than 2 days) and increase \mdot by about 0.5\accrateunits or a factor of 2. 

$HST$ captured the beginning of the larger burst at MJD=59509 in at least 2 visits, but due to synchronization problems with the command flow went into safe mode soon after and was unable to capture the end of the burst. The accretion rate increased from a quiescent rate of $\sim$0.43--2.15\accrateunits within 4 days, a factor of 5.0 (0.70 dex) increase. This burst (and the smaller ones discussed above) was driven by the very high density accretion column, for which $\dot{m}_{\mathrm{Very\ High}}$ increased by about a factor of 3.5 over that time.

\subsection{UV Spectral Lines and Features} \label{sec: UV Luminosities}

\begin{deluxetable}{c c c}[htp]
\setlength{\tabcolsep}{10pt}
\tablecaption{UV Features \label{tab: UV Lines}}
\centering
\tablehead{
\colhead{$\ \ \ \ \ $Line$\ \ \ \ \ $} & \colhead{$\ \ \ \ \ \ \lambda_0\ \ \ \ \ \ $} & \colhead{$\ \ \ \ $Approx.$\ \ \ \ $} \\
\colhead{} & \colhead{[\AA]} & \colhead{Width [\AA]}
}
\startdata
Mg~{\sc ii}$^*$ & 2798 & 20 \\
Al~{\sc iii]} & 2670 & 10 \\
C~{\sc iii]} & 1909 & 10 \\
Si~{\sc iii]} & 1892 & 5 \\
Si~{\sc ii} & 1808 & 10 \\
O~{\sc iii]} & 1666 & 5 \\
He~{\sc ii} & 1640 & 5 \\
C~{\sc iv}$^{*\dagger}$ & 1548 & 8 \\
C~{\sc i}$^{\dagger}$ & 1463 & 2 \\
Si~{\sc iv}$^{\dagger}$ & 1403 & 4 \\
Si~{\sc iii}$^{\dagger}$ & 1399 & 2 \\
\hh Bump & 1600 & 200 \\
\enddata
\tablenotetext{*}{We do not separate the doublet components of the Mg~{\sc ii} and C~{\sc iv} lines. Central wavelengths noted here are the short wavelength component of the doublet.}
\tablenotetext{\dagger}{These lines are blended with \hh lines \citep{France2012}.}
\end{deluxetable}

UV line luminosities have previously been found to correlate with mass accretion \citep{Calvet2004, Yang2012, Ingleby2013, Ardila2013, RE19}. Additionally, a broad region near 1600 {\AA} dubbed the ``H$_2$ bump,'' \citep[e.g.,][]{Herczeg2004, Bergin2004} has also been linked with accretion \citep[see][]{Yang2012, Ingleby2009, Ingleby2012, France2017, Thanathibodee2018, Espaillat2019, France2023}.  Here we measure the luminosities of selected UV lines, the FUV and NUV continua, and the H$_2$ bump. 

\subsubsection{UV Line Luminosities}

While these COS spectra are filled with many emission lines, we focus on 11 of the brightest lines previously linked with accretion (see Table \ref{tab: UV Lines}). Note that many of these lines are blended and only approximate line centers are provided in Table \ref{tab: UV Lines}. Additionally, Mg~{\sc ii} is affected by wind absorption \citep{Ardila2002, Xu2021}. To measure the line luminosities, we first subtract the continuum from the dereddened spectra. Within $\sim$50 {\AA} of the line center, we mask outlying points above or below the median flux of the spectrum, where the outlier threshold vary per line and per object. We then fit a continuum to the remaining data using the \textit{specutils} function \textit{fit\_generic\_continuum} and subtract this from the original spectrum. We then integrate the flux of the continuum-subtracted spectrum within a narrow region that varies by line. Finally, we convert these fluxes to luminosities. 

\begin{figure*}[ht]
    \centering
    \includegraphics[width=0.975\textwidth]{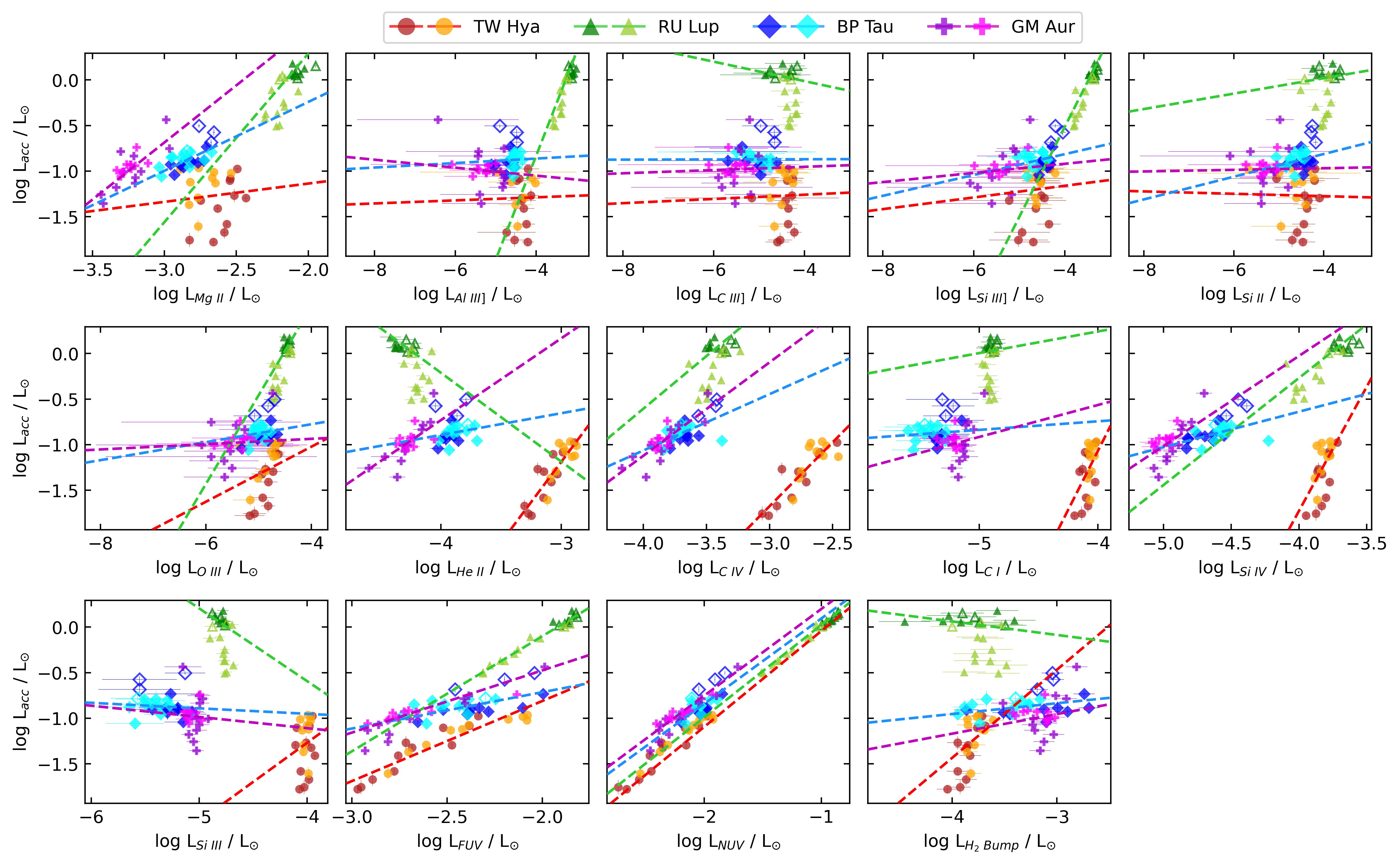}
    \caption{Accretion luminosity vs luminosity for various UV lines/regions. Red/orange circles, green/lime triangles, blue/cyan diamonds, and purple/pink pluses are TW~Hya, RU~Lup, BP~Tau, GM~Aur, respectively. Open points are visits without contemporaneous photometry. Colored, dashed lines are the intra-object log-log linear fits for all observations of each target (see Table \ref{tab: UV Correlations}). For all UV features besides the \hh~Bump, we exclude data with SNR$<$2. For the \hh~Bump, we only exclude data where the bump was not visually detected, independent of SNR.}
    \label{fig: UV Luminosities}  
\end{figure*}

Figure \ref{fig: UV Luminosities} shows our UV luminosities along with log-log linear relationships with \lacc (Equation \ref{eq: Lacc}) derived here. We exclude non-detections (here defined as SNR$<$2) in Figure \ref{fig: UV Luminosities} and in the associated linear fits. UV luminosities, fit coefficients, and associated uncertainties are shown in Tables \ref{tab: UV Luminosities, TW Hya}--\ref{tab: UV Luminosities, GM Aur} and Table \ref{tab: UV Correlations}.

\begin{equation} \label{eq: Lacc}
    L_{acc} = \frac{GM_*\dot{M}}{R_*}(1-\frac{1}{R_i}),  R_i=5 R_*
\end{equation}

Looking at all objects together, only three lines exhibit moderate correlations to \lacc: Mg {\sc ii} ($r$=0.66), Al {\sc iii]} ($r$=0.55), and Si {\sc iii]} ($r$=0.61), though Si {\sc iii]} is often only marginally detected, despite this global correlation. The other lines show weak/no correlation and are often skewed by either TW~Hya and/or RU~Lup. \lacc in TW~Hya is comparable to BP~Tau and GM~Aur, but in many cases it exhibits much brighter emission lines. \lacc is much higher in RU~Lup, but its lines are often significantly dimmer than would be expected given its accretion rate (particularly He {\sc ii}, C {\sc iv}), rendering those global correlations very weak.

If instead we focus on the inter-object trends, many object-line pairs show strong correlations that differ from the global trend or exist when a global trend does not. TW~Hya shows weak to moderate correlations in all lines except He~{\sc ii} ($r$=0.85) and C~{\sc iv} ($r$=0.86), which are strongly correlated to \lacc. In RU~Lup, some lines (Al~{\sc iii]}, Si~{\sc iii]}, Si~{\sc iv}) show strong correlations to \lacc, but others show no or even anti-correlation. He~{\sc ii} is actually moderately anti-correlated with \lacc ($r$=$-$0.49) and is exceptionally dim relative to other lines and targets. BP~Tau in all cases shows weak correlations between \lacc and \lline, though Mg~{\sc ii} is moderately correlated ($r$=0.59). The UV lines in GM~Aur do not show any particularly strong correlations with \lacc, with Mg~{\sc ii}, He~{\sc ii} and C~{\sc iv} showing the strongest correlations at $r$=0.75, 0.67, and 0.81, respectively. The others are weakly correlated or have large uncertainties due to so many visits being discarded as non-detections (e.g., Al~{\sc iii]}, C~{\sc iii}, Si~{\sc iii]}, O~{\sc iii]}).

\subsubsection{FUV and NUV Continuum Luminosities}

For the FUV and NUV fluxes, we integrate the flux of the fitted, dereddened continuum spectra between 1400--2100 and 2500--3200 {\AA}, respectively. The FUV and NUV luminosities and fit coefficients are listed in Tables \ref{tab: UV Luminosities, TW Hya}--\ref{tab: UV Luminosities, GM Aur} in the Appendix and in Figure \ref{fig: UV Luminosities}.

TW~Hya shows a wide range of FUV luminosities, ranging between 0.005--0.040 \lstar in both epochs, with a median of 0.014 \lstar (0.003 \lsun). The NUV varies between 0.009--0.056 \lstar, with a median of 0.028 \lstar (0.006 \lsun). RU~Lup is generally brighter than TW~Hya, and varies between 0.008--0.026 \lstar (median 0.020 \lstar, 0.015 \lsun) in the FUV and 0.052--0.238 \lstar (median 0.169 \lstar, 0.124 \lsun) in the NUV. BP~Tau shows the least degree of continuum variability, varying between 0.002--0.011 \lstar (median 0.004 \lstar, 0.003 \lsun) in the FUV and 0.007--0.016 \lstar (median 0.010 \lstar, 0.008 \lsun) in the NUV.  L$_{FUV}$/L$_{NUV}$ is on average higher in E1 (0.52) than E2 (0.39), reinforcing the denser flow in E1 predicted by the shock model. GM~Aur is generally the dimmest amongst the four targets, ranging between 0.0009--0.008 \lstar (median 0.0014 \lstar, 0.002 \lsun) in the FUV and 0.003--0.010 \lstar (median 0.005 \lstar, 0.006 \lsun) in the NUV.

The FUV/NUV luminosities also show stronger time variability than the accretion rates, readily varying by a factor of 2 on day-long timescales. TW~Hya exhibits a factor of $\sim$5.1 increase in FUV flux within just 2 days, from 7.3$\times$10$^{-3}$ \lstar in Visit 2.1 to 37.2$\times$10$^{-3}$ \lstar in Visit 2.2. The FUV and NUV continua rise considerably in GM~Aur during the burst near MJD=59509, by a factor of 8.5 and 4.7, respectively, in about 4 days.

The FUV/NUV continuum luminosities show far stronger correlations with \lacc than any of the lines, though some of the same trends persist. Globally, both L$_{FUV}$ and L$_{NUV}$ are strongly correlated with \lacc, with $r$=0.85 and 0.96, respectively. In the FUV, RU~Lup again skews the fit, having either elevated accretion rates or diminished FUV luminosities. In the NUV all 4 targets exhibit similar slopes, though the y-intercept of the fitted line varies by target. Individually, each target (with the exception of BP~Tau) shows fairly strong correlations with \lacc: r$\ge$0.88 in all cases. BP~Tau is moderately correlated (r$_{FUV/NUV}$=0.55/0.66), but shows the same linear slope as the other targets in the NUV.

\subsubsection{\texorpdfstring{H$_2$}{H2} Bump Luminosities}  

For the \hh~Bump near 1600 {\AA}, we follow a similar method as \citet{France2017} to extract the bump spectrum and calculate its luminosity. We first hand-pick $\sim$110 points between 1400--1800 {\AA} selected to avoid strong atomic and molecular lines. The lines present in each target vary, and so we change the selected points for each target. From these $\sim$110 points, we define a binned spectrum whose flux and uncertainty are the mean and standard deviation of the original spectrum within 0.75 {\AA} of the selected points. This binned spectrum is thus a pseudo-continuum, composed of the true, underlying continuum plus the broad \hh~Bump feature and so does not include bright emission lines.

From this binned spectrum, we fit a continuum using a second-order polynomial using the points away from the \hh~Bump, between 1405--1490 {\AA}~and 1690--1800 {\AA}. Next we subtract this continuum from the binned spectrum. Finally, we integrate this continuum-subtracted binned spectrum between 1490--1690 {\AA} to calculate the \hh~Bump flux and convert to luminosity. The \hh~Bump luminosities are shown in Tables \ref{tab: UV Luminosities, TW Hya}--\ref{tab: UV Luminosities, GM Aur} and Figure \ref{fig: UV Luminosities} along with fit coefficients given in the Appendix. For these correlations, we exclude data where the bump was not visually distinct from the underlying continuum, regardless of the calculated flux or SNR.

Note that while the measured uncertainties on the \hh~Bump luminosities are low (typically $<$5\%), the true uncertainties are likely higher. The determination of the continuum in each spectrum not only depends heavily on the number and location of the chosen binned continuum points, but can also be affected by any spurious emission lines. For this reason, we apply an additional 10\% uncertainty added in quadrature to the measured uncertainty for TW~Hya, BP~Tau, and GM~Aur. RU~Lup exhibits uniquely strong, wide, and numerous emission/absorption features, making its continuum particularly uncertain, so we apply a 20\% uncertainty.

The \hh~Bump shows similar trends as the other lines. TW~Hya is the only target that shows a moderately positive correlation ($r$=0.54). Here, it is not an outlier and shows very similar bump luminosities as BP~Tau and GM~Aur. RU~Lup shows a very weakly negative correlation and possesses much lower bump luminosities than the other targets with comparatively high \lacc, though in most cases the bump is not detected. Even ignoring Visit 1.7, with a particularly low bump luminosity, there is still only a weakly positive correlation in RU~Lup. BP~Tau shows a wide range of bump luminosities ($\sim$1 dex) with no correlation to \lacc, similar to many of the emission lines. The \hh~Bump in GM~Aur shows a very weakly positive correlation with \lacc, with far more spread in \lacc than L$_{H2\ Bump}$. Globally, we see a weakly negative correlation between \lacc and L$_{H2\ Bump}$.

\section{Discussion} \label{sec: Discussion}

Our shock model fitting finds diversity in accretion column structures between objects, but these structures remain roughly consistent for individual objects over the course of at least a year (i.e., the longest timescale probed in this work) while the accretion rate changes on short timescales of about a day. In TW~Hya, the shock model predicts a large hotspot dominated by the low-density accretion column while in RU~Lup the accretion is dominated by the medium density accretion column. GM~Aur is primarily dominated by the low and medium density columns, while BP~Tau is marginally dominated by the medium density column but sees strong contributions from the other columns at various times.  Besides the fraction of the star covered by each column, there are no significant changes in these structures during the course of the $HST$ observations, even when the accretion rate itself changes.  Despite this relative stability in the accretion structure, there is variability in the accretion rate up to a factor of four within two days.  This is consistent with previous reports that accretion is variable on all timescales probed  \citep{Costigan2014, Hartmann2016, Zsidi2022, Herczeg2023}.  Specifically, \citet{Venuti2014} find that accretion variability is typically about 0.5 dex on timescales of days-weeks and our results are consistent with this. Our accretion rates are also within the scatter seen in accretion rate in larger samples \citep[e.g.,][]{Manara2021}.  We see an average total surface coverage of about 15\%, in line with magnetospheric accretion models that suggest total filling factors of about 20\% \citep{Zhu2023a}.

TW~Hya and GM~Aur have been studied by other researchers with datasets overlapping the timing of the $HST$ observations presented here, allowing us to directly compare our results. \citet{Herczeg2023} find that in over 25 yrs of TW~Hya optical spectroscopic data, with some that overlap the times of our dataset, that the average accretion rate is 0.25\accrateunits, which flickers over timescales of hours but is roughly stable over 25 yrs and varies between 0.05 to 0.9\accrateunits. Our results are consistent and we find that the average accretion rate of TW~Hya is about 0.3\accrateunits with a range of $\sim$0.1--0.5\accrateunits. GM~Aur was observed by \citet{Bouvier2023} who performed an optical and NIR spectrophotometric monitoring campaign overlapping some of the $HST$ observations, probing timescales of days to months over a period of six months and covering 30 rotation periods.  \citet{Bouvier2023} find evidence of rotational modulation of accretion and a stable accretion funnel flow and accretion shock.  Our results are consistent with this study.

Some UV lines, such as the C~{\sc iv} line, have been seen to correlate with accretion rate in larger samples \citep[e.g.,][]{Calvet2004, Yang2012, Ardila2013, Ingleby2013, RE19}, but here we caution against using these lines alone to characterize the accretion rate. Like \citetalias{RE19} and \citet{Calvet2004}, we do generally find a moderate, positive correlation between accretion rate and UV line luminosities. That said, certain targets (i.e., TW~Hya, RU~Lup) often deviate from these patterns. For some lines, the relationship between L$_{line}$ and \lacc in TW~Hya is consistent with the other targets, but in others it is completely removed from the other targets and the relationship described in \citetalias{RE19}, and is self-consistent between E1 and E2. In these cases, either the accretion rates are lower than would be expected or the UV luminosity is higher than expected. As for RU~Lup, in only some cases (Mg~{\sc ii}, Al~{\sc iii]}, Si~{\sc iii]}, Si~{\sc iv}) does it trend with other targets and previous studies. Even then, RU~Lup exhibits a much steeper relationship than either BP~Tau, GM~Aur, or the empirical relationships of \citet{Calvet2004}, \citet{Yang2012}, or \citetalias{RE19}. The bright He~{\sc ii} line is interesting as it shows moderate anti-correlation and is rather dim when considering that RU~Lup is comparatively bright in most lines and in \lacc. TW~Hya and RU~Lup are outliers and this may be linked to their viewing geometry, which are both seen closer to face-on (Table~\ref{tab: Targets}). Ultimately, while moderate correlations between \lacc and L$_{UV}$ are generally found, we emphasize not only that there is considerable scatter in these relationships, but that the variability trends in individual targets often differ from the global trend. They should not be used to estimate accretion rates, at least on their own. 

In the majority of cases, the correlation between observed flux (line or continuum) and accretion rate is positive. As more energy is available from a larger accretion rate, more power is radiated in the continuum and the line. If the structure of the accretion shock does not change with accretion rate, one might expect that the observed flux and accretion rate are directly proportional. However, in RU~Lup, some line fluxes are observed to be anti-correlated with the accretion rate. Detailed multi-dimensional modeling with radiative transfer is beyond the scope of this work, but here we speculate on how this fits into our knowledge of accretion shock emission. Decreasing fluxes with increasing accretion luminosity are clearly seen in He~{\sc ii} and Si~{\sc iii} and less clearly in the \hh Bump and C~{\sc iii]}. Thus, the conditions present in the medium density accretion column (temperature, density, physical size, etc.) may promote a high optical depth for these lines. When the accretion rate increases in RU~Lup, so does the size of the medium density column, which may in turn increase the optical depth in these moderate temperature lines, resulting in the negative relationships we see. Other lines, formed at lower (C~{\sc i}, Mg~{\sc ii}) and higher (C~{\sc iv}) temperatures than the Si~{\sc iii} and He~{\sc ii} lines would form in physically distinct locations and different conditions and may not be subject to same optical depth effects as Si~{\sc iii} and He~{\sc ii}.

We also compare our results on the \hh~Bump to those in the literature. The \hh~Bump has been suggested to be driven by collisional excitation of \hh via X-ray photons \citep{Bergin2004}, Ly$\alpha$ dissociation of H$_2$O in the inner disk \citep{France2017}, linked to the surface density of the inner disk \citep{Espaillat2019}, or due to the H~{\sc i} 2-photon continuum \citep{Bottorff2006}. \citet{Espaillat2019} found that there was no correlation between L$_{H2\ Bump}$ and contemporaneous X-ray data and \citet{France2017} show that the \hh Bump spectrum itself is inconsistent with electron-impact excitation, ruling out the first scenario. For the latter two scenarios, there is an expected correlation between the \hh~Bump luminosity and L$_{acc}$, since accretion should produce Ly$\alpha$ photons \citep[e.g.,][]{Alencar2012}. \citet{France2023} find strong correlations between the \hh~Bump and both Ly$\alpha$ and \emph{WISE} W3-W4 colors, which they suggest reinforces that the \hh~Bump originates in the inner disk, where Ly$\alpha$ photons can reach and dissociate the H$_2$O content of the inner disk. \citet{Espaillat2019} suggest that the \hh~Bump is driven by changes in the surface density in the inner disk that propagate through the accretion column and lead to a change in the accretion rate. They find a strong correlation ($r$=0.7) in a sample of 7 CTTSs (27 observations). Looking at all four targets in our sample together (89 observations), we do not recover a strong (or even positive) correlation ($r$=-0.42) between the \hh~Bump and L$_{acc}$. Individually, the only target that shows even moderate correlation is TW~Hya ($r$=0.54). The other 3 targets show very weak correlations ($r\le0.31$). Therefore, we do not find strong evidence supporting that the \hh~Bump is due to Ly$\alpha$ dissociation of H$_2$O or surface density changes.  

Lastly, the FUV luminosity can change by a factor of 2--5 on day timescales and correlates well with accretion rate. While BP~Tau shows comparatively little variability, the other 3 targets exhibit strong FUV/NUV continuum variability, about a factor of 10 peak-to-peak. The median FUV luminosity in our sample is 2.0$\times10^{31}$ erg s$^{-1}$, which is what is usually used in FUV photoevaporation models \citep[e.g.,][]{Gorti2009, Pascucci2023}.  FUV radiation has also been shown to affect disk chemistry near planet-forming radii \citep{Bergin2004, Yang2012, Adamkovics2016}. The FUV variability seen here implies that accretion-driven FUV radiation fields can reach levels 10$\times$ those assumed in \citet{Bergin2004}, showing that accretion variability can drive disk chemistry dynamics.

\section{Summary} \label{sec: Conclusion}

We studied a multi-epoch $HST$ UV spectral monitoring campaign of four CTTSs: TW~Hya, RU~Lup, BP~Tau, and GM~Aur. We use a magnetospheric accretion shock model to estimate the accretion rates and accretion column filling factors for roughly 24 observations of each target. From those same spectra we measure luminosities of various UV lines, the 1600 {\AA} \hh~Bump, and the FUV/NUV continuum. Our main findings are as follows:

\begin{enumerate}
\itemsep0em 
    \item Our accretion shock modeling reveals an array of accretion structures in our four targets. Accretion in TW~Hya is primarily driven by the low density column, but some visits see strong contribution from other columns. RU~Lup is dominated by medium density accretion columns, with little accretion coming from other columns. BP~Tau sees roughly equal contribution from all four columns, with a preference for the low and medium density columns. GM~Aur is dominated by the low and medium density columns, with some contribution from the very high density column when the accretion rate increases.
    \item All four targets show accretion variability of at least a factor of 3 (0.48 dex) during our monitoring. TW~Hya and GM~Aur exhibit variability in their accretion rate of a factor of 3 or greater within two days. RU~Lup, on the other hand, shows a sustained decrease of a factor of 2.2 between 2021 and 2022, showing that CTTSs can undergo a sustained decrease in accretion rate on timescales of over one year. BP~Tau exhibited the least accretion variability in our sample, varying by about a factor of 3.6 peak-to-peak across all observations.
    \item UV line luminosities are generally only moderately correlated with accretion rate, but TW~Hya, RU~Lup, and BP~Tau often show scatter both with respect to other targets in the sample and within a single object. In TW~Hya, the UV line luminosities are higher than expected given its low accretion rate. RU~Lup is very inconsistent, showing no, strongly positive, and weakly negative correlations with accretion. We suggest this is due to its strong accretion-driven outflow and/or near face-on inclination. BP~Tau generally shows little correlation to UV line luminosities; the accretion rate tends to vary significantly less than the line luminosities.
    \item FUV and NUV continuum luminosities show strong correlations to accretion rate. TW~Hya is again an outlier, with low accretion rates and high UV luminosities. Together, RU~Lup, BP~Tau, and GM~Aur trend closely with one another. In general, the FUV/NUV continua are even more variable than the overall accretion rate, with both varying by factors of 2--5 within 1--2 days in all four targets.
    \item We find generally comparable FUV luminosities to those used in previous works to estimate the effect of FUV radiation on disk dynamics, but note that accretion induced variability can exceed this by a factor of $\sim$10 in some cases.
    \item With the exception of TW~Hya, we do not find strong correlations between accretion luminosity and the 1600 {\AA} \hh~Bump. 
\end{enumerate}

Our study further demonstrates that accretion in CTTSs is highly variable and reinforces the importance of simultaneity in future multi-instrument/multi-wavelength observations of CTTSs. Additionally, we show that each star in our sample exhibits unique relationships between accretion rate and other UV accretion tracers. These empirical relationships should not be utilized when studying an individual CTTSs and should be used cautiously for small samples.

\begin{acknowledgments}
This work is supported by $HST$ AR-16129 from the Space Telescope Science Institute, which is operated by AURA, Inc. This material is based upon work supported by the National Science Foundation under Grant Number AST-2108446. This work benefited from discussions with the ODYSSEUS team (\url{https://sites.bu.edu/odysseus/}); see \cite{Espaillat2022} for an overview of the ODYSSEUS survey. This work is supported by the ANID FONDECYT Postdoctoral program No. 3220029. We acknowledge support by ANID, -- Millennium Science Initiative Program -- NCN19\_171. This work has been supported by the project PRIN‐INAF 2019 “Spectroscopically Tracing the Disk Dispersal Evolution (STRADE)” and by the INAF Large Grant 2022 “YSOs Outflow, Disks and Accretion (YODA)”. This work was also supported by the NKFIH excellence grant TKP2021-NKTA-64. IM is funded by grant PID2022-138366NA-I00, by the Spanish Ministry of Science and Innovation/State Agency of Research MCIN/AEI/10.13039/501100011033 and the European Union, and by a Ram\'on y Cajal fellowship RyC2019-026992-I. We thank Jerome Bouvier for his suggestions that improved this manuscript. We thank the ULLYSES team at STScI for planning and implementing the $HST$ observations and creating data products, with guidance from the research community.
\end{acknowledgments}

\vspace{5mm}
\facilities{$HST$, LCOGT, $TESS$, AAVSO, ZTF, ASAS-SN, Konkoly Observatory, CrAO}

\software{Python, Specutils, AstroPy, emcee}

\bibliography{main}

\appendix

\section{Shock Modeling of RU~Lup} \label{appendix: RU Lup}

As discussed in Section \ref{sec: Shock Model Results}, our choice of template for RU~Lup is complicated by our lack of contemporaneous veiling estimates. The photospheric template we employ in our modeling (F$_{\rm{Phot,\ RU\ Lup}}$) is obtained by scaling the raw WTTS spectrum to the radius and distance of RU~Lup. While this can introduce additional uncertainty from the stellar radii and distances, we believe this is a reasonable estimate and is within upper and lower limits for the true photosphere.

As a lower limit on the photospheric contribution (F$_{\rm{Phot,\ Low}}$), we can assume that all observed emission originates from accretion by imposing F$_{\rm{Phot,\ Low}}=0$. As an upper limit on the photospheric contribution (F$_{\rm{Phot,\ High}}$), we scale the raw WTTS template to the dimmest observation of our $V$ light curve of RU~Lup (see Paper II). This assumes that all emission at $V$ originates from the photosphere, implying no contribution from accretion at $V$. RU~Lup is likely accreting at all times, even if only very weakly, meaning the photospheric contribution to the observed spectrum is likely smaller.

\begin{figure}[htbp]
    \centering
    \includegraphics{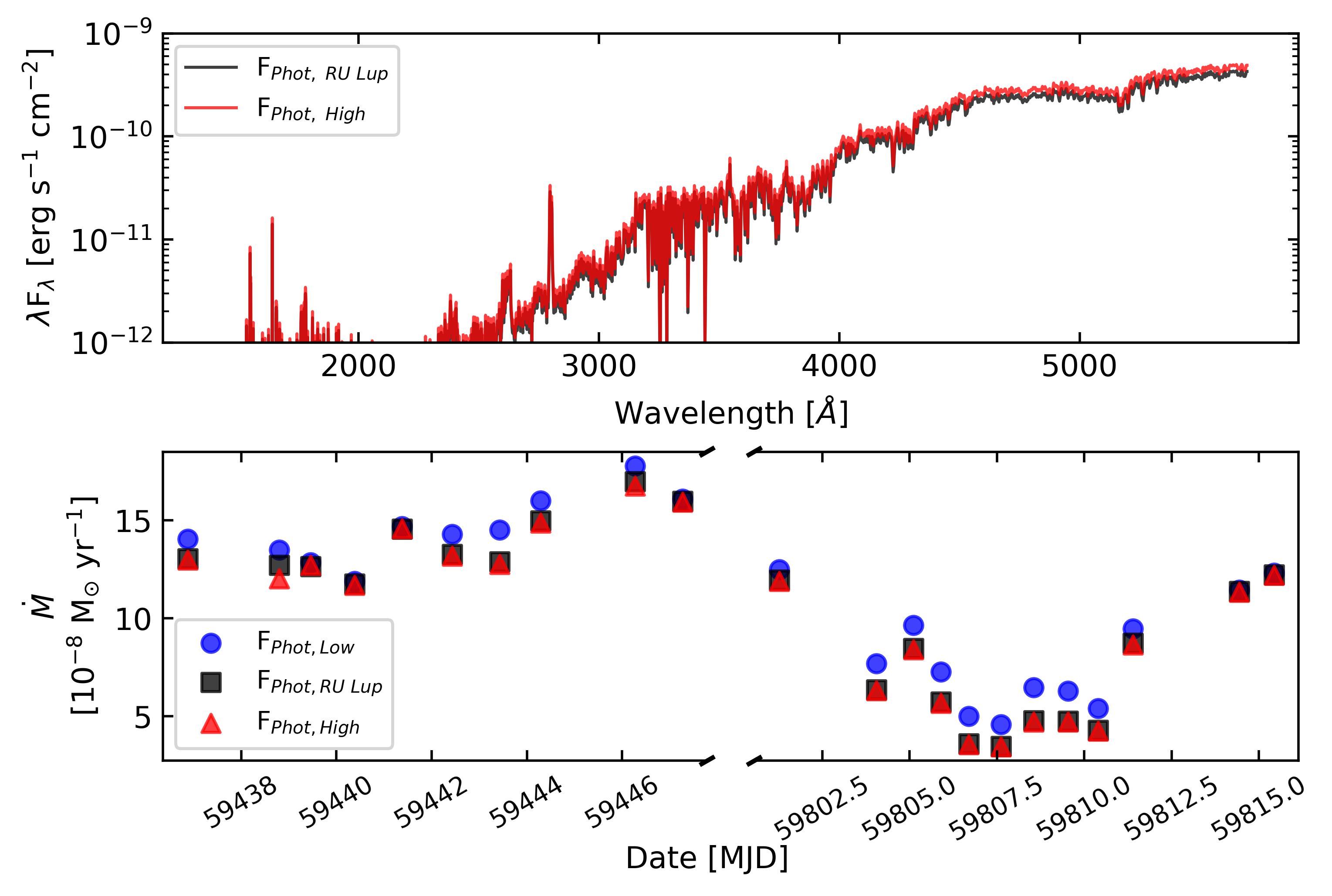}
    \caption{Top: Comparison between F$_{\rm{Phot,\ High}}$ (red) and F$_{\rm{Phot,\ RU\ Lup}}$ (black). Recall that F$_{\rm{Phot,\ Low}}$ is zero, i.e., the photosphere does not contribute to the observed spectrum. F$_{\rm{Phot,\ RU\ Lup}}$ lays between F$_{\rm{Phot,\ High}}$ and F$_{\rm{Phot,\ Low}}$. Bottom: Accretion rates for RU~Lup when using F$_{\rm{Phot,\ High}}$ (red triangles), F$_{\rm{Phot,\ RU\ Lup}}$ (black squares), and F$_{\rm{Phot,\ Low}}$ (blue circles) as photospheric templates.}
    \label{fig: RU Lup Template Comparison}
\end{figure}

Figure \ref{fig: RU Lup Template Comparison} shows the comparison between F$_{\rm{Phot,\ High}}$ and F$_{\rm{Phot,\ RU\ Lup}}$, along with accretion rates derived using each template. In all cases, the accretion rates obtained using F$_{\rm{Phot,\ RU\ Lup}}$ fall between those derived from the F$_{\rm{Phot,\ High}}$ and F$_{\rm{Phot,\ Low}}$.

\section{Effect of No Contemporaneous Photometry on Shock Model Fits} \label{appendix: No Photometry Fits}

\begin{figure}[htbp]
    \centering
    \includegraphics[width=\textwidth]{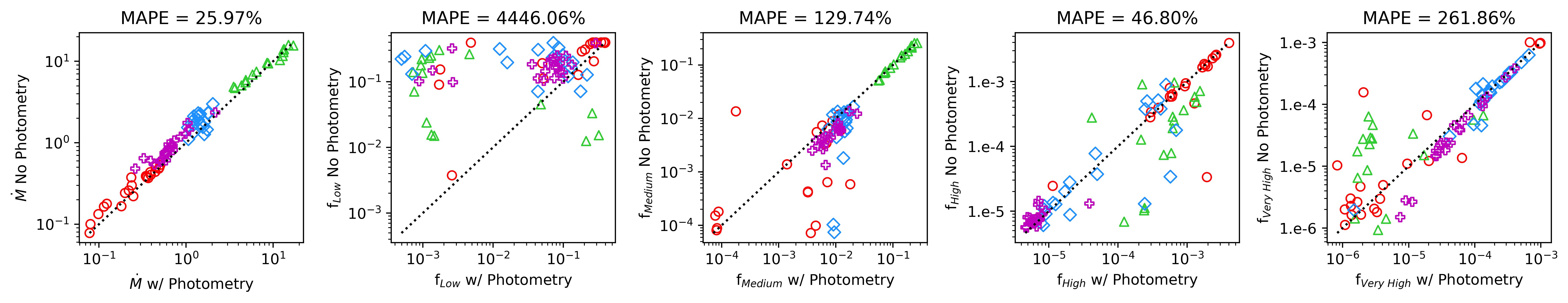}
    \caption{Comparison of accretion shock model results when removing optical photometry points from fitting. Left to right: \mdot, f$_{Low}$, f$_{Medium}$, f$_{High}$, and f$_{Very\ High}$. Red circles, green triangles, blue diamonds, and purple pluses are TW~Hya, RU~Lup, BP~Tau, and GM~Aur, respectively. Mean Absolute Percent Error (MAPE) values between the parameters with and without photometry are listed above each panel.}
    \label{fig: Phot-No Phot Comparison}
\end{figure}

10 of the 89 $HST$ visits do not have contemporaneous $BgV$ photometry. The optical portion of the shock model fit is therefore less constrained and here we discuss what effect that may have on the results for those visits. To gauge this effect, we re-run our shock models for all visits but remove the photometry, effectively fitting only the UV COS spectrum from $\sim$1800--3200 \AA. A comparison between some of the resulting accretion parameters (accretion rate and filling factors) for visits that have contemporaneous photometry is shown in Figure \ref{fig: Phot-No Phot Comparison}.

In general, when no optical photometry is present, the filling factors are accurate to no better than about 50\%. Fits without photometry show larger filling factors for the low density column, with a Mean Absolute Percent Error (MAPE) of 4584\%. Much of this strong deviation comes from visits that are normally fit with particularly low f$_{Low}$ (near 0\%), but jump to the upper limit of f$_{Low}$=40\% when the photometry is removed. In turn, this higher f$_{Low}$ results in generally lower f$_{Medium}$ (MAPE=130\%). The high and very high density columns are generally consistent to within 47\% and 263\%, though for the very high density column the strongest deviations occur when f$_{Very\ High}$ is very low and contributes little to the total mass accretion.

\begin{figure}[hb]
    \centering
    \includegraphics[width=0.75\textwidth]{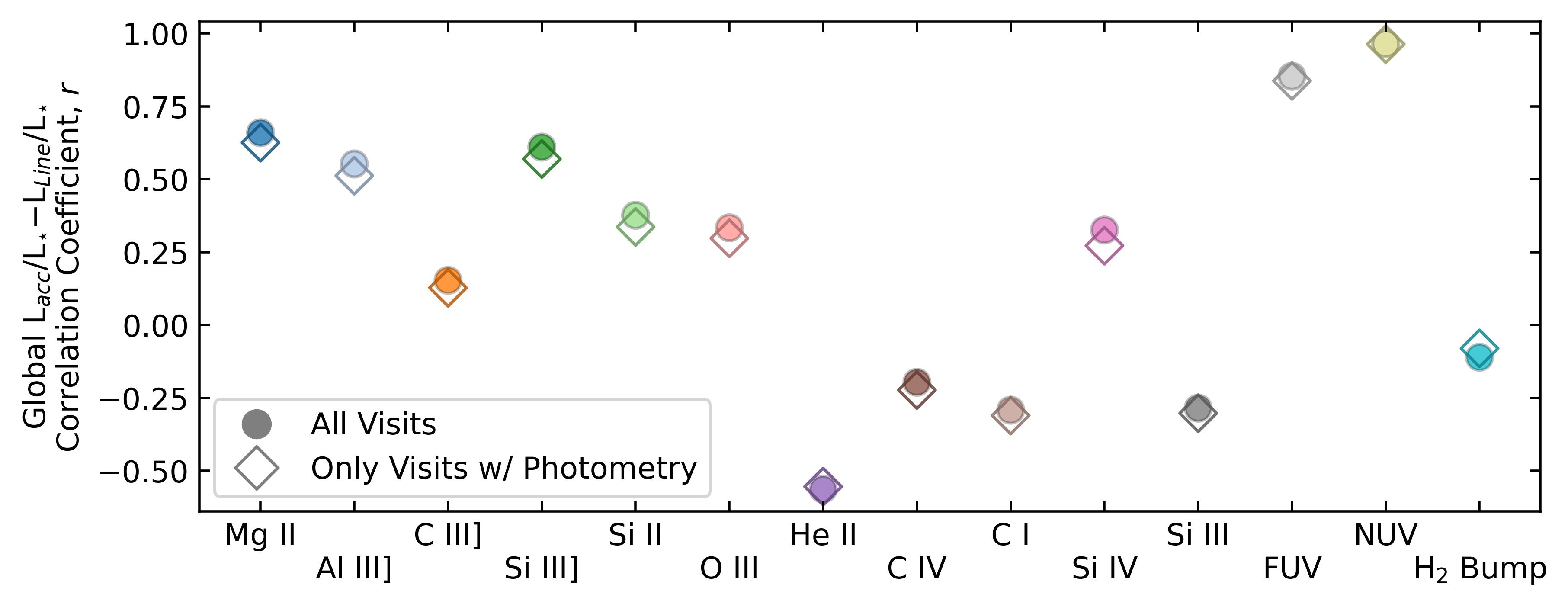}
    \caption{Comparison of \lacc/\lstar vs \lline/\lstar correlation coefficients. Filled circles are for all visits, regardless of whether they have contemporaneous optical photometry (what is shown in Figure \ref{fig: UV Luminosities}). Open diamonds are for only visits with contemporaneous photometry.}
    \label{fig: Lacc vs Lline with(out) photometry}
\end{figure}

While the filling factors are largely inaccurate, the overall mass accretion rate is typically consistent to within about 26\%, generally being overestimated due to the unconstrained f$_{Low}$. These results suggest that while some optical data is very important to constrain the filling factors, the total accretion rate is fairly well constrained. Thus, for visits with no contemporaneous optical photometry (which are noted in Tables \ref{tab: Shock Model Results, TW Hya/RU Lup}-\ref{tab: Shock Model Results, BP Tau/GM Aur}), the accretion rates are likely overestimated by up to about 30\%. The associated filling factors are likely inaccurate, with little consistency as to how over/underestimated they are. Figure \ref{fig: Lacc vs Lline with(out) photometry} shows that the correlations between \lacc and \lline are largely consistent whether or not visits without contemporaneous photometry are included.

\section{Shock Model Results} \label{appendix: Shock Model} \label{appendix: Shock Model Results}  \label{appendix: Shock Model Fits}

\begin{figure}[ht]
    \centering
    \includegraphics[width=0.95\textwidth]{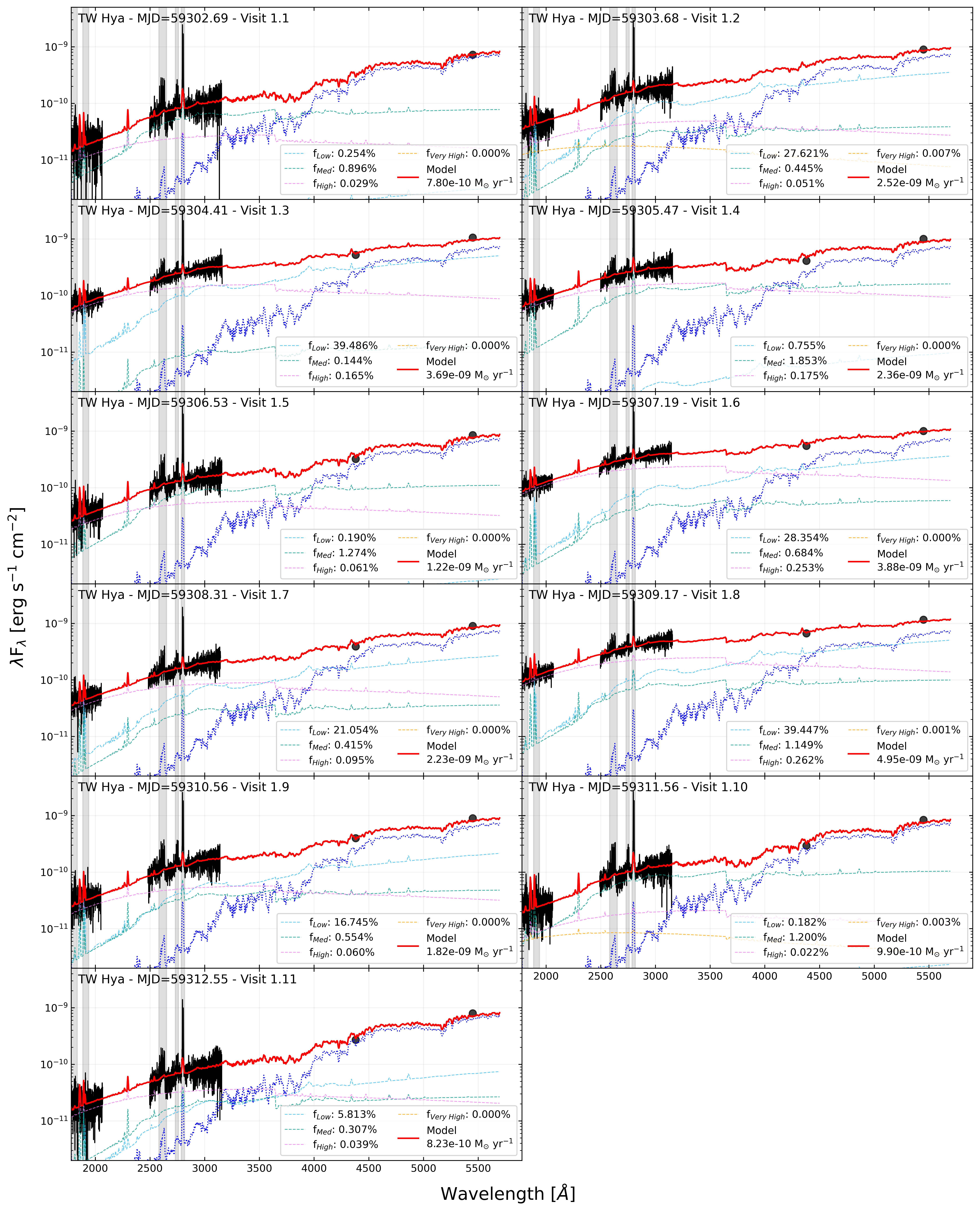}
    \caption{Shock model fits for TW~Hya, Epoch 1. Black, red, and dotted blue lines are the data, total model, and scaled WTTS template (F$_{Phot,\ Obs}$), respectively. Black circles are contemporaneous photometry points. Cyan, green, pink, and orange lines are the low, medium, high, and very high density accretion columns. In some cases the contribution from an individual column is low and does not appear in the plot. Filled grey regions are masked in the fitting due to the presence of bright emission lines.}
    \label{fig: Shock Model Fits - TW Hya - E1}
\end{figure}

\begin{figure}[h]
    \centering
    \includegraphics[width=0.95\textwidth]{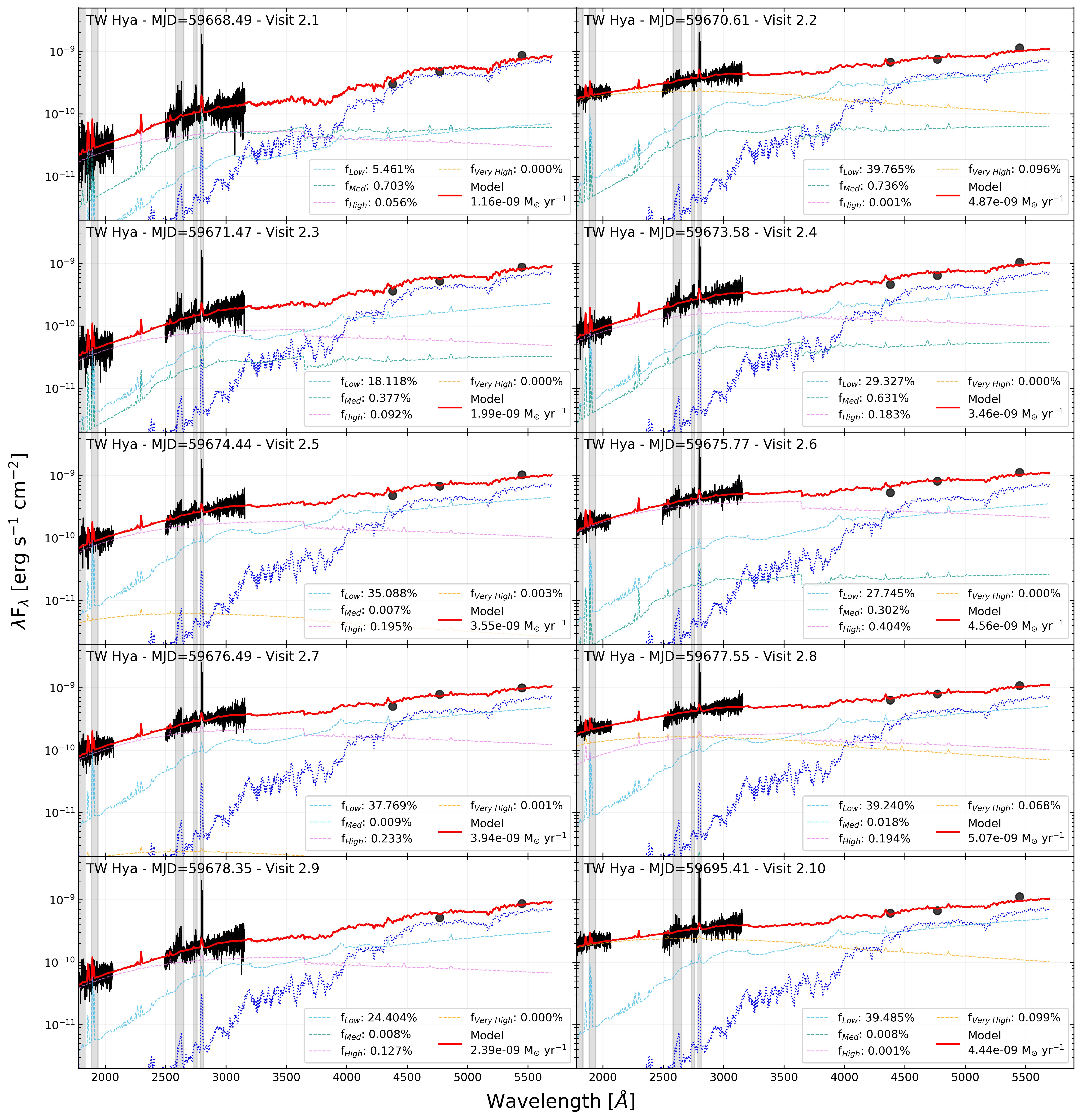}
    \caption{Same as Figure \ref{fig: Shock Model Fits - TW Hya - E1}, but for TW~Hya, Epoch 2.}
    \label{fig: Shock Model Fits - TW Hya - E2}
\end{figure}

\begin{figure}[p]
    \centering
    \includegraphics[width=0.95\textwidth]{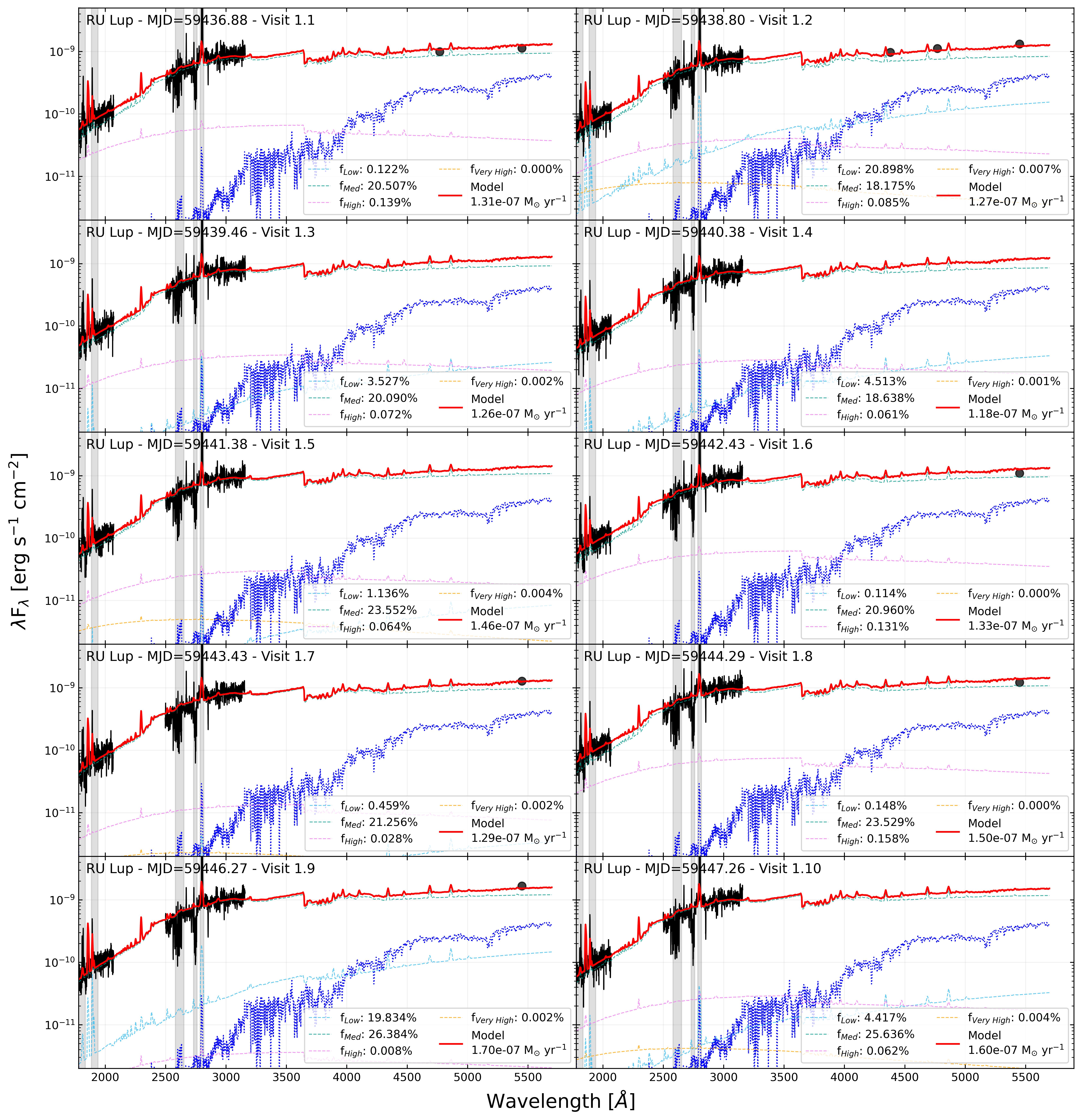}
    \caption{Same as Figure \ref{fig: Shock Model Fits - TW Hya - E1}, but for RU~Lup, Epoch 1.}
    \label{fig: Shock Model Fits - RU Lup - E1}
\end{figure}

\begin{figure}[p]
    \centering
    \includegraphics[width=0.95\textwidth]{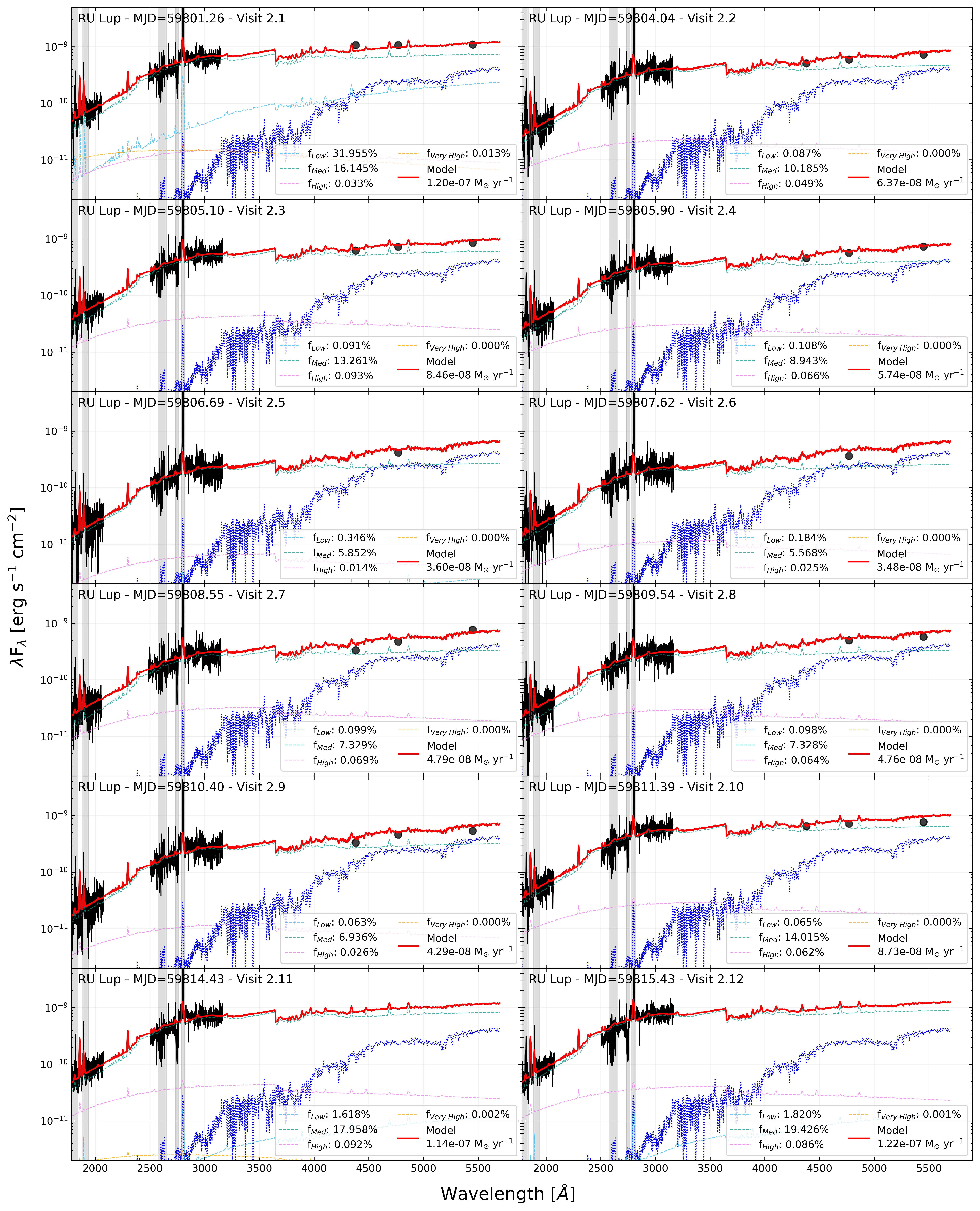}
    \caption{Same as Figure \ref{fig: Shock Model Fits - TW Hya - E1}, but for RU~Lup, Epoch 2.}
    \label{fig: Shock Model Fits - RU Lup - E2}
\end{figure}

\begin{figure}[p]
    \centering
    \includegraphics[width=0.95\textwidth]{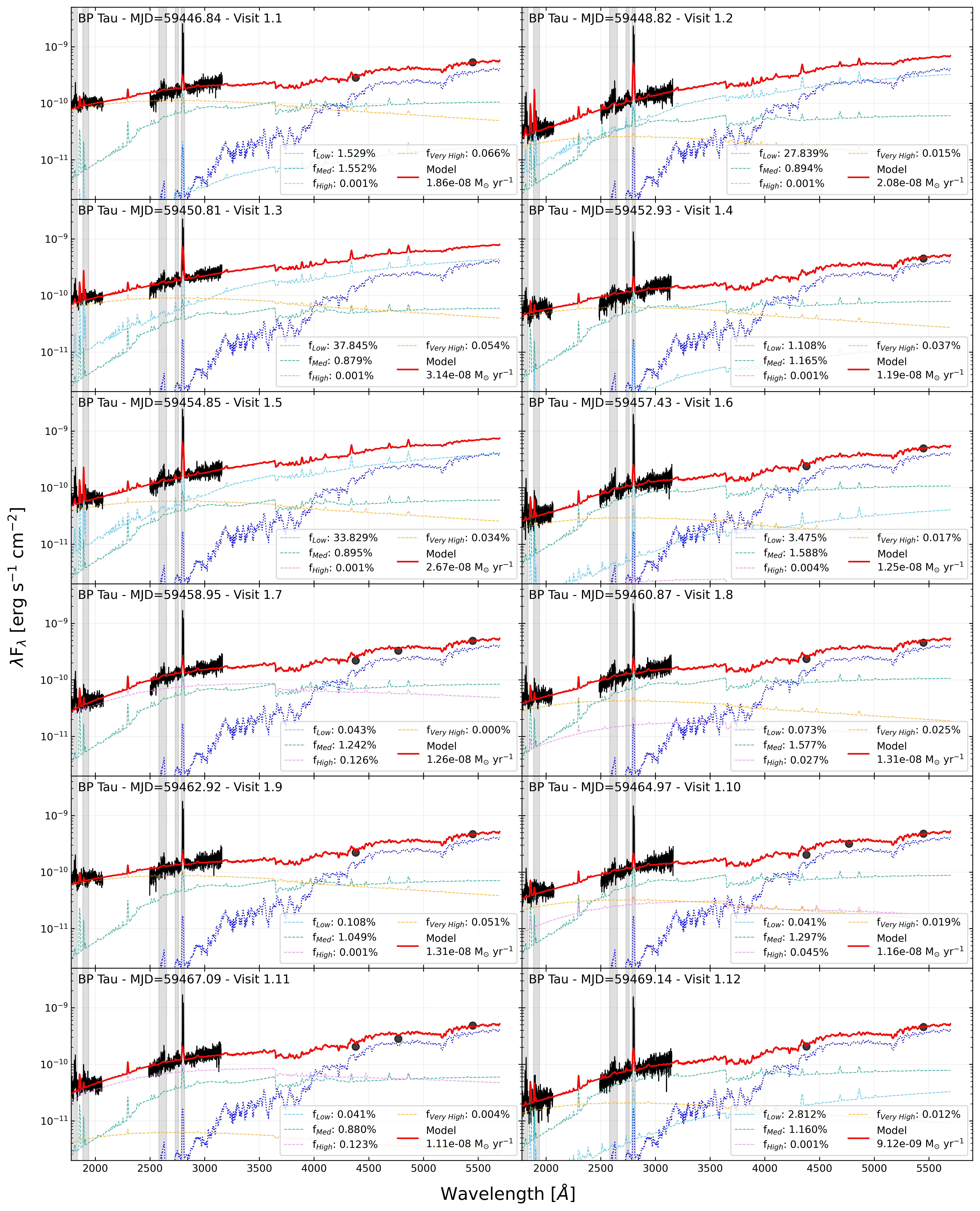}
    \caption{Same as Figure \ref{fig: Shock Model Fits - TW Hya - E1}, but for BP~Tau, Epoch 1.}
    \label{fig: Shock Model Fits - BP Tau - E1}
\end{figure}

\begin{figure}[p]
    \centering
    \includegraphics[width=0.95\textwidth]{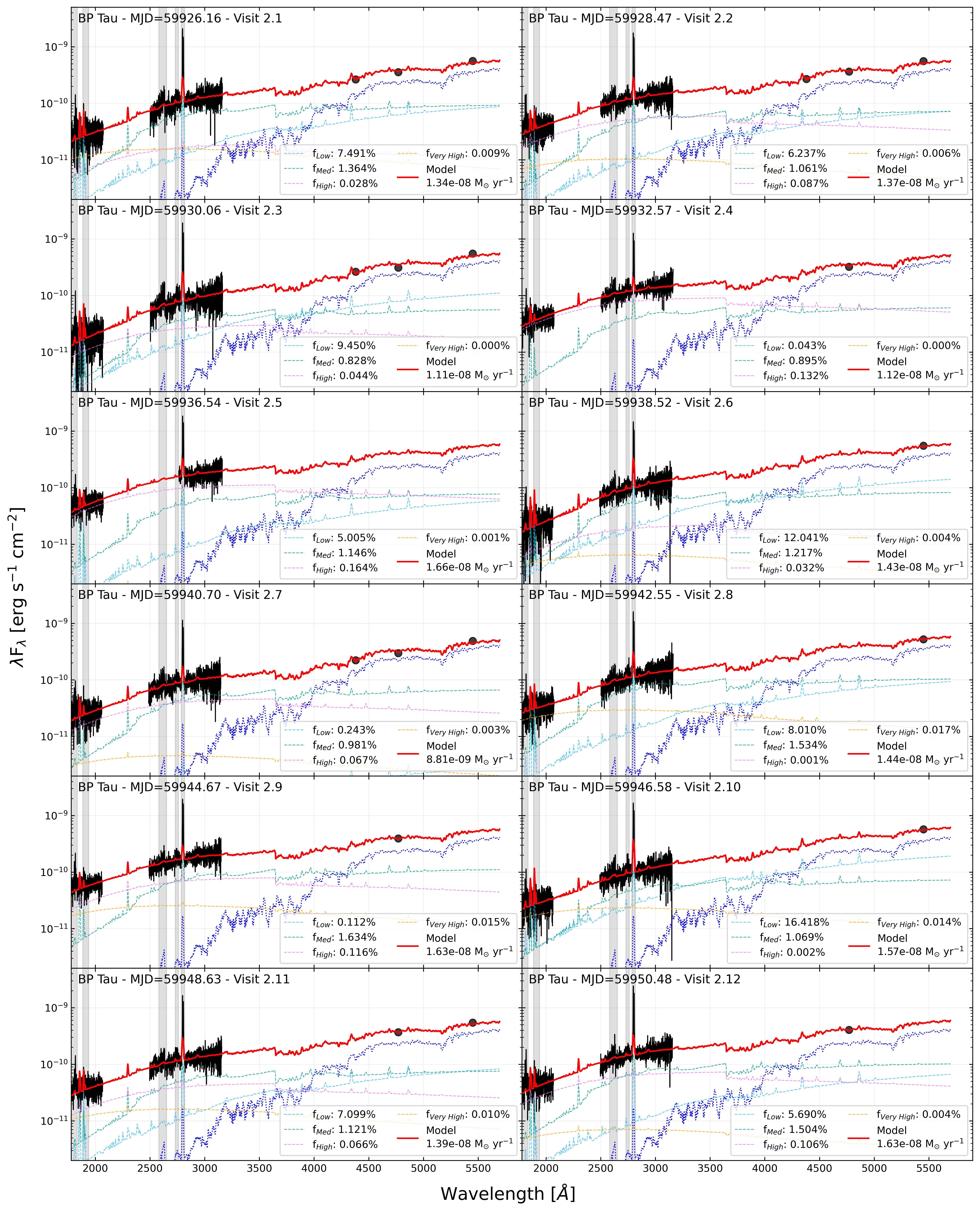}
    \caption{Same as Figure \ref{fig: Shock Model Fits - TW Hya - E1}, but for BP~Tau, Epoch 2.}
    \label{fig: Shock Model Fits - BP Tau - E2}
\end{figure}

\begin{figure}[p]
    \centering
    \includegraphics[width=0.95\textwidth]{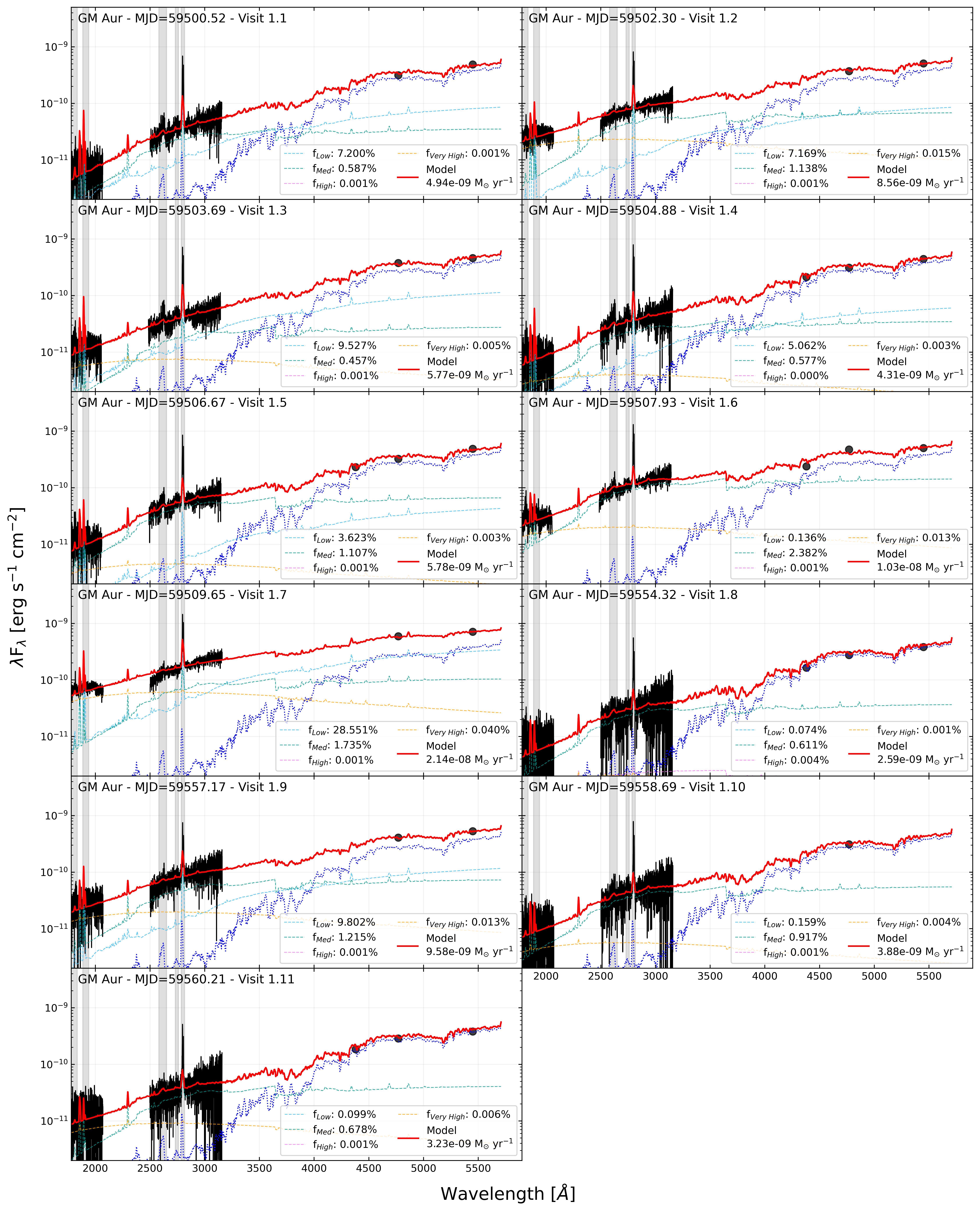}
    \caption{Same as Figure \ref{fig: Shock Model Fits - TW Hya - E1}, but for GM~Aur, Epoch 1.}
    \label{fig: Shock Model Fits - GM Aur - E1}
\end{figure}

\begin{figure}[p]
    \centering
    \includegraphics[width=0.95\textwidth]{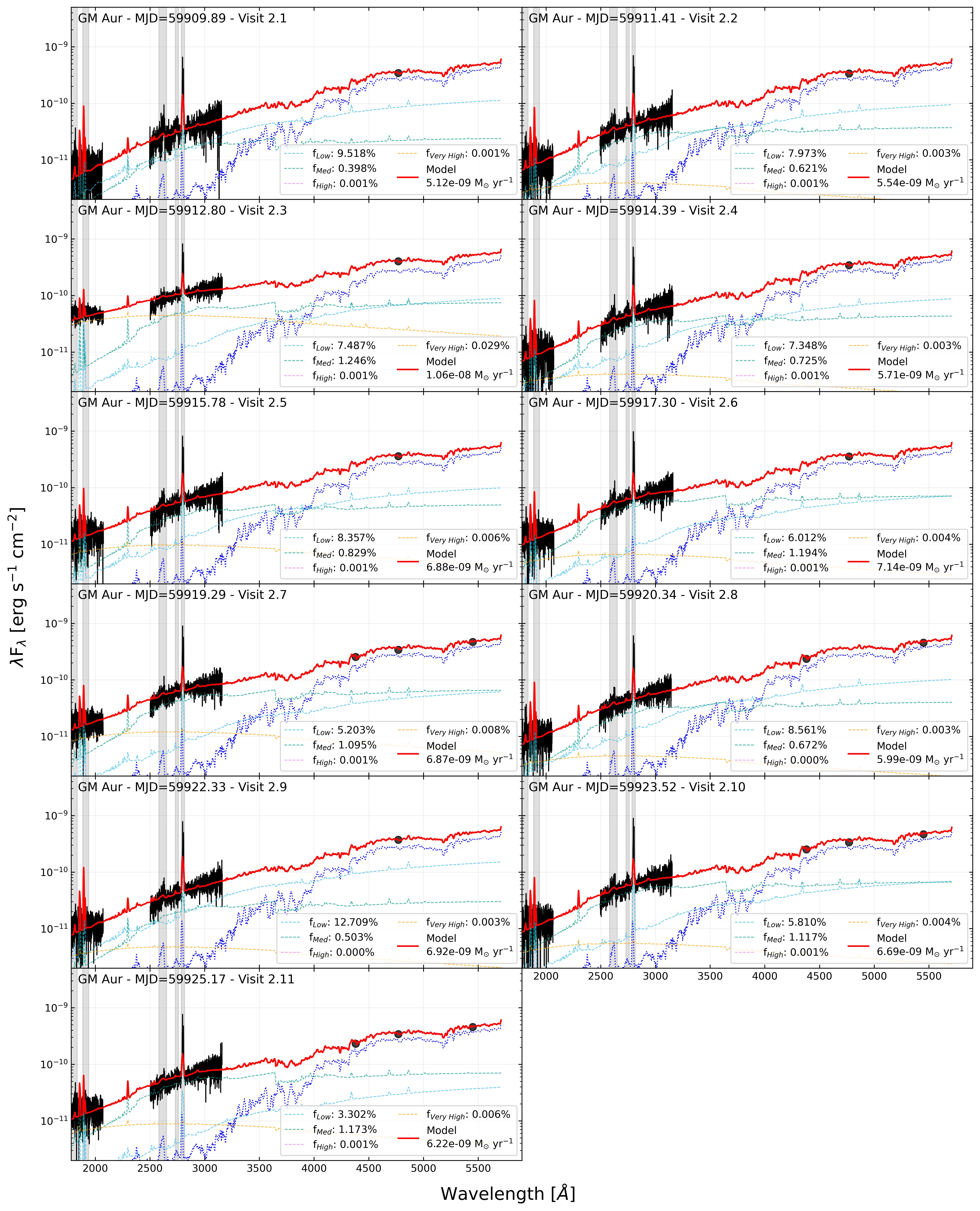}
    \caption{Same as Figure \ref{fig: Shock Model Fits - TW Hya - E1}, but for GM~Aur, Epoch 2.}
    \label{fig: Shock Model Fits - GM Aur - E2}
\end{figure}

Tables \ref{tab: Shock Model Results, TW Hya/RU Lup} and \ref{tab: Shock Model Results, BP Tau/GM Aur} show the mass accretion rates and filling factors of the four accretion columns obtained from our shock model fitting of TW~Hya, RU~Lup, BP~Tau, and GM~Aur. Figures \ref{fig: Shock Model Fits - TW Hya - E1}--\ref{fig: Shock Model Fits - GM Aur - E2} show the shock model fits for each $HST$ visit.

For TW~Hya, BP~Tau, and GM~Aur, the upper/lower uncertainties are the average differences between the 84th/16th and 50th percentiles in the resulting posterior distributions, except where the standard deviation of results from all F$_{Phot,\ True}$ is larger. For RU~Lup, we only cite the difference between the 84th/16th and 50th percentiles in the resulting posterior distributions, as only one F$_{Phot,\ True}$ was used. These uncertainties thus underestimate the true uncertainty in the model parameters given the uncertainty in the true, underlying F$_{Phot,\ True}$. Additionally, the uncertainties for all four targets only describe the widths of the posterior distributions of each parameter. They do not properly account for non-systematic uncertainties, such as R$_*$, M$_*$, and A$_V$. \citetalias{RE19} note that the true uncertainties are likely to be at least 10\% when accounting for other sources of error. 

\begin{deluxetable}{c c c c c c c c}[htp] \label{tab: Shock Model Results, TW Hya/RU Lup}
\tablecaption{Shock model results for TW Hya and RU Lup.}
\tablehead{
\colhead{Object} & \colhead{Visit} & \colhead{Date} & \colhead{\mdot} & \colhead{$f_{Low}$} & \colhead{$f_{Medium}$} & \colhead{$f_{High}$} & \colhead{$f_{Very\ High}$} \\
\colhead{} & \colhead{} & \colhead{[MJD]} & \colhead{[10$^{-8}$ \accrate]} & \colhead{[\%]} & \colhead{[\%]} & \colhead{[\%]} & \colhead{[\%]}
}
\startdata
TW Hya & 1.1 & 59302.69 & 0.078$^{+0.012}_{-0.007}$ & 0.3$^{+1.7}_{-0.2}$ & 0.9$^{+0.05}_{-0.08}$ & 0.029$^{+0.002}_{-0.006}$ & 0.0002$^{+0.0011}_{-0.0002}$ \\ 
TW Hya & 1.2 & 59303.68 & 0.25$^{+0.04}_{-0.05}$ & 27.6$^{+3.4}_{-4.1}$ & 0.4$^{+0.4}_{-0.4}$ & 0.05$^{+0.04}_{-0.04}$ & 0.007$^{+0.008}_{-0.007}$ \\ 
TW Hya & 1.3 & 59304.41 & 0.368$^{+0.009}_{-0.011}$ & 39.5$^{+0.4}_{-1.0}$ & 0.14$^{+0.11}_{-0.08}$ & 0.165$^{+0.004}_{-0.01}$ & 0.0002$^{+0.002}_{-0.0002}$ \\ 
TW Hya & 1.4 & 59305.47 & 0.24$^{+0.03}_{-0.02}$ & 0.8$^{+4.6}_{-0.7}$ & 1.9$^{+0.2}_{-0.2}$ & 0.175$^{+0.006}_{-0.023}$ & 0.0004$^{+0.0046}_{-0.0004}$ \\ 
TW Hya & 1.5 & 59306.53 & 0.12$^{+0.01}_{-0.01}$ & 0.2$^{+1.2}_{-0.2}$ & 1.27$^{+0.08}_{-0.08}$ & 0.061$^{+0.003}_{-0.015}$ & 0.0003$^{+0.003}_{-0.0003}$ \\ 
TW Hya & 1.6 & 59307.19 & 0.39$^{+0.03}_{-0.04}$ & 28.4$^{+4.1}_{-4.9}$ & 0.7$^{+0.3}_{-0.3}$ & 0.253$^{+0.006}_{-0.008}$ & 0.0001$^{+0.0013}_{-0.0001}$ \\ 
TW Hya & 1.7 & 59308.31 & 0.22$^{+0.02}_{-0.02}$ & 21.1$^{+3.1}_{-3.5}$ & 0.4$^{+0.2}_{-0.2}$ & 0.095$^{+0.003}_{-0.005}$ & 0.0001$^{+0.0009}_{-0.0001}$ \\ 
TW Hya & 1.8 & 59309.17 & 0.49$^{+0.02}_{-0.02}$ & 39.4$^{+0.4}_{-1.1}$ & 1.1$^{+0.2}_{-0.1}$ & 0.262$^{+0.007}_{-0.028}$ & 0.0005$^{+0.0057}_{-0.0005}$ \\ 
TW Hya & 1.9 & 59310.56 & 0.18$^{+0.02}_{-0.02}$ & 16.7$^{+2.4}_{-2.5}$ & 0.6$^{+0.1}_{-0.1}$ & 0.06$^{+0.003}_{-0.005}$ & 0.0001$^{+0.001}_{-0.0001}$ \\ 
TW Hya & 1.10 & 59311.56 & 0.1$^{+0.02}_{-0.02}$ & 0.2$^{+1.1}_{-0.2}$ & 1.2$^{+0.1}_{-0.1}$ & 0.02$^{+0.02}_{-0.02}$ & 0.003$^{+0.004}_{-0.003}$ \\ 
TW Hya & 1.11 & 59312.55 & 0.08$^{+0.02}_{-0.02}$ & 5.8$^{+2.0}_{-2.0}$ & 0.3$^{+0.1}_{-0.1}$ & 0.039$^{+0.003}_{-0.013}$ & 0.0004$^{+0.0025}_{-0.0003}$ \\ 
\hline 
TW Hya & 2.1 & 59668.49 & 0.12$^{+0.02}_{-0.02}$ & 5.5$^{+2.3}_{-2.9}$ & 0.7$^{+0.2}_{-0.1}$ & 0.056$^{+0.002}_{-0.004}$ & 0.0001$^{+0.0007}_{-0.0001}$ \\ 
TW Hya & 2.2 & 59670.61 & 0.487$^{+0.008}_{-0.007}$ & 39.8$^{+0.2}_{-0.5}$ & 0.74$^{+0.06}_{-0.07}$ & 0.001$^{+0.009}_{-0.001}$ & 0.096$^{+0.001}_{-0.002}$ \\ 
TW Hya & 2.3 & 59671.47 & 0.2$^{+0.02}_{-0.02}$ & 18.1$^{+2.7}_{-3.0}$ & 0.4$^{+0.2}_{-0.1}$ & 0.092$^{+0.003}_{-0.004}$ & 0.0001$^{+0.0005}_{-0.0001}$ \\ 
TW Hya & 2.4 & 59673.58 & 0.35$^{+0.03}_{-0.04}$ & 29.3$^{+4.5}_{-5.0}$ & 0.6$^{+0.3}_{-0.3}$ & 0.183$^{+0.005}_{-0.007}$ & 0.0001$^{+0.001}_{-0.0001}$ \\ 
TW Hya & 2.5 & 59674.44 & 0.35$^{+0.02}_{-0.02}$ & 35.1$^{+2.2}_{-1.7}$ & 0.007$^{+0.031}_{-0.005}$ & 0.19$^{+0.01}_{-0.02}$ & 0.003$^{+0.005}_{-0.002}$ \\ 
TW Hya & 2.6 & 59675.77 & 0.46$^{+0.04}_{-0.06}$ & 27.7$^{+4.8}_{-7.7}$ & 0.3$^{+0.5}_{-0.3}$ & 0.404$^{+0.008}_{-0.012}$ & 0.0002$^{+0.0013}_{-0.0001}$ \\ 
TW Hya & 2.7 & 59676.49 & 0.39$^{+0.01}_{-0.01}$ & 37.8$^{+1.3}_{-1.5}$ & 0.009$^{+0.047}_{-0.007}$ & 0.233$^{+0.006}_{-0.014}$ & 0.001$^{+0.0032}_{-0.0009}$ \\ 
TW Hya & 2.8 & 59677.55 & 0.51$^{+0.02}_{-0.02}$ & 39.2$^{+0.6}_{-1.2}$ & 0.02$^{+0.13}_{-0.02}$ & 0.19$^{+0.01}_{-0.02}$ & 0.068$^{+0.005}_{-0.004}$ \\ 
TW Hya & 2.9 & 59678.35 & 0.239$^{+0.009}_{-0.009}$ & 24.4$^{+1.1}_{-1.2}$ & 0.008$^{+0.042}_{-0.006}$ & 0.127$^{+0.003}_{-0.008}$ & 0.0002$^{+0.0017}_{-0.0002}$ \\ 
TW Hya & 2.10 & 59695.41 & 0.444$^{+0.005}_{-0.006}$ & 39.5$^{+0.4}_{-0.9}$ & 0.008$^{+0.032}_{-0.006}$ & 0.0007$^{+0.003}_{-0.0005}$ & 0.099$^{+0.001}_{-0.001}$ \\ 
\hline
\hline
RU Lup & 1.1 & 59436.88 & 13.1$^{+0.3}_{-0.3}$ & 0.1$^{+0.6}_{-0.1}$ & 20.5$^{+0.4}_{-0.4}$ & 0.14$^{+0.02}_{-0.02}$ & 0.0002$^{+0.003}_{-0.0002}$ \\ 
RU Lup & 1.2 & 59438.80 & 12.7$^{+0.7}_{-0.9}$ & 20.9$^{+7.9}_{-10.7}$ & 18.2$^{+0.7}_{-0.6}$ & 0.09$^{+0.04}_{-0.08}$ & 0.007$^{+0.016}_{-0.007}$ \\ 
RU Lup & 1.3 & 59439.46$^a$ & 12.6$^{+1.2}_{-0.6}$ & 3.5$^{+18.8}_{-3.4}$ & 20.1$^{+0.6}_{-0.7}$ & 0.07$^{+0.03}_{-0.07}$ & 0.002$^{+0.015}_{-0.002}$ \\ 
RU Lup & 1.4 & 59440.38$^a$ & 11.8$^{+1.4}_{-0.6}$ & 4.5$^{+21.7}_{-4.4}$ & 18.6$^{+0.5}_{-0.7}$ & 0.06$^{+0.03}_{-0.06}$ & 0.001$^{+0.012}_{-0.001}$ \\ 
RU Lup & 1.5 & 59441.38$^a$ & 14.6$^{+0.9}_{-0.6}$ & 1.1$^{+13.6}_{-1.1}$ & 23.6$^{+0.6}_{-0.7}$ & 0.06$^{+0.04}_{-0.06}$ & 0.004$^{+0.012}_{-0.004}$ \\ 
RU Lup & 1.6 & 59442.43 & 13.3$^{+0.3}_{-0.3}$ & 0.11$^{+0.7}_{-0.09}$ & 21.0$^{+0.4}_{-0.4}$ & 0.13$^{+0.02}_{-0.02}$ & 0.0002$^{+0.0023}_{-0.0002}$ \\ 
RU Lup & 1.7 & 59443.43 & 12.9$^{+0.5}_{-0.3}$ & 0.5$^{+6.0}_{-0.4}$ & 21.3$^{+0.4}_{-0.4}$ & 0.03$^{+0.02}_{-0.03}$ & 0.002$^{+0.006}_{-0.002}$ \\ 
RU Lup & 1.8 & 59444.29 & 15.0$^{+0.3}_{-0.3}$ & 0.1$^{+1.1}_{-0.1}$ & 23.5$^{+0.5}_{-0.5}$ & 0.16$^{+0.02}_{-0.02}$ & 0.0002$^{+0.0028}_{-0.0002}$ \\ 
RU Lup & 1.9 & 59446.27 & 17.0$^{+0.4}_{-0.6}$ & 19.8$^{+3.0}_{-7.7}$ & 26.4$^{+0.5}_{-0.5}$ & 0.008$^{+0.026}_{-0.007}$ & 0.002$^{+0.005}_{-0.001}$ \\ 
RU Lup & 1.10 & 59447.26$^a$ & 16.0$^{+1.1}_{-0.7}$ & 4.4$^{+16.6}_{-4.3}$ & 25.6$^{+0.7}_{-0.9}$ & 0.06$^{+0.04}_{-0.06}$ & 0.004$^{+0.013}_{-0.004}$ \\ 
\hline 
RU Lup & 2.1 & 59801.26 & 12.0$^{+0.5}_{-0.5}$ & 32.0$^{+1.6}_{-3.2}$ & 16.1$^{+0.4}_{-0.5}$ & 0.03$^{+0.07}_{-0.03}$ & 0.013$^{+0.007}_{-0.013}$ \\ 
RU Lup & 2.2 & 59804.04 & 6.4$^{+0.1}_{-0.1}$ & 0.09$^{+0.44}_{-0.07}$ & 10.2$^{+0.2}_{-0.2}$ & 0.049$^{+0.008}_{-0.012}$ & 0.0002$^{+0.0019}_{-0.0002}$ \\ 
RU Lup & 2.3 & 59805.10 & 8.5$^{+0.2}_{-0.2}$ & 0.09$^{+0.59}_{-0.07}$ & 13.3$^{+0.2}_{-0.3}$ & 0.09$^{+0.01}_{-0.01}$ & 0.0002$^{+0.0014}_{-0.0001}$ \\ 
RU Lup & 2.4 & 59805.90 & 5.7$^{+0.1}_{-0.1}$ & 0.11$^{+0.82}_{-0.09}$ & 8.9$^{+0.2}_{-0.2}$ & 0.066$^{+0.009}_{-0.009}$ & 0.0002$^{+0.0013}_{-0.0001}$ \\ 
RU Lup & 2.5 & 59806.69 & 3.6$^{+0.2}_{-0.1}$ & 0.3$^{+2.9}_{-0.3}$ & 5.9$^{+0.2}_{-0.2}$ & 0.014$^{+0.009}_{-0.012}$ & 0.0004$^{+0.0022}_{-0.0004}$ \\ 
RU Lup & 2.6 & 59807.62 & 3.5$^{+0.2}_{-0.2}$ & 0.2$^{+1.4}_{-0.2}$ & 5.6$^{+0.2}_{-0.2}$ & 0.025$^{+0.009}_{-0.018}$ & 0.0003$^{+0.0032}_{-0.0003}$ \\ 
RU Lup & 2.7 & 59808.55 & 4.8$^{+0.1}_{-0.1}$ & 0.1$^{+0.6}_{-0.08}$ & 7.3$^{+0.2}_{-0.2}$ & 0.069$^{+0.009}_{-0.009}$ & 0.0001$^{+0.0011}_{-0.0001}$ \\ 
RU Lup & 2.8 & 59809.54 & 4.8$^{+0.1}_{-0.1}$ & 0.1$^{+0.57}_{-0.08}$ & 7.3$^{+0.2}_{-0.2}$ & 0.064$^{+0.007}_{-0.008}$ & 0.0001$^{+0.001}_{-0.0001}$ \\ 
RU Lup & 2.9 & 59810.40 & 4.3$^{+0.1}_{-0.1}$ & 0.06$^{+0.25}_{-0.04}$ & 6.9$^{+0.1}_{-0.1}$ & 0.026$^{+0.006}_{-0.007}$ & 0.0002$^{+0.0012}_{-0.0001}$ \\ 
RU Lup & 2.10 & 59811.39 & 8.7$^{+0.2}_{-0.2}$ & 0.06$^{+0.3}_{-0.05}$ & 14.0$^{+0.2}_{-0.3}$ & 0.06$^{+0.01}_{-0.01}$ & 0.0002$^{+0.0011}_{-0.0001}$ \\ 
RU Lup & 2.11 & 59814.43$^a$ & 11.4$^{+1.1}_{-0.6}$ & 1.6$^{+16.3}_{-1.6}$ & 18.0$^{+0.6}_{-0.6}$ & 0.09$^{+0.03}_{-0.09}$ & 0.002$^{+0.017}_{-0.002}$ \\ 
RU Lup & 2.12 & 59815.43$^a$ & 12.2$^{+1.2}_{-0.6}$ & 1.8$^{+19.0}_{-1.8}$ & 19.4$^{+0.6}_{-0.7}$ & 0.09$^{+0.03}_{-0.08}$ & 0.001$^{+0.016}_{-0.001}$ \\ 
\hline
\enddata
\tablenotetext{a}{No contemporaneous photometry}
\end{deluxetable}

\begin{deluxetable}{c c c c c c c c}[htp] \label{tab: Shock Model Results, BP Tau/GM Aur}
\tablecaption{Shock model results for BP Tau and GM Aur.}
\tablehead{
\colhead{Object} & \colhead{Visit} & \colhead{Date} & \colhead{\mdot} & \colhead{$f_{Low}$} & \colhead{$f_{Medium}$} & \colhead{$f_{High}$} & \colhead{$f_{Very\ High}$} \\
\colhead{} & \colhead{} & \colhead{[MJD]} & \colhead{[10$^{-8}$ \accrate]} & \colhead{[\%]} & \colhead{[\%]} & \colhead{[\%]} & \colhead{[\%]}
}
\startdata
BP Tau & 1.1 & 59446.84 & 1.86$^{+0.09}_{-0.08}$ & 1.5$^{+1.5}_{-1.1}$ & 1.55$^{+0.08}_{-0.1}$ & 0.0009$^{+0.0043}_{-0.0007}$ & 0.066$^{+0.001}_{-0.001}$ \\ 
BP Tau & 1.2 & 59448.82$^a$ & 2.1$^{+0.2}_{-0.2}$ & 27.8$^{+3.3}_{-3.2}$ & 0.9$^{+0.1}_{-0.1}$ & 0.001$^{+0.006}_{-0.001}$ & 0.0153$^{+0.0006}_{-0.0011}$ \\ 
BP Tau & 1.3 & 59450.81$^a$ & 3.1$^{+0.1}_{-0.2}$ & 37.8$^{+1.6}_{-3.2}$ & 0.88$^{+0.12}_{-0.09}$ & 0.0011$^{+0.0062}_{-0.0008}$ & 0.054$^{+0.001}_{-0.001}$ \\ 
BP Tau & 1.4 & 59452.93 & 1.19$^{+0.09}_{-0.07}$ & 1.1$^{+1.6}_{-1.0}$ & 1.16$^{+0.07}_{-0.08}$ & 0.0008$^{+0.0038}_{-0.0006}$ & 0.0365$^{+0.0009}_{-0.001}$ \\ 
BP Tau & 1.5 & 59454.85$^a$ & 2.7$^{+0.2}_{-0.2}$ & 33.8$^{+3.9}_{-4.4}$ & 0.9$^{+0.2}_{-0.2}$ & 0.001$^{+0.01}_{-0.001}$ & 0.034$^{+0.001}_{-0.002}$ \\ 
BP Tau & 1.6 & 59457.43 & 1.2$^{+0.1}_{-0.1}$ & 3.5$^{+1.0}_{-1.0}$ & 1.59$^{+0.07}_{-0.15}$ & 0.004$^{+0.022}_{-0.003}$ & 0.0174$^{+0.0008}_{-0.0039}$ \\ 
BP Tau & 1.7 & 59458.95 & 1.26$^{+0.05}_{-0.06}$ & 0.04$^{+0.12}_{-0.03}$ & 1.24$^{+0.07}_{-0.05}$ & 0.126$^{+0.004}_{-0.01}$ & 0.0002$^{+0.002}_{-0.0002}$ \\ 
BP Tau & 1.8 & 59460.87 & 1.3$^{+0.3}_{-0.2}$ & 0.07$^{+0.29}_{-0.05}$ & 1.6$^{+0.2}_{-0.3}$ & 0.03$^{+0.05}_{-0.02}$ & 0.025$^{+0.005}_{-0.008}$ \\ 
BP Tau & 1.9 & 59462.92 & 1.31$^{+0.05}_{-0.04}$ & 0.11$^{+0.51}_{-0.09}$ & 1.05$^{+0.05}_{-0.07}$ & 0.001$^{+0.007}_{-0.001}$ & 0.051$^{+0.001}_{-0.002}$ \\ 
BP Tau & 1.10 & 59464.97 & 1.2$^{+0.3}_{-0.2}$ & 0.04$^{+0.12}_{-0.03}$ & 1.3$^{+0.2}_{-0.3}$ & 0.04$^{+0.04}_{-0.03}$ & 0.019$^{+0.005}_{-0.008}$ \\ 
BP Tau & 1.11 & 59467.09 & 1.1$^{+0.2}_{-0.2}$ & 0.04$^{+0.1}_{-0.03}$ & 0.9$^{+0.2}_{-0.1}$ & 0.12$^{+0.02}_{-0.03}$ & 0.004$^{+0.006}_{-0.004}$ \\ 
BP Tau & 1.12 & 59469.14 & 0.91$^{+0.06}_{-0.06}$ & 2.8$^{+0.9}_{-0.9}$ & 1.16$^{+0.05}_{-0.06}$ & 0.0011$^{+0.0056}_{-0.0009}$ & 0.0123$^{+0.0004}_{-0.001}$ \\ 
\hline 
BP Tau & 2.1 & 59926.16 & 1.3$^{+0.3}_{-0.3}$ & 7.5$^{+1.2}_{-1.2}$ & 1.4$^{+0.2}_{-0.3}$ & 0.03$^{+0.05}_{-0.03}$ & 0.009$^{+0.005}_{-0.009}$ \\ 
BP Tau & 2.2 & 59928.47 & 1.4$^{+0.3}_{-0.3}$ & 6.2$^{+1.2}_{-1.3}$ & 1.1$^{+0.3}_{-0.2}$ & 0.09$^{+0.03}_{-0.04}$ & 0.006$^{+0.008}_{-0.006}$ \\ 
BP Tau & 2.3 & 59930.06 & 1.11$^{+0.06}_{-0.06}$ & 9.5$^{+1.0}_{-0.9}$ & 0.83$^{+0.06}_{-0.05}$ & 0.044$^{+0.002}_{-0.005}$ & 0.0001$^{+0.0009}_{-0.0001}$ \\ 
BP Tau & 2.4 & 59932.57 & 1.12$^{+0.04}_{-0.04}$ & 0.04$^{+0.12}_{-0.03}$ & 0.9$^{+0.06}_{-0.05}$ & 0.132$^{+0.004}_{-0.007}$ & 0.0002$^{+0.0012}_{-0.0001}$ \\ 
BP Tau & 2.5 & 59936.54$^a$ & 1.7$^{+0.3}_{-0.3}$ & 5.0$^{+4.4}_{-4.3}$ & 1.1$^{+0.3}_{-0.2}$ & 0.164$^{+0.009}_{-0.05}$ & 0.001$^{+0.009}_{-0.001}$ \\ 
BP Tau & 2.6 & 59938.52 & 1.4$^{+0.2}_{-0.2}$ & 12.0$^{+1.3}_{-1.4}$ & 1.2$^{+0.2}_{-0.2}$ & 0.03$^{+0.02}_{-0.03}$ & 0.004$^{+0.005}_{-0.004}$ \\ 
BP Tau & 2.7 & 59940.70 & 0.9$^{+0.1}_{-0.2}$ & 0.2$^{+0.7}_{-0.2}$ & 1.0$^{+0.2}_{-0.1}$ & 0.07$^{+0.01}_{-0.03}$ & 0.003$^{+0.005}_{-0.003}$ \\ 
BP Tau & 2.8 & 59942.55 & 1.44$^{+0.09}_{-0.08}$ & 8.0$^{+1.4}_{-1.3}$ & 1.53$^{+0.07}_{-0.09}$ & 0.001$^{+0.009}_{-0.001}$ & 0.0174$^{+0.0006}_{-0.0015}$ \\ 
BP Tau & 2.9 & 59944.67 & 1.6$^{+0.3}_{-0.3}$ & 0.11$^{+0.58}_{-0.09}$ & 1.6$^{+0.2}_{-0.3}$ & 0.12$^{+0.04}_{-0.04}$ & 0.015$^{+0.007}_{-0.008}$ \\ 
BP Tau & 2.10 & 59946.58 & 1.57$^{+0.1}_{-0.09}$ & 16.4$^{+1.3}_{-1.3}$ & 1.07$^{+0.07}_{-0.1}$ & 0.002$^{+0.013}_{-0.002}$ & 0.0138$^{+0.0007}_{-0.0023}$ \\ 
BP Tau & 2.11 & 59948.63 & 1.4$^{+0.3}_{-0.3}$ & 7.1$^{+1.5}_{-1.5}$ & 1.1$^{+0.3}_{-0.3}$ & 0.07$^{+0.05}_{-0.04}$ & 0.01$^{+0.008}_{-0.008}$ \\ 
BP Tau & 2.12 & 59950.48 & 1.6$^{+0.4}_{-0.5}$ & 5.7$^{+1.6}_{-2.1}$ & 1.5$^{+0.6}_{-0.2}$ & 0.11$^{+0.02}_{-0.09}$ & 0.004$^{+0.016}_{-0.004}$ \\ 
\hline
\hline
GM Aur & 1.1 & 59500.52 & 0.49$^{+0.04}_{-0.04}$ & 7.2$^{+0.9}_{-0.9}$ & 0.59$^{+0.04}_{-0.04}$ & 0.0007$^{+0.0025}_{-0.0005}$ & 0.0008$^{+0.0003}_{-0.0005}$ \\ 
GM Aur & 1.2 & 59502.30 & 0.86$^{+0.05}_{-0.05}$ & 7.2$^{+1.1}_{-1.1}$ & 1.14$^{+0.05}_{-0.06}$ & 0.0007$^{+0.0031}_{-0.0005}$ & 0.0152$^{+0.0004}_{-0.0007}$ \\ 
GM Aur & 1.3 & 59503.69 & 0.58$^{+0.03}_{-0.03}$ & 9.5$^{+0.8}_{-0.8}$ & 0.46$^{+0.03}_{-0.04}$ & 0.0006$^{+0.0022}_{-0.0004}$ & 0.0049$^{+0.0003}_{-0.0004}$ \\ 
GM Aur & 1.4 & 59504.88 & 0.43$^{+0.03}_{-0.03}$ & 5.1$^{+0.7}_{-0.7}$ & 0.58$^{+0.03}_{-0.03}$ & 0.0005$^{+0.0014}_{-0.0003}$ & 0.0026$^{+0.0002}_{-0.0003}$ \\ 
GM Aur & 1.5 & 59506.67 & 0.58$^{+0.04}_{-0.04}$ & 3.6$^{+0.8}_{-0.8}$ & 1.11$^{+0.04}_{-0.05}$ & 0.0007$^{+0.0036}_{-0.0006}$ & 0.003$^{+0.0003}_{-0.0006}$ \\ 
GM Aur & 1.6 & 59507.93 & 1.04$^{+0.03}_{-0.02}$ & 0.1$^{+0.6}_{-0.1}$ & 2.38$^{+0.04}_{-0.05}$ & 0.0007$^{+0.0034}_{-0.0006}$ & 0.0132$^{+0.0005}_{-0.0007}$ \\ 
GM Aur & 1.7 & 59509.65 & 2.15$^{+0.08}_{-0.08}$ & 28.6$^{+1.8}_{-1.8}$ & 1.7$^{+0.1}_{-0.1}$ & 0.0008$^{+0.0043}_{-0.0006}$ & 0.04$^{+0.001}_{-0.001}$ \\ 
GM Aur & 1.8 & 59554.32 & 0.26$^{+0.03}_{-0.02}$ & 0.07$^{+0.27}_{-0.05}$ & 0.61$^{+0.03}_{-0.04}$ & 0.004$^{+0.007}_{-0.004}$ & 0.0013$^{+0.0007}_{-0.0013}$ \\ 
GM Aur & 1.9 & 59557.17 & 0.96$^{+0.06}_{-0.06}$ & 9.8$^{+1.3}_{-1.3}$ & 1.21$^{+0.06}_{-0.07}$ & 0.0009$^{+0.0054}_{-0.0007}$ & 0.013$^{+0.0006}_{-0.001}$ \\ 
GM Aur & 1.10 & 59558.69 & 0.39$^{+0.03}_{-0.02}$ & 0.2$^{+0.7}_{-0.1}$ & 0.92$^{+0.03}_{-0.04}$ & 0.0006$^{+0.0023}_{-0.0005}$ & 0.0037$^{+0.0003}_{-0.0005}$ \\ 
GM Aur & 1.11 & 59560.21 & 0.32$^{+0.02}_{-0.01}$ & 0.1$^{+0.37}_{-0.08}$ & 0.68$^{+0.02}_{-0.03}$ & 0.0007$^{+0.0031}_{-0.0006}$ & 0.006$^{+0.0003}_{-0.0006}$ \\ 
\hline 
GM Aur & 2.1 & 59909.89 & 0.51$^{+0.04}_{-0.04}$ & 9.5$^{+1.0}_{-1.1}$ & 0.4$^{+0.04}_{-0.04}$ & 0.0007$^{+0.0029}_{-0.0005}$ & 0.001$^{+0.0003}_{-0.0005}$ \\ 
GM Aur & 2.2 & 59911.41 & 0.55$^{+0.05}_{-0.04}$ & 8.0$^{+1.1}_{-1.1}$ & 0.62$^{+0.05}_{-0.05}$ & 0.0006$^{+0.0019}_{-0.0004}$ & 0.0026$^{+0.0003}_{-0.0004}$ \\ 
GM Aur & 2.3 & 59912.80 & 1.07$^{+0.05}_{-0.06}$ & 7.5$^{+1.2}_{-1.3}$ & 1.25$^{+0.07}_{-0.07}$ & 0.0007$^{+0.0031}_{-0.0005}$ & 0.0293$^{+0.0007}_{-0.0008}$ \\ 
GM Aur & 2.4 & 59914.39 & 0.57$^{+0.05}_{-0.05}$ & 7.3$^{+1.1}_{-1.1}$ & 0.73$^{+0.05}_{-0.05}$ & 0.0006$^{+0.0025}_{-0.0005}$ & 0.0027$^{+0.0003}_{-0.0005}$ \\ 
GM Aur & 2.5 & 59915.78 & 0.69$^{+0.06}_{-0.06}$ & 8.4$^{+1.4}_{-1.3}$ & 0.83$^{+0.06}_{-0.07}$ & 0.0006$^{+0.0029}_{-0.0005}$ & 0.0064$^{+0.0004}_{-0.0006}$ \\ 
GM Aur & 2.6 & 59917.30 & 0.71$^{+0.05}_{-0.05}$ & 6.0$^{+1.3}_{-1.3}$ & 1.19$^{+0.06}_{-0.06}$ & 0.0007$^{+0.0034}_{-0.0006}$ & 0.0044$^{+0.0004}_{-0.0006}$ \\ 
GM Aur & 2.7 & 59919.29 & 0.69$^{+0.03}_{-0.03}$ & 5.2$^{+0.8}_{-0.8}$ & 1.1$^{+0.04}_{-0.04}$ & 0.0005$^{+0.0018}_{-0.0004}$ & 0.008$^{+0.0003}_{-0.0004}$ \\ 
GM Aur & 2.8 & 59920.34 & 0.6$^{+0.03}_{-0.03}$ & 8.6$^{+0.8}_{-0.8}$ & 0.67$^{+0.03}_{-0.03}$ & 0.0004$^{+0.0013}_{-0.0003}$ & 0.003$^{+0.0002}_{-0.0003}$ \\ 
GM Aur & 2.9 & 59922.33 & 0.69$^{+0.05}_{-0.05}$ & 12.7$^{+1.1}_{-1.2}$ & 0.5$^{+0.05}_{-0.05}$ & 0.0005$^{+0.0015}_{-0.0003}$ & 0.0031$^{+0.0003}_{-0.0003}$ \\ 
GM Aur & 2.10 & 59923.52 & 0.67$^{+0.04}_{-0.04}$ & 5.8$^{+0.9}_{-0.9}$ & 1.12$^{+0.04}_{-0.04}$ & 0.0005$^{+0.0017}_{-0.0004}$ & 0.0036$^{+0.0003}_{-0.0003}$ \\ 
GM Aur & 2.11 & 59925.17 & 0.62$^{+0.04}_{-0.04}$ & 3.3$^{+1.0}_{-1.0}$ & 1.17$^{+0.04}_{-0.05}$ & 0.0006$^{+0.0024}_{-0.0004}$ & 0.0059$^{+0.0003}_{-0.0004}$ \\ 
\hline
\enddata
\tablenotetext{a}{No contemporaneous photometry}
\end{deluxetable}






\section{UV Luminosities}

In Tables \ref{tab: UV Luminosities, TW Hya}--\ref{tab: UV Luminosities, GM Aur} are the UV feature luminosities, including 11 FUV-NUV lines, FUV/NUV continua, and the \hh~Bump. Table \ref{tab: UV Correlations} gives the log-log linear fit and Spearman correlation coefficients between L$_{acc}$ and L$_{UV}$ for each UV feature. 

\begin{longrotatetable}
\centerwidetable
\begin{deluxetable}{c c | c c c c c c c c c c c c c c} \label{tab: UV Luminosities, TW Hya}
\tablecaption{UV line, FUV/NUV continuum, and H$_2$ bump luminosities for TW Hya in units of 10$^{29}$ erg s$^{-1}$.}
\tabletypesize{\footnotesize}
\centering
\tablehead{\colhead{Object} & \colhead{Visit} & \colhead{Si III} & \colhead{Si IV} &\colhead{C I} & \colhead{C IV} & \colhead{He II} & \colhead{O III]} & \colhead{Si II} & \colhead{Si III]} & \colhead{C III]} & \colhead{Al III]} & \colhead{Mg II} & \colhead{FUV} & \colhead{NUV} & \colhead{Bump}}
\startdata
TW~Hya & 1.1 & 3.2$\pm$0.1 & 4.3$\pm$0.1 & 2.4$\pm$0.1 & 37.6$\pm$0.3 & 21.6$\pm$0.3 & 0.3$\pm$0.1 & 1.3$\pm$0.4 & 0.8$\pm$0.3 & 1.0$\pm$0.3 & 2.4$\pm$0.3 & 83.8$\pm$1.5 & 40.9$\pm$0.1 & 83.1$\pm$0.2 & 3.5$\pm$0.4 \\ 
TW~Hya & 1.2 & 3.7$\pm$0.1 & 5.3$\pm$0.1 & 2.8$\pm$0.1 & 47.9$\pm$0.3 & 24.4$\pm$0.3 & 0.5$\pm$0.1 & 1.4$\pm$0.4 & 1.2$\pm$0.3 & 2.0$\pm$0.4 & 2.8$\pm$0.6 & 117.0$\pm$1.8 & 73.3$\pm$0.2 & 165.0$\pm$0.4 & 4.3$\pm$0.4 \\ 
TW~Hya & 1.3 & 3.1$\pm$0.1 & 6.4$\pm$0.1 & 3.2$\pm$0.1 & 93.7$\pm$0.4 & 32.8$\pm$0.4 & 0.9$\pm$0.1 & 2.4$\pm$0.6 & 1.2$\pm$0.5 & 1.6$\pm$0.5 & 1.3$\pm$0.6 & 109.0$\pm$1.7 & 115.0$\pm$0.3 & 261.0$\pm$0.6 & 6.7$\pm$0.7 \\ 
TW~Hya & 1.4 & 3.0$\pm$0.1 & 5.2$\pm$0.1 & 2.7$\pm$0.1 & 64.6$\pm$0.3 & 27.4$\pm$0.3 & 0.7$\pm$0.1 & 3.1$\pm$0.6 & 1.4$\pm$0.6 & 0.3$\pm$0.5 & 1.7$\pm$0.8 & 141.0$\pm$2.3 & 128.0$\pm$0.3 & 268.0$\pm$0.7 & 5.4$\pm$0.6 \\ 
TW~Hya & 1.5 & 3.3$\pm$0.1 & 6.1$\pm$0.1 & 3.0$\pm$0.1 & 57.1$\pm$0.4 & 27.2$\pm$0.3 & 0.5$\pm$0.1 & 1.4$\pm$0.4 & 1.7$\pm$0.4 & 1.7$\pm$0.4 & 2.2$\pm$0.5 & 104.0$\pm$1.5 & 63.8$\pm$0.2 & 135.0$\pm$0.3 & 4.3$\pm$0.4 \\ 
TW~Hya & 1.6 & 4.0$\pm$0.1 & 6.3$\pm$0.1 & 3.0$\pm$0.1 & 96.0$\pm$0.4 & 39.4$\pm$0.3 & 0.8$\pm$0.1 & 0.3$\pm$0.5 & 1.0$\pm$0.6 & 2.1$\pm$0.5 & 1.5$\pm$0.8 & 110.0$\pm$2.1 & 158.0$\pm$0.4 & 343.0$\pm$0.8 & 7.9$\pm$0.8 \\ 
TW~Hya & 1.7 & 4.2$\pm$0.1 & 5.6$\pm$0.1 & 3.2$\pm$0.1 & 69.5$\pm$0.4 & 35.5$\pm$0.4 & 0.4$\pm$0.1 & 1.7$\pm$0.4 & 0.2$\pm$0.3 & 1.5$\pm$0.4 & 1.7$\pm$0.6 & 68.5$\pm$1.4 & 79.1$\pm$0.2 & 169.0$\pm$0.4 & 5.2$\pm$0.5 \\ 
TW~Hya & 1.8 & 4.3$\pm$0.1 & 7.3$\pm$0.1 & 3.7$\pm$0.1 & 131.0$\pm$0.5 & 49.0$\pm$0.4 & 1.0$\pm$0.1 & 2.4$\pm$0.7 & 1.8$\pm$0.5 & 1.7$\pm$0.6 & 2.4$\pm$0.7 & 123.0$\pm$1.7 & 176.0$\pm$0.4 & 407.0$\pm$1.1 & 7.4$\pm$0.8 \\ 
TW~Hya & 1.9 & 4.5$\pm$0.1 & 6.4$\pm$0.1 & 3.6$\pm$0.1 & 74.6$\pm$0.4 & 31.7$\pm$0.3 & 0.6$\pm$0.1 & 1.6$\pm$0.3 & 0.9$\pm$0.3 & 1.7$\pm$0.3 & 1.9$\pm$0.5 & 89.6$\pm$1.1 & 66.9$\pm$0.2 & 131.0$\pm$0.3 & 5.7$\pm$0.6 \\ 
TW~Hya & 1.10 & 4.0$\pm$0.1 & 5.6$\pm$0.1 & 3.2$\pm$0.1 & 43.7$\pm$0.2 & 18.9$\pm$0.3 & 0.6$\pm$0.1 & 1.0$\pm$0.4 & 0.4$\pm$0.3 & 2.3$\pm$0.4 & 0.7$\pm$0.5 & 100.0$\pm$1.6 & 49.0$\pm$0.1 & 106.0$\pm$0.3 & 5.2$\pm$0.5 \\ 
TW~Hya & 1.11 & 3.6$\pm$0.1 & 5.1$\pm$0.1 & 2.8$\pm$0.1 & 33.5$\pm$0.2 & 20.3$\pm$0.3 & 0.3$\pm$0.1 & 0.8$\pm$0.3 & 0.4$\pm$0.2 & 1.1$\pm$0.3 & 1.1$\pm$0.3 & 57.3$\pm$1.1 & 42.7$\pm$0.1 & 70.5$\pm$0.2 & 4.6$\pm$0.5 \\ 
\hline 
TW~Hya & 2.1 & 3.9$\pm$0.1 & 5.2$\pm$0.1 & 3.3$\pm$0.1 & 58.8$\pm$0.4 & 29.3$\pm$0.4 & 0.3$\pm$0.1 & 0.4$\pm$0.3 & -0.0$\pm$0.3 & 0.8$\pm$0.3 & 1.3$\pm$0.5 & 65.7$\pm$1.3 & 59.1$\pm$0.2 & 109.0$\pm$0.3 & 5.8$\pm$0.6 \\ 
TW~Hya & 2.2 & 4.1$\pm$0.1 & 5.6$\pm$0.1 & 3.2$\pm$0.1 & 81.2$\pm$0.5 & 36.7$\pm$0.5 & 0.7$\pm$0.2 & 0.1$\pm$0.8 & 0.6$\pm$0.7 & 1.1$\pm$0.8 & -0.0$\pm$0.7 & 49.1$\pm$1.5 & 301.0$\pm$0.7 & 385.0$\pm$1.0 & 6.6$\pm$0.7 \\ 
TW~Hya & 2.3 & 3.5$\pm$0.1 & 5.0$\pm$0.1 & 3.1$\pm$0.1 & 65.4$\pm$0.4 & 31.1$\pm$0.5 & 0.4$\pm$0.1 & 1.5$\pm$0.4 & 0.9$\pm$0.3 & 1.2$\pm$0.4 & 1.2$\pm$0.5 & 56.7$\pm$1.2 & 75.5$\pm$0.2 & 159.0$\pm$0.4 & 5.3$\pm$0.5 \\ 
TW~Hya & 2.4 & 3.5$\pm$0.1 & 6.2$\pm$0.1 & 3.3$\pm$0.1 & 134.0$\pm$0.5 & 47.0$\pm$0.4 & 0.7$\pm$0.1 & 0.9$\pm$0.4 & 1.1$\pm$0.5 & 2.2$\pm$0.5 & 3.7$\pm$0.7 & 86.4$\pm$1.5 & 126.0$\pm$0.3 & 283.0$\pm$0.7 & 7.7$\pm$0.8 \\ 
TW~Hya & 2.5 & 4.0$\pm$0.1 & 6.5$\pm$0.1 & 3.6$\pm$0.1 & 102.0$\pm$0.5 & 49.5$\pm$0.6 & 0.8$\pm$0.1 & 1.3$\pm$0.6 & 1.1$\pm$0.5 & 0.8$\pm$0.5 & 0.9$\pm$0.5 & 57.8$\pm$1.8 & 145.0$\pm$0.4 & 274.0$\pm$0.7 & 7.3$\pm$0.7 \\ 
TW~Hya & 2.6 & 3.3$\pm$0.1 & 5.2$\pm$0.1 & 3.3$\pm$0.1 & 108.0$\pm$0.6 & 51.2$\pm$0.6 & 0.8$\pm$0.2 & 1.7$\pm$0.7 & 0.9$\pm$0.7 & 2.3$\pm$0.7 & 0.4$\pm$0.8 & 84.4$\pm$1.9 & 253.0$\pm$0.6 & 453.0$\pm$1.1 & 5.3$\pm$0.6 \\ 
TW~Hya & 2.7 & 3.8$\pm$0.1 & 5.6$\pm$0.1 & 3.2$\pm$0.1 & 90.2$\pm$0.5 & 41.8$\pm$0.5 & 0.7$\pm$0.1 & 2.3$\pm$0.5 & 0.7$\pm$0.5 & 1.6$\pm$0.6 & 3.0$\pm$0.7 & 82.1$\pm$1.8 & 157.0$\pm$0.4 & 311.0$\pm$0.7 & 5.4$\pm$0.6 \\ 
TW~Hya & 2.8 & 3.9$\pm$0.1 & 6.5$\pm$0.1 & 3.7$\pm$0.1 & 100.0$\pm$0.5 & 45.5$\pm$0.6 & 0.9$\pm$0.2 & 0.7$\pm$0.8 & 1.1$\pm$0.6 & 1.2$\pm$0.7 & -0.0$\pm$0.8 & 66.2$\pm$1.9 & 319.0$\pm$0.8 & 448.0$\pm$1.2 & 6.8$\pm$0.7 \\ 
TW~Hya & 2.9 & 3.6$\pm$0.1 & 5.6$\pm$0.1 & 3.4$\pm$0.1 & 68.6$\pm$0.5 & 34.5$\pm$0.5 & 0.5$\pm$0.1 & 0.9$\pm$0.4 & 0.9$\pm$0.4 & 0.7$\pm$0.4 & 2.0$\pm$0.5 & 65.6$\pm$1.4 & 94.0$\pm$0.2 & 186.0$\pm$0.5 & 6.3$\pm$0.6 \\ 
TW~Hya & 2.10 & 4.1$\pm$0.1 & 6.6$\pm$0.1 & 3.5$\pm$0.1 & 78.5$\pm$0.4 & 39.8$\pm$0.5 & 0.2$\pm$0.2 & 0.4$\pm$0.7 & 2.4$\pm$0.8 & 1.4$\pm$0.8 & -1.6$\pm$0.8 & 111.0$\pm$1.7 & 317.0$\pm$0.8 & 346.0$\pm$0.9 & 9.4$\pm$1.0 \\ 
\hline
\enddata
\end{deluxetable}
\end{longrotatetable}

\begin{longrotatetable}
\centerwidetable
\begin{deluxetable}{c c | c c c c c c c c c c c c c c} \label{tab: UV Luminosities, RU Lup}
\tablecaption{Same as Table \ref{tab: UV Luminosities, TW Hya}, but for RU Lup}
\tabletypesize{\footnotesize}
\centering
\tablehead{\colhead{Object} & \colhead{Visit} & \colhead{Si III} & \colhead{Si IV} &\colhead{C I} & \colhead{C IV} & \colhead{He II} & \colhead{O III} & \colhead{Si II} & \colhead{Si III]} & \colhead{C III]} & \colhead{Al III]} & \colhead{Mg II} & \colhead{FUV} & \colhead{NUV} & \colhead{Bump}}
\startdata
RU~Lup & 1.1 & 0.7$\pm$0.0 & 7.8$\pm$0.1 & 0.5$\pm$0.0 & 12.7$\pm$0.2 & 1.6$\pm$0.1 & 1.1$\pm$0.1 & 3.5$\pm$0.6 & 18.1$\pm$0.8 & 0.8$\pm$0.4 & 30.7$\pm$1.2 & 338.0$\pm$2.6 & 523.0$\pm$1.5 & 4670.0$\pm$12.8 & 3.1$\pm$0.6 \\ 
RU~Lup & 1.2 & 0.6$\pm$0.0 & 8.1$\pm$0.1 & 0.5$\pm$0.0 & 14.9$\pm$0.2 & 1.8$\pm$0.1 & 1.3$\pm$0.1 & 2.7$\pm$0.5 & 12.8$\pm$0.8 & 0.5$\pm$0.5 & 23.3$\pm$0.9 & 289.0$\pm$2.8 & 501.0$\pm$1.4 & 3890.0$\pm$10.1 & 4.6$\pm$1.0 \\ 
RU~Lup & 1.3 & 0.6$\pm$0.0 & 8.9$\pm$0.1 & 0.5$\pm$0.0 & 16.5$\pm$0.2 & 2.3$\pm$0.2 & 1.3$\pm$0.1 & 8.6$\pm$0.8 & 17.5$\pm$1.0 & 1.9$\pm$0.5 & 24.6$\pm$1.2 & 314.0$\pm$2.5 & 521.0$\pm$1.4 & 4250.0$\pm$11.1 & 7.0$\pm$1.4 \\ 
RU~Lup & 1.4 & 0.7$\pm$0.0 & 8.8$\pm$0.1 & 0.5$\pm$0.0 & 17.3$\pm$0.2 & 2.0$\pm$0.1 & 1.2$\pm$0.1 & 6.7$\pm$0.7 & 16.7$\pm$0.9 & 0.9$\pm$0.4 & 29.2$\pm$1.1 & 322.0$\pm$2.7 & 467.0$\pm$1.3 & 3880.0$\pm$9.3 & 12.3$\pm$2.5 \\ 
RU~Lup & 1.5 & 0.6$\pm$0.0 & 10.6$\pm$0.1 & 0.5$\pm$0.0 & 20.5$\pm$0.2 & 2.3$\pm$0.1 & 1.5$\pm$0.1 & 8.8$\pm$0.6 & 20.0$\pm$1.0 & 2.0$\pm$0.5 & 26.9$\pm$1.1 & 319.0$\pm$3.1 & 582.0$\pm$1.6 & 4900.0$\pm$12.0 & 6.3$\pm$1.3 \\ 
RU~Lup & 1.6 & 0.6$\pm$0.0 & 7.2$\pm$0.1 & 0.5$\pm$0.0 & 12.9$\pm$0.2 & 1.7$\pm$0.1 & 1.3$\pm$0.1 & 6.5$\pm$0.6 & 15.7$\pm$0.8 & 0.6$\pm$0.4 & 33.6$\pm$1.3 & 295.0$\pm$2.6 & 529.0$\pm$1.5 & 4760.0$\pm$12.1 & 15.0$\pm$3.0 \\ 
RU~Lup & 1.7 & 0.7$\pm$0.0 & 6.6$\pm$0.1 & 0.4$\pm$0.0 & 11.9$\pm$0.2 & 1.6$\pm$0.1 & 1.2$\pm$0.1 & 4.8$\pm$0.6 & 13.0$\pm$0.9 & 1.7$\pm$0.4 & 29.1$\pm$1.1 & 308.0$\pm$2.7 & 435.0$\pm$1.3 & 4290.0$\pm$10.9 & 1.4$\pm$0.3 \\ 
RU~Lup & 1.8 & 0.5$\pm$0.0 & 7.0$\pm$0.1 & 0.5$\pm$0.0 & 12.9$\pm$0.2 & 1.6$\pm$0.1 & 1.4$\pm$0.1 & 2.2$\pm$0.5 & 17.0$\pm$0.9 & 0.7$\pm$0.5 & 37.8$\pm$1.4 & 359.0$\pm$3.2 & 558.0$\pm$1.7 & 5360.0$\pm$13.5 & 3.5$\pm$0.7 \\ 
RU~Lup & 1.9 & 0.6$\pm$0.0 & 7.6$\pm$0.1 & 0.5$\pm$0.0 & 14.0$\pm$0.2 & 1.5$\pm$0.1 & 1.2$\pm$0.1 & 6.0$\pm$0.6 & 17.2$\pm$0.9 & 1.3$\pm$0.5 & 29.4$\pm$1.2 & 297.0$\pm$3.1 & 545.0$\pm$1.5 & 5330.0$\pm$13.8 & 10.3$\pm$2.1 \\ 
RU~Lup & 1.10 & 0.5$\pm$0.0 & 9.5$\pm$0.1 & 0.5$\pm$0.0 & 19.0$\pm$0.2 & 2.0$\pm$0.1 & 1.5$\pm$0.1 & 3.0$\pm$0.6 & 19.7$\pm$0.9 & 2.6$\pm$0.5 & 24.4$\pm$1.2 & 430.0$\pm$3.5 & 574.0$\pm$1.6 & 5320.0$\pm$13.4 & 4.8$\pm$1.0 \\ 
\hline 
RU~Lup & 2.1 & 0.8$\pm$0.0 & 8.0$\pm$0.1 & 0.5$\pm$0.0 & 22.4$\pm$0.3 & 3.2$\pm$0.1 & 1.6$\pm$0.1 & 5.0$\pm$0.6 & 13.9$\pm$0.7 & 1.5$\pm$0.4 & 20.7$\pm$0.9 & 263.0$\pm$2.1 & 492.0$\pm$1.3 & 3440.0$\pm$8.9 & 11.3$\pm$2.3 \\ 
RU~Lup & 2.2 & 0.7$\pm$0.0 & 5.1$\pm$0.1 & 0.4$\pm$0.0 & 13.3$\pm$0.2 & 2.7$\pm$0.1 & 0.8$\pm$0.1 & 5.5$\pm$0.6 & 9.6$\pm$0.6 & 2.5$\pm$0.4 & 16.9$\pm$0.8 & 257.0$\pm$2.7 & 284.0$\pm$0.7 & 2180.0$\pm$5.5 & 9.1$\pm$1.8 \\ 
RU~Lup & 2.3 & 0.5$\pm$0.0 & 5.3$\pm$0.1 & 0.4$\pm$0.0 & 13.6$\pm$0.2 & 2.4$\pm$0.1 & 0.9$\pm$0.1 & 4.6$\pm$0.6 & 10.7$\pm$0.6 & 1.4$\pm$0.4 & 17.3$\pm$1.0 & 326.0$\pm$2.5 & 372.0$\pm$1.0 & 2940.0$\pm$7.5 & 7.5$\pm$1.5 \\ 
RU~Lup & 2.4 & 0.6$\pm$0.0 & 5.6$\pm$0.1 & 0.5$\pm$0.0 & 15.8$\pm$0.2 & 3.0$\pm$0.1 & 1.2$\pm$0.1 & 2.5$\pm$0.4 & 9.3$\pm$0.7 & 1.6$\pm$0.3 & 15.0$\pm$0.8 & 262.0$\pm$2.4 & 272.0$\pm$0.7 & 1930.0$\pm$4.8 & 6.3$\pm$1.3 \\ 
RU~Lup & 2.5 & 0.7$\pm$0.0 & 4.2$\pm$0.1 & 0.5$\pm$0.0 & 12.2$\pm$0.2 & 2.8$\pm$0.1 & 0.8$\pm$0.1 & 3.7$\pm$0.4 & 6.2$\pm$0.4 & 1.7$\pm$0.3 & 10.1$\pm$0.7 & 242.0$\pm$2.4 & 186.0$\pm$0.5 & 1180.0$\pm$3.0 & 5.0$\pm$1.0 \\ 
RU~Lup & 2.6 & 0.7$\pm$0.0 & 3.5$\pm$0.1 & 0.5$\pm$0.0 & 10.4$\pm$0.1 & 2.2$\pm$0.1 & 0.7$\pm$0.1 & 3.2$\pm$0.4 & 6.0$\pm$0.4 & 1.9$\pm$0.3 & 11.5$\pm$0.5 & 229.0$\pm$2.4 & 186.0$\pm$0.5 & 1220.0$\pm$3.0 & 10.6$\pm$2.1 \\ 
RU~Lup & 2.7 & 0.7$\pm$0.0 & 4.1$\pm$0.1 & 0.5$\pm$0.0 & 12.6$\pm$0.2 & 2.3$\pm$0.1 & 0.7$\pm$0.1 & 4.9$\pm$0.5 & 7.7$\pm$0.6 & 2.6$\pm$0.4 & 16.4$\pm$0.6 & 233.0$\pm$2.8 & 230.0$\pm$0.6 & 1620.0$\pm$4.1 & 6.3$\pm$1.3 \\ 
RU~Lup & 2.8 & 0.6$\pm$0.0 & 3.6$\pm$0.1 & 0.4$\pm$0.0 & 11.2$\pm$0.2 & 2.4$\pm$0.1 & 0.9$\pm$0.1 & 4.4$\pm$0.5 & 5.5$\pm$0.6 & 2.4$\pm$0.3 & 15.4$\pm$0.9 & 193.0$\pm$2.1 & 243.0$\pm$0.6 & 1680.0$\pm$4.4 & 8.9$\pm$1.8 \\ 
RU~Lup & 2.9 & 0.8$\pm$0.0 & 3.7$\pm$0.1 & 0.5$\pm$0.0 & 12.2$\pm$0.2 & 2.4$\pm$0.1 & 0.9$\pm$0.1 & 3.9$\pm$0.4 & 6.6$\pm$0.5 & 1.8$\pm$0.3 & 10.2$\pm$0.6 & 191.0$\pm$2.2 & 202.0$\pm$0.5 & 1480.0$\pm$3.8 & 12.6$\pm$2.5 \\ 
RU~Lup & 2.10 & 0.6$\pm$0.0 & 3.1$\pm$0.1 & 0.4$\pm$0.0 & 8.8$\pm$0.1 & 1.9$\pm$0.1 & 0.7$\pm$0.1 & 1.2$\pm$0.5 & 6.6$\pm$0.7 & 2.6$\pm$0.5 & 14.9$\pm$0.9 & 213.0$\pm$2.7 & 330.0$\pm$0.9 & 3110.0$\pm$7.9 & 6.7$\pm$1.3 \\ 
RU~Lup & 2.11 & 0.5$\pm$0.0 & 6.3$\pm$0.1 & 0.5$\pm$0.0 & 15.8$\pm$0.2 & 2.2$\pm$0.1 & 1.5$\pm$0.1 & 1.4$\pm$0.5 & 16.5$\pm$0.8 & 1.9$\pm$0.5 & 21.1$\pm$1.1 & 242.0$\pm$2.6 & 496.0$\pm$1.4 & 3810.0$\pm$9.6 & 3.8$\pm$0.8 \\ 
RU~Lup & 2.12 & 0.7$\pm$0.0 & 7.5$\pm$0.1 & 0.5$\pm$0.0 & 16.8$\pm$0.2 & 2.2$\pm$0.1 & 1.5$\pm$0.1 & 4.9$\pm$0.7 & 15.5$\pm$0.8 & 1.4$\pm$0.5 & 22.4$\pm$1.0 & 252.0$\pm$2.8 & 528.0$\pm$1.5 & 4170.0$\pm$10.8 & 7.2$\pm$1.5 \\ 
\hline
\enddata
\end{deluxetable}
\end{longrotatetable}

\begin{longrotatetable}
\centerwidetable
\begin{deluxetable}{c c | c c c c c c c c c c c c c c} \label{tab: UV Luminosities, BP Tau}
\tablecaption{Same as Table \ref{tab: UV Luminosities, TW Hya}, but for BP Tau}
\tabletypesize{\footnotesize}
\centering
\tablehead{\colhead{Object} & \colhead{Visit} & \colhead{Si III} & \colhead{Si IV} &\colhead{C I} & \colhead{C IV} & \colhead{He II} & \colhead{O III} & \colhead{Si II} & \colhead{Si III]} & \colhead{C III]} & \colhead{Al III]} & \colhead{Mg II} & \colhead{FUV} & \colhead{NUV} & \colhead{Bump}}
\startdata
BP~Tau & 1.1 & 0.2$\pm$0.0 & 1.0$\pm$0.1 & 0.3$\pm$0.1 & 8.1$\pm$0.2 & 4.7$\pm$0.3 & 0.6$\pm$0.1 & 1.2$\pm$0.2 & 2.0$\pm$0.2 & 0.2$\pm$0.2 & 0.9$\pm$0.2 & 75.2$\pm$0.5 & 386.0$\pm$0.9 & 543.0$\pm$1.3 & 69.5$\pm$7.0 \\ 
BP~Tau & 1.2 & 0.1$\pm$0.0 & 1.1$\pm$0.1 & 0.2$\pm$0.0 & 10.4$\pm$0.2 & 5.2$\pm$0.3 & 0.3$\pm$0.1 & 2.5$\pm$0.1 & 1.9$\pm$0.2 & 0.8$\pm$0.1 & 1.3$\pm$0.1 & 80.9$\pm$0.6 & 133.0$\pm$0.3 & 344.0$\pm$0.9 & 24.8$\pm$2.5 \\ 
BP~Tau & 1.3 & 0.3$\pm$0.1 & 1.4$\pm$0.1 & 0.2$\pm$0.1 & 14.6$\pm$0.2 & 6.2$\pm$0.2 & 0.8$\pm$0.1 & 2.1$\pm$0.2 & 2.4$\pm$0.2 & 0.4$\pm$0.2 & 0.5$\pm$0.2 & 66.6$\pm$0.5 & 347.0$\pm$0.8 & 575.0$\pm$1.4 & 34.5$\pm$3.5 \\ 
BP~Tau & 1.4 & 0.3$\pm$0.0 & 0.6$\pm$0.0 & 0.2$\pm$0.0 & 6.4$\pm$0.2 & 4.0$\pm$0.2 & 0.3$\pm$0.1 & 0.9$\pm$0.2 & 1.8$\pm$0.2 & -0.0$\pm$0.1 & 0.3$\pm$0.2 & 42.3$\pm$0.4 & 199.0$\pm$0.5 & 340.0$\pm$0.8 & 48.2$\pm$4.9 \\ 
BP~Tau & 1.5 & 0.1$\pm$0.0 & 1.6$\pm$0.1 & 0.2$\pm$0.0 & 14.3$\pm$0.2 & 3.4$\pm$0.2 & 0.6$\pm$0.1 & 2.0$\pm$0.2 & 3.5$\pm$0.2 & 0.8$\pm$0.1 & 1.3$\pm$0.2 & 84.5$\pm$0.7 & 242.0$\pm$0.6 & 471.0$\pm$1.2 & 35.3$\pm$3.6 \\ 
BP~Tau & 1.6 & 0.2$\pm$0.0 & 0.8$\pm$0.0 & 0.2$\pm$0.0 & 8.5$\pm$0.2 & 4.1$\pm$0.2 & 0.4$\pm$0.1 & 1.4$\pm$0.1 & 1.5$\pm$0.2 & 0.9$\pm$0.1 & 1.0$\pm$0.1 & 65.6$\pm$0.6 & 115.0$\pm$0.3 & 327.0$\pm$0.8 & 5.2$\pm$0.6 \\ 
BP~Tau & 1.7 & 0.1$\pm$0.0 & 1.1$\pm$0.1 & 0.2$\pm$0.0 & 10.5$\pm$0.2 & 4.9$\pm$0.2 & 0.6$\pm$0.1 & 2.2$\pm$0.1 & 1.2$\pm$0.1 & 0.5$\pm$0.1 & 0.8$\pm$0.1 & 59.1$\pm$0.5 & 160.0$\pm$0.4 & 394.0$\pm$1.0 & 28.8$\pm$3.0 \\ 
BP~Tau & 1.8 & 0.2$\pm$0.0 & 0.9$\pm$0.1 & 0.2$\pm$0.0 & 8.4$\pm$0.2 & 4.5$\pm$0.2 & 0.9$\pm$0.1 & 1.6$\pm$0.1 & 1.2$\pm$0.2 & 0.1$\pm$0.1 & 0.9$\pm$0.2 & 71.9$\pm$0.5 & 178.0$\pm$0.4 & 381.0$\pm$1.0 & 38.7$\pm$3.9 \\ 
BP~Tau & 1.9 & 0.2$\pm$0.0 & 0.7$\pm$0.1 & 0.2$\pm$0.0 & 8.1$\pm$0.2 & 4.7$\pm$0.2 & 0.6$\pm$0.1 & 1.9$\pm$0.2 & 1.9$\pm$0.2 & 0.2$\pm$0.1 & 0.9$\pm$0.2 & 58.1$\pm$0.5 & 301.0$\pm$0.7 & 399.0$\pm$1.0 & 77.4$\pm$7.8 \\ 
BP~Tau & 1.10 & 0.2$\pm$0.0 & 0.6$\pm$0.0 & 0.2$\pm$0.0 & 7.5$\pm$0.2 & 4.2$\pm$0.2 & 0.4$\pm$0.1 & 0.5$\pm$0.1 & 1.1$\pm$0.2 & 0.4$\pm$0.1 & 0.4$\pm$0.2 & 49.7$\pm$0.5 & 158.0$\pm$0.4 & 356.0$\pm$0.9 & 11.8$\pm$1.2 \\ 
BP~Tau & 1.11 & 0.3$\pm$0.0 & 0.9$\pm$0.1 & 0.2$\pm$0.0 & 8.6$\pm$0.2 & 5.3$\pm$0.2 & 0.6$\pm$0.1 & 1.6$\pm$0.1 & 1.6$\pm$0.2 & 0.4$\pm$0.1 & 1.1$\pm$0.2 & 52.0$\pm$0.5 & 154.0$\pm$0.4 & 356.0$\pm$0.9 & 13.6$\pm$1.4 \\ 
BP~Tau & 1.12 & 0.3$\pm$0.0 & 0.6$\pm$0.0 & 0.2$\pm$0.0 & 6.2$\pm$0.2 & 3.6$\pm$0.2 & 0.2$\pm$0.1 & 1.3$\pm$0.1 & 1.0$\pm$0.1 & 0.4$\pm$0.1 & 0.7$\pm$0.1 & 44.9$\pm$0.4 & 84.8$\pm$0.2 & 241.0$\pm$0.6 & 30.2$\pm$3.1 \\ 
\hline 
BP~Tau & 2.1 & 0.1$\pm$0.0 & 1.0$\pm$0.0 & 0.1$\pm$0.0 & 6.9$\pm$0.1 & 4.9$\pm$0.2 & 0.4$\pm$0.1 & 0.7$\pm$0.2 & 1.6$\pm$0.3 & 0.6$\pm$0.2 & 1.7$\pm$0.3 & 72.6$\pm$0.8 & 112.0$\pm$0.3 & 313.0$\pm$0.8 & 5.0$\pm$0.6 \\ 
BP~Tau & 2.2 & 0.2$\pm$0.0 & 1.1$\pm$0.1 & 0.2$\pm$0.0 & 7.2$\pm$0.2 & 4.3$\pm$0.2 & 0.5$\pm$0.1 & 1.4$\pm$0.2 & 1.5$\pm$0.2 & 0.7$\pm$0.2 & 0.9$\pm$0.3 & 54.5$\pm$0.7 & 140.0$\pm$0.3 & 355.0$\pm$0.9 & 15.6$\pm$1.6 \\ 
BP~Tau & 2.3 & 0.2$\pm$0.0 & 0.9$\pm$0.0 & 0.1$\pm$0.0 & 6.2$\pm$0.1 & 3.7$\pm$0.2 & 0.3$\pm$0.1 & 0.6$\pm$0.2 & 0.7$\pm$0.2 & 0.5$\pm$0.2 & 1.2$\pm$0.2 & 57.3$\pm$0.8 & 71.2$\pm$0.2 & 240.0$\pm$0.6 & 4.3$\pm$0.5 \\ 
BP~Tau & 2.4 & 0.3$\pm$0.0 & 2.3$\pm$0.1 & 0.2$\pm$0.0 & 16.0$\pm$0.3 & 7.6$\pm$0.2 & 0.5$\pm$0.1 & 0.4$\pm$0.2 & 0.9$\pm$0.2 & 0.7$\pm$0.2 & 1.0$\pm$0.2 & 35.2$\pm$0.6 & 155.0$\pm$0.4 & 359.0$\pm$0.9 & 14.4$\pm$1.5 \\ 
BP~Tau & 2.5 & 0.1$\pm$0.0 & 0.9$\pm$0.1 & 0.1$\pm$0.0 & 7.6$\pm$0.1 & 6.4$\pm$0.2 & 0.4$\pm$0.1 & 1.4$\pm$0.3 & 0.7$\pm$0.3 & -0.4$\pm$0.3 & 0.4$\pm$0.2 & 55.0$\pm$0.9 & 192.0$\pm$0.4 & 300.0$\pm$0.9 & 15.4$\pm$1.6 \\ 
BP~Tau & 2.6 & 0.1$\pm$0.0 & 1.0$\pm$0.0 & 0.1$\pm$0.0 & 7.6$\pm$0.2 & 6.2$\pm$0.2 & 0.3$\pm$0.1 & 1.2$\pm$0.2 & 0.7$\pm$0.2 & 0.3$\pm$0.1 & 1.2$\pm$0.2 & 42.8$\pm$0.7 & 81.1$\pm$0.2 & 303.0$\pm$0.7 & 5.9$\pm$0.7 \\ 
BP~Tau & 2.7 & 0.1$\pm$0.0 & 0.7$\pm$0.0 & 0.1$\pm$0.0 & 5.5$\pm$0.1 & 4.5$\pm$0.2 & 0.2$\pm$0.1 & 0.9$\pm$0.2 & 0.6$\pm$0.3 & -0.3$\pm$0.2 & 1.1$\pm$0.2 & 36.5$\pm$0.6 & 94.1$\pm$0.2 & 264.0$\pm$0.7 & 7.0$\pm$0.7 \\ 
BP~Tau & 2.8 & 0.1$\pm$0.0 & 0.6$\pm$0.0 & 0.1$\pm$0.0 & 5.9$\pm$0.1 & 4.7$\pm$0.2 & 0.3$\pm$0.1 & 0.3$\pm$0.2 & 0.4$\pm$0.2 & -0.3$\pm$0.2 & 1.0$\pm$0.2 & 43.5$\pm$0.7 & 117.0$\pm$0.3 & 343.0$\pm$0.9 & 20.5$\pm$2.1 \\ 
BP~Tau & 2.9 & 0.1$\pm$0.0 & 1.2$\pm$0.1 & 0.2$\pm$0.0 & 9.1$\pm$0.2 & 7.3$\pm$0.2 & 0.4$\pm$0.1 & 0.8$\pm$0.3 & 1.2$\pm$0.3 & 0.2$\pm$0.3 & 1.1$\pm$0.3 & 58.1$\pm$0.8 & 217.0$\pm$0.5 & 491.0$\pm$1.2 & 26.4$\pm$2.7 \\ 
BP~Tau & 2.10 & 0.2$\pm$0.0 & 1.1$\pm$0.0 & 0.1$\pm$0.0 & 7.5$\pm$0.1 & 4.8$\pm$0.2 & 0.3$\pm$0.1 & 1.3$\pm$0.2 & 0.5$\pm$0.2 & 0.2$\pm$0.2 & 1.1$\pm$0.2 & 52.2$\pm$0.9 & 101.0$\pm$0.2 & 299.0$\pm$0.7 & 17.8$\pm$1.8 \\ 
BP~Tau & 2.11 & 0.2$\pm$0.0 & 1.1$\pm$0.0 & 0.2$\pm$0.0 & 7.3$\pm$0.1 & 5.2$\pm$0.2 & 0.6$\pm$0.1 & 0.7$\pm$0.2 & 0.6$\pm$0.2 & 0.4$\pm$0.2 & 1.2$\pm$0.2 & 50.1$\pm$0.8 & 148.0$\pm$0.4 & 357.0$\pm$0.9 & 21.2$\pm$2.2 \\ 
BP~Tau & 2.12 & 0.2$\pm$0.0 & 1.1$\pm$0.0 & 0.1$\pm$0.0 & 8.4$\pm$0.1 & 5.2$\pm$0.2 & 0.4$\pm$0.1 & 1.5$\pm$0.3 & 1.3$\pm$0.3 & 0.1$\pm$0.2 & 1.6$\pm$0.3 & 81.4$\pm$0.9 & 157.0$\pm$0.4 & 437.0$\pm$1.1 & 8.2$\pm$0.8 \\ 
\hline
\enddata
\end{deluxetable}
\end{longrotatetable}

\begin{longrotatetable}
\centerwidetable
\begin{deluxetable}{c c | c c c c c c c c c c c c c c} \label{tab: UV Luminosities, GM Aur}
\tablecaption{Same as Table \ref{tab: UV Luminosities, TW Hya}, but for GM Aur}
\tabletypesize{\footnotesize}
\centering
\tablehead{\colhead{Object} & \colhead{Visit} & \colhead{Si III} & \colhead{Si IV} &\colhead{C I} & \colhead{C IV} & \colhead{He II} & \colhead{O III} & \colhead{Si II} & \colhead{Si III]} & \colhead{C III]} & \colhead{Al III]} & \colhead{Mg II} & \colhead{FUV} & \colhead{NUV} & \colhead{Bump}}
\startdata
GM~Aur & 1.1 & 0.4$\pm$0.0 & 0.4$\pm$0.0 & 0.3$\pm$0.0 & 5.3$\pm$0.1 & 1.9$\pm$0.1 & 0.1$\pm$0.1 & 0.2$\pm$0.1 & 0.1$\pm$0.1 & 0.0$\pm$0.1 & 0.6$\pm$0.1 & 24.0$\pm$0.3 & 52.7$\pm$0.1 & 161.0$\pm$0.4 & 22.4$\pm$2.3 \\ 
GM~Aur & 1.2 & 0.4$\pm$0.0 & 0.5$\pm$0.0 & 0.3$\pm$0.0 & 6.3$\pm$0.1 & 2.7$\pm$0.2 & 0.2$\pm$0.1 & 0.0$\pm$0.1 & 0.2$\pm$0.1 & 0.4$\pm$0.1 & 0.1$\pm$0.1 & 23.5$\pm$0.4 & 150.0$\pm$0.4 & 343.0$\pm$0.9 & 11.7$\pm$1.3 \\ 
GM~Aur & 1.3 & 0.5$\pm$0.0 & 0.4$\pm$0.0 & 0.3$\pm$0.0 & 4.4$\pm$0.1 & 1.8$\pm$0.1 & 0.1$\pm$0.1 & 0.3$\pm$0.1 & 0.1$\pm$0.1 & 0.1$\pm$0.1 & 0.3$\pm$0.1 & 20.0$\pm$0.3 & 76.9$\pm$0.2 & 179.0$\pm$0.4 & 32.2$\pm$3.3 \\ 
GM~Aur & 1.4 & 0.4$\pm$0.0 & 0.4$\pm$0.0 & 0.3$\pm$0.0 & 3.1$\pm$0.1 & 1.3$\pm$0.2 & 0.0$\pm$0.1 & 0.2$\pm$0.1 & 0.0$\pm$0.1 & 0.1$\pm$0.1 & 0.6$\pm$0.1 & 23.9$\pm$0.3 & 46.2$\pm$0.1 & 157.0$\pm$0.4 & 23.5$\pm$2.4 \\ 
GM~Aur & 1.5 & 0.4$\pm$0.0 & 0.5$\pm$0.0 & 0.3$\pm$0.0 & 6.3$\pm$0.1 & 2.0$\pm$0.1 & 0.0$\pm$0.1 & 0.3$\pm$0.1 & 0.2$\pm$0.1 & 0.1$\pm$0.1 & 0.5$\pm$0.1 & 26.3$\pm$0.3 & 67.2$\pm$0.2 & 236.0$\pm$0.6 & 22.7$\pm$2.4 \\ 
GM~Aur & 1.6 & 0.4$\pm$0.0 & 0.6$\pm$0.0 & 0.3$\pm$0.0 & 11.8$\pm$0.2 & 3.1$\pm$0.1 & 0.0$\pm$0.1 & 0.6$\pm$0.1 & 0.3$\pm$0.1 & 0.1$\pm$0.1 & 0.6$\pm$0.1 & 41.5$\pm$0.4 & 157.0$\pm$0.4 & 487.0$\pm$1.2 & 37.2$\pm$3.8 \\ 
GM~Aur & 1.7 & 0.3$\pm$0.0 & 0.8$\pm$0.0 & 0.4$\pm$0.0 & 12.4$\pm$0.2 & 3.3$\pm$0.2 & 0.7$\pm$0.1 & 0.4$\pm$0.1 & 0.7$\pm$0.1 & 0.2$\pm$0.1 & 0.0$\pm$0.2 & 39.3$\pm$0.4 & 393.0$\pm$0.9 & 737.0$\pm$1.9 & 58.9$\pm$6.1 \\ 
GM~Aur & 1.8 & 0.4$\pm$0.0 & 0.4$\pm$0.0 & 0.3$\pm$0.0 & 4.0$\pm$0.1 & 1.7$\pm$0.1 & 0.1$\pm$0.1 & 0.2$\pm$0.1 & 0.1$\pm$0.1 & 0.1$\pm$0.1 & 0.2$\pm$0.2 & 14.4$\pm$0.5 & 44.8$\pm$0.1 & 132.0$\pm$0.3 & 26.5$\pm$2.7 \\ 
GM~Aur & 1.9 & 0.4$\pm$0.0 & 0.5$\pm$0.0 & 0.3$\pm$0.0 & 7.3$\pm$0.2 & 2.7$\pm$0.1 & 0.1$\pm$0.1 & 0.1$\pm$0.2 & 0.3$\pm$0.1 & 0.2$\pm$0.2 & 0.1$\pm$0.3 & 19.0$\pm$0.6 & 148.0$\pm$0.4 & 360.0$\pm$0.9 & 47.0$\pm$4.8 \\ 
GM~Aur & 1.10 & 0.3$\pm$0.0 & 0.4$\pm$0.0 & 0.3$\pm$0.0 & 5.2$\pm$0.1 & 1.8$\pm$0.1 & 0.1$\pm$0.1 & 0.6$\pm$0.2 & 0.0$\pm$0.1 & 0.1$\pm$0.1 & 0.6$\pm$0.2 & 17.7$\pm$0.5 & 60.0$\pm$0.1 & 191.0$\pm$0.5 & 35.6$\pm$3.6 \\ 
GM~Aur & 1.11 & 0.3$\pm$0.0 & 0.3$\pm$0.0 & 0.2$\pm$0.0 & 2.6$\pm$0.1 & 1.0$\pm$0.1 & 0.1$\pm$0.1 & 0.2$\pm$0.2 & 0.3$\pm$0.1 & 0.3$\pm$0.1 & 0.2$\pm$0.2 & 13.9$\pm$0.5 & 59.1$\pm$0.1 & 156.0$\pm$0.4 & 28.8$\pm$2.9 \\ 
\hline 
GM~Aur & 2.1 & 0.3$\pm$0.0 & 0.3$\pm$0.0 & 0.2$\pm$0.0 & 4.8$\pm$0.1 & 2.0$\pm$0.1 & 0.2$\pm$0.1 & 0.1$\pm$0.1 & 0.1$\pm$0.1 & 0.1$\pm$0.1 & 0.1$\pm$0.1 & 19.0$\pm$0.2 & 46.2$\pm$0.1 & 149.0$\pm$0.4 & 30.5$\pm$3.1 \\ 
GM~Aur & 2.2 & 0.4$\pm$0.0 & 0.3$\pm$0.0 & 0.3$\pm$0.0 & 5.0$\pm$0.1 & 2.3$\pm$0.1 & 0.1$\pm$0.1 & 0.2$\pm$0.1 & 0.1$\pm$0.1 & 0.1$\pm$0.1 & 0.2$\pm$0.1 & 19.5$\pm$0.3 & 59.0$\pm$0.1 & 187.0$\pm$0.5 & 30.3$\pm$3.1 \\ 
GM~Aur & 2.3 & 0.4$\pm$0.0 & 0.5$\pm$0.0 & 0.2$\pm$0.0 & 5.9$\pm$0.2 & 2.3$\pm$0.2 & 0.2$\pm$0.1 & 0.4$\pm$0.2 & 0.6$\pm$0.1 & 0.1$\pm$0.1 & -0.1$\pm$0.1 & 24.5$\pm$0.4 & 280.0$\pm$0.7 & 449.0$\pm$1.2 & 29.2$\pm$3.0 \\ 
GM~Aur & 2.4 & 0.3$\pm$0.0 & 0.4$\pm$0.0 & 0.2$\pm$0.0 & 5.9$\pm$0.2 & 1.9$\pm$0.1 & 0.1$\pm$0.1 & 0.2$\pm$0.1 & 0.2$\pm$0.1 & 0.1$\pm$0.1 & 0.2$\pm$0.1 & 19.5$\pm$0.3 & 60.0$\pm$0.1 & 201.0$\pm$0.5 & 31.1$\pm$3.2 \\ 
GM~Aur & 2.5 & 0.3$\pm$0.0 & 0.4$\pm$0.0 & 0.2$\pm$0.0 & 5.0$\pm$0.1 & 1.7$\pm$0.2 & 0.2$\pm$0.1 & 0.3$\pm$0.1 & 0.1$\pm$0.1 & 0.0$\pm$0.1 & -0.0$\pm$0.1 & 22.2$\pm$0.4 & 83.2$\pm$0.2 & 246.0$\pm$0.7 & 31.1$\pm$3.2 \\ 
GM~Aur & 2.6 & 0.3$\pm$0.0 & 0.4$\pm$0.0 & 0.2$\pm$0.0 & 6.2$\pm$0.1 & 2.2$\pm$0.2 & 0.2$\pm$0.1 & 0.2$\pm$0.1 & -0.0$\pm$0.1 & 0.1$\pm$0.1 & 0.4$\pm$0.1 & 29.4$\pm$0.4 & 75.9$\pm$0.2 & 288.0$\pm$0.7 & 20.0$\pm$2.1 \\ 
GM~Aur & 2.7 & 0.3$\pm$0.0 & 0.4$\pm$0.0 & 0.2$\pm$0.0 & 5.1$\pm$0.1 & 2.3$\pm$0.2 & 0.3$\pm$0.1 & 0.1$\pm$0.1 & 0.2$\pm$0.1 & 0.2$\pm$0.1 & 0.3$\pm$0.1 & 22.4$\pm$0.3 & 100.0$\pm$0.2 & 280.0$\pm$0.7 & 24.6$\pm$2.6 \\ 
GM~Aur & 2.8 & 0.3$\pm$0.0 & 0.3$\pm$0.0 & 0.2$\pm$0.0 & 5.0$\pm$0.1 & 1.6$\pm$0.1 & 0.1$\pm$0.1 & 0.3$\pm$0.1 & 0.1$\pm$0.1 & 0.3$\pm$0.1 & 0.1$\pm$0.1 & 17.8$\pm$0.3 & 64.8$\pm$0.2 & 194.0$\pm$0.5 & 36.3$\pm$3.7 \\ 
GM~Aur & 2.9 & 0.3$\pm$0.0 & 0.4$\pm$0.0 & 0.2$\pm$0.0 & 6.1$\pm$0.1 & 2.3$\pm$0.1 & 0.1$\pm$0.1 & 0.1$\pm$0.1 & 0.4$\pm$0.1 & 0.1$\pm$0.1 & 0.0$\pm$0.1 & 20.9$\pm$0.3 & 71.8$\pm$0.2 & 205.0$\pm$0.5 & 18.3$\pm$1.9 \\ 
GM~Aur & 2.10 & 0.4$\pm$0.0 & 0.4$\pm$0.0 & 0.2$\pm$0.0 & 6.0$\pm$0.1 & 3.4$\pm$0.2 & 0.2$\pm$0.1 & 0.2$\pm$0.1 & 0.0$\pm$0.1 & 0.1$\pm$0.1 & 0.2$\pm$0.1 & 23.9$\pm$0.4 & 62.6$\pm$0.2 & 258.0$\pm$0.7 & 14.6$\pm$1.5 \\ 
GM~Aur & 2.11 & 0.3$\pm$0.0 & 0.3$\pm$0.0 & 0.2$\pm$0.0 & 4.5$\pm$0.1 & 2.0$\pm$0.1 & 0.1$\pm$0.1 & 0.3$\pm$0.1 & 0.2$\pm$0.1 & 0.2$\pm$0.1 & 0.1$\pm$0.1 & 22.5$\pm$0.3 & 84.6$\pm$0.2 & 262.0$\pm$0.6 & 15.3$\pm$1.6 \\ 
\hline
\enddata
\end{deluxetable}
\end{longrotatetable}

\begin{deluxetable*}{c c | c c c c | c}[h]
\setlength{\tabcolsep}{10pt}
\tablecaption{L$_{acc}$--L$_{UV}$ Log-log Linear Fit Coefficients} \label{tab: UV Correlations}
\tabletypesize{\scriptsize}
\tablehead{
\colhead{Line} & \colhead{Coefficient} & \colhead{TW~Hya} & \colhead{RU~Lup} & \colhead{BP~Tau} & \colhead{GM~Aur} & \colhead{Global}
}
\startdata
Mg II & $m$ & 0.20$^{0.10}_{0.10}$ & 1.85$^{0.08}_{0.07}$ & 0.76$^{0.10}_{0.10}$ & 1.25$^{0.07}_{0.07}$ & 0.78$^{0.01}_{0.01}$ \\ 
Mg II & $b$ & -0.75$^{0.27}_{0.26}$ & 3.88$^{0.16}_{0.16}$ & 1.31$^{0.29}_{0.30}$ & 3.08$^{0.24}_{0.22}$ & 1.33$^{0.02}_{0.02}$ \\ 
Mg II & $r$ & 0.10$^{0.05}_{0.05}$ & 0.73$^{0.02}_{0.02}$ & 0.59$^{0.07}_{0.07}$ & 0.75$^{0.03}_{0.03}$ & 0.66$^{0.01}_{0.01}$ \\ 
Mg II & $p$ & 0.68$^{0.15}_{0.14}$ & $<$0.01$^{0.00}_{0.00}$ & $<$0.01$^{0.01}_{0.00}$ & $<$0.01$^{0.00}_{0.00}$ & $<$0.01$^{0.00}_{0.00}$ \\ 
\hline 
Al III] & $m$ & 0.02$^{0.15}_{0.10}$ & 1.14$^{0.08}_{0.08}$ & 0.02$^{0.10}_{0.10}$ & -0.05$^{0.08}_{0.05}$ & 0.30$^{0.07}_{0.07}$ \\ 
Al III] & $b$ & -1.23$^{0.64}_{0.46}$ & 3.63$^{0.26}_{0.25}$ & -0.78$^{0.47}_{0.45}$ & -1.22$^{0.43}_{0.27}$ & 0.51$^{0.27}_{0.27}$ \\ 
Al III] & $r$ & 0.04$^{0.29}_{0.25}$ & 0.87$^{0.03}_{0.03}$ & 0.04$^{0.20}_{0.21}$ & -0.24$^{0.41}_{0.26}$ & 0.55$^{0.06}_{0.07}$ \\ 
Al III] & $p$ & 0.44$^{0.36}_{0.30}$ & $<$0.01$^{0.00}_{0.00}$ & 0.51$^{0.34}_{0.32}$ & 0.20$^{0.49}_{0.17}$ & $<$0.01$^{0.00}_{0.00}$ \\ 
\hline 
C III] & $m$ & 0.03$^{0.09}_{0.07}$ & -0.10$^{0.07}_{0.08}$ & 2.5e-3$^{0.03}_{0.03}$ & 0.01$^{0.05}_{0.04}$ & 0.08$^{0.04}_{0.03}$ \\ 
C III] & $b$ & -1.16$^{0.40}_{0.33}$ & -0.52$^{0.32}_{0.34}$ & -0.83$^{0.16}_{0.14}$ & -0.90$^{0.25}_{0.23}$ & -0.37$^{0.20}_{0.17}$ \\ 
C III] & $r$ & 0.06$^{0.15}_{0.15}$ & -0.21$^{0.16}_{0.14}$ & 0.01$^{0.16}_{0.16}$ & 0.06$^{0.20}_{0.21}$ & 0.15$^{0.07}_{0.06}$ \\ 
C III] & $p$ & 0.63$^{0.24}_{0.33}$ & 0.35$^{0.38}_{0.23}$ & 0.64$^{0.26}_{0.27}$ & 0.51$^{0.34}_{0.33}$ & 0.16$^{0.24}_{0.12}$ \\ 
\hline 
Si III] & $m$ & 0.07$^{0.11}_{0.09}$ & 0.98$^{0.10}_{0.09}$ & 0.12$^{0.07}_{0.05}$ & 0.05$^{0.05}_{0.04}$ & 0.33$^{0.06}_{0.06}$ \\ 
Si III] & $b$ & -0.95$^{0.48}_{0.42}$ & 3.36$^{0.33}_{0.30}$ & -0.32$^{0.32}_{0.24}$ & -0.70$^{0.27}_{0.22}$ & 0.70$^{0.26}_{0.25}$ \\ 
Si III] & $r$ & 0.15$^{0.21}_{0.21}$ & 0.88$^{0.03}_{0.04}$ & 0.29$^{0.13}_{0.12}$ & 0.25$^{0.19}_{0.20}$ & 0.61$^{0.05}_{0.06}$ \\ 
Si III] & $p$ & 0.44$^{0.37}_{0.34}$ & $<$0.01$^{0.00}_{0.00}$ & 0.16$^{0.27}_{0.13}$ & 0.27$^{0.45}_{0.22}$ & $<$0.01$^{0.00}_{0.00}$ \\ 
\hline 
Si II & $m$ & -0.02$^{0.09}_{0.06}$ & 0.09$^{0.12}_{0.09}$ & 0.12$^{0.07}_{0.05}$ & 0.01$^{0.06}_{0.04}$ & 0.24$^{0.07}_{0.06}$ \\ 
Si II & $b$ & -1.34$^{0.40}_{0.28}$ & 0.28$^{0.47}_{0.37}$ & -0.28$^{0.29}_{0.25}$ & -0.92$^{0.33}_{0.23}$ & 0.30$^{0.32}_{0.26}$ \\ 
Si II & $r$ & -0.05$^{0.24}_{0.20}$ & 0.12$^{0.11}_{0.12}$ & 0.30$^{0.11}_{0.11}$ & 0.03$^{0.24}_{0.19}$ & 0.38$^{0.06}_{0.06}$ \\ 
Si II & $p$ & 0.50$^{0.35}_{0.32}$ & 0.59$^{0.27}_{0.29}$ & 0.16$^{0.23}_{0.11}$ & 0.53$^{0.33}_{0.35}$ & $<$0.01$^{0.00}_{0.00}$ \\ 
\hline 
O III] & $m$ & 0.30$^{0.25}_{0.26}$ & 0.97$^{0.20}_{0.17}$ & 0.10$^{0.08}_{0.09}$ & 0.03$^{0.04}_{0.03}$ & 0.24$^{0.08}_{0.07}$ \\ 
O III] & $b$ & 0.17$^{1.21}_{1.24}$ & 4.32$^{0.94}_{0.77}$ & -0.37$^{0.40}_{0.45}$ & -0.81$^{0.21}_{0.18}$ & 0.41$^{0.37}_{0.33}$ \\ 
O III] & $r$ & 0.37$^{0.22}_{0.29}$ & 0.65$^{0.09}_{0.11}$ & 0.22$^{0.18}_{0.21}$ & 0.15$^{0.17}_{0.17}$ & 0.33$^{0.06}_{0.06}$ \\ 
O III] & $p$ & 0.10$^{0.61}_{0.09}$ & $<$0.01$^{0.01}_{0.00}$ & 0.29$^{0.48}_{0.24}$ & 0.47$^{0.35}_{0.32}$ & $<$0.01$^{0.01}_{0.00}$ \\ 
\hline 
He II & $m$ & 1.76$^{0.10}_{0.09}$ & -0.98$^{0.20}_{0.21}$ & 0.24$^{0.17}_{0.15}$ & 0.90$^{0.15}_{0.14}$ & -0.54$^{0.02}_{0.01}$ \\ 
He II & $b$ & 4.11$^{0.31}_{0.28}$ & -4.24$^{0.87}_{0.92}$ & 0.09$^{0.67}_{0.60}$ & 2.87$^{0.64}_{0.61}$ & -2.90$^{0.06}_{0.05}$ \\ 
He II & $r$ & 0.85$^{0.02}_{0.03}$ & -0.49$^{0.10}_{0.08}$ & 0.18$^{0.11}_{0.11}$ & 0.67$^{0.06}_{0.07}$ & -0.56$^{0.01}_{0.01}$ \\ 
He II & $p$ & $<$0.01$^{0.00}_{0.00}$ & 0.02$^{0.05}_{0.02}$ & 0.41$^{0.33}_{0.24}$ & $<$0.01$^{0.00}_{0.00}$ & $<$0.01$^{0.00}_{0.00}$ \\ 
\hline 
C IV & $m$ & 1.40$^{0.08}_{0.07}$ & 1.16$^{0.07}_{0.08}$ & 0.62$^{0.07}_{0.07}$ & 1.03$^{0.05}_{0.05}$ & -0.21$^{0.01}_{0.01}$ \\ 
C IV & $b$ & 2.53$^{0.20}_{0.19}$ & 3.89$^{0.25}_{0.28}$ & 1.41$^{0.25}_{0.24}$ & 3.00$^{0.20}_{0.20}$ & -1.52$^{0.03}_{0.03}$ \\ 
C IV & $r$ & 0.86$^{0.02}_{0.02}$ & 0.51$^{0.03}_{0.03}$ & 0.55$^{0.05}_{0.05}$ & 0.81$^{0.03}_{0.03}$ & -0.20$^{0.01}_{0.01}$ \\ 
C IV & $p$ & $<$0.01$^{0.00}_{0.00}$ & 0.02$^{0.01}_{0.01}$ & $<$0.01$^{0.01}_{0.00}$ & $<$0.01$^{0.00}_{0.00}$ & 0.07$^{0.01}_{0.01}$ \\ 
\hline 
C I & $m$ & 2.52$^{0.55}_{0.51}$ & 0.24$^{0.56}_{0.52}$ & 0.09$^{0.11}_{0.13}$ & 0.35$^{0.23}_{0.21}$ & -0.27$^{0.02}_{0.02}$ \\ 
C I & $b$ & 9.02$^{2.21}_{2.10}$ & 1.11$^{2.74}_{2.57}$ & -0.34$^{0.60}_{0.68}$ & 0.84$^{1.21}_{1.08}$ & -2.14$^{0.11}_{0.11}$ \\ 
C I & $r$ & 0.50$^{0.09}_{0.10}$ & 0.08$^{0.19}_{0.18}$ & 0.16$^{0.20}_{0.21}$ & 0.25$^{0.15}_{0.16}$ & -0.29$^{0.02}_{0.02}$ \\ 
C I & $p$ & 0.02$^{0.05}_{0.02}$ & 0.53$^{0.34}_{0.34}$ & 0.39$^{0.39}_{0.31}$ & 0.26$^{0.39}_{0.19}$ & $<$0.01$^{0.00}_{0.00}$ \\ 
\hline 
Si IV & $m$ & 2.73$^{0.38}_{0.35}$ & 1.18$^{0.05}_{0.04}$ & 0.39$^{0.08}_{0.08}$ & 0.99$^{0.22}_{0.19}$ & 0.30$^{0.01}_{0.01}$ \\ 
Si IV & $b$ & 9.16$^{1.45}_{1.32}$ & 4.39$^{0.17}_{0.16}$ & 0.93$^{0.38}_{0.35}$ & 3.95$^{1.08}_{0.95}$ & 0.52$^{0.04}_{0.03}$ \\ 
Si IV & $r$ & 0.57$^{0.06}_{0.07}$ & 0.84$^{0.01}_{0.02}$ & 0.43$^{0.07}_{0.09}$ & 0.61$^{0.09}_{0.11}$ & 0.33$^{0.01}_{0.01}$ \\ 
Si IV & $p$ & $<$0.01$^{0.01}_{0.00}$ & $<$0.01$^{0.00}_{0.00}$ & 0.04$^{0.07}_{0.03}$ & $<$0.01$^{0.01}_{0.00}$ & $<$0.01$^{0.00}_{0.00}$ \\ 
\hline 
Si III & $m$ & 0.88$^{0.48}_{0.47}$ & -0.79$^{0.43}_{0.39}$ & -0.06$^{0.10}_{0.09}$ & -0.13$^{0.43}_{0.35}$ & -0.26$^{0.02}_{0.02}$ \\ 
Si III & $b$ & 2.23$^{1.93}_{1.87}$ & -3.88$^{2.07}_{1.88}$ & -1.18$^{0.54}_{0.48}$ & -1.62$^{2.20}_{1.79}$ & -2.06$^{0.10}_{0.11}$ \\ 
Si III & $r$ & 0.17$^{0.09}_{0.09}$ & -0.27$^{0.14}_{0.13}$ & -0.12$^{0.20}_{0.18}$ & -0.08$^{0.24}_{0.21}$ & -0.28$^{0.02}_{0.02}$ \\ 
Si III & $p$ & 0.45$^{0.28}_{0.20}$ & 0.23$^{0.34}_{0.16}$ & 0.43$^{0.37}_{0.29}$ & 0.46$^{0.37}_{0.31}$ & $<$0.01$^{0.00}_{0.00}$ \\ 
\hline 
FUV & $m$ & 0.88$^{0.04}_{0.04}$ & 1.26$^{0.02}_{0.02}$ & 0.40$^{0.04}_{0.04}$ & 0.68$^{0.02}_{0.02}$ & 1.23$^{0.01}_{0.01}$ \\ 
FUV & $b$ & 0.95$^{0.09}_{0.08}$ & 2.44$^{0.04}_{0.05}$ & 0.10$^{0.09}_{0.10}$ & 0.86$^{0.05}_{0.05}$ & 2.15$^{0.02}_{0.02}$ \\ 
FUV & $r$ & 0.92$^{0.01}_{0.01}$ & 0.98$^{2.7e-3}_{2.9e-3}$ & 0.55$^{0.04}_{0.05}$ & 0.88$^{0.01}_{0.02}$ & 0.85$^{3.3e-3}_{3.3e-3}$ \\ 
FUV & $p$ & $<$0.01$^{0.00}_{0.00}$ & $<$0.01$^{0.00}_{0.00}$ & $<$0.01$^{0.01}_{0.00}$ & $<$0.01$^{0.00}_{0.00}$ & $<$0.01$^{0.00}_{0.00}$ \\ 
\hline 
NUV & $m$ & 1.05$^{0.04}_{0.04}$ & 1.01$^{0.02}_{0.02}$ & 0.94$^{0.08}_{0.09}$ & 0.95$^{0.03}_{0.03}$ & 0.92$^{0.01}_{0.01}$ \\ 
NUV & $b$ & 1.10$^{0.09}_{0.09}$ & 1.02$^{0.02}_{0.02}$ & 1.06$^{0.17}_{0.18}$ & 1.13$^{0.06}_{0.06}$ & 0.94$^{0.01}_{0.01}$ \\ 
NUV & $r$ & 0.96$^{0.01}_{0.01}$ & 0.99$^{1.9e-3}_{2.5e-3}$ & 0.66$^{0.05}_{0.05}$ & 0.92$^{0.01}_{0.01}$ & 0.96$^{2.1e-3}_{2.3e-3}$ \\ 
NUV & $p$ & $<$0.01$^{0.00}_{0.00}$ & $<$0.01$^{0.00}_{0.00}$ & $<$0.01$^{0.00}_{0.00}$ & $<$0.01$^{0.00}_{0.00}$ & $<$0.01$^{0.00}_{0.00}$ \\ 
\hline 
H$_2$ Bump & $m$ & 0.99$^{0.26}_{0.26}$ & -0.14$^{0.10}_{0.10}$ & 0.12$^{0.03}_{0.03}$ & 0.22$^{0.12}_{0.12}$ & -0.13$^{0.05}_{0.05}$ \\ 
H$_2$ Bump & $b$ & 2.50$^{0.99}_{0.99}$ & -0.63$^{0.37}_{0.39}$ & -0.46$^{0.11}_{0.10}$ & -0.28$^{0.38}_{0.38}$ & -1.26$^{0.19}_{0.19}$ \\ 
H$_2$ Bump & $r$ & 0.54$^{0.13}_{0.14}$ & -0.20$^{0.13}_{0.14}$ & 0.31$^{0.08}_{0.08}$ & 0.22$^{0.11}_{0.12}$ & -0.11$^{0.05}_{0.04}$ \\ 
H$_2$ Bump & $p$ & 0.01$^{0.06}_{0.01}$ & 0.37$^{0.36}_{0.25}$ & 0.14$^{0.15}_{0.08}$ & 0.32$^{0.33}_{0.19}$ & 0.30$^{0.24}_{0.16}$ \\ 
\enddata
\tablecomments{Fit takes the form of $log_{10}(L_{acc}/L_{\odot})=m\cdot log_{10}(L_{UV}/L_{\odot})+b$. $r$ is the Pearson correlation coefficient and $p$ is the p-value of the null hypothesis. See Figure \ref{fig: UV Luminosities}.}
\end{deluxetable*}


\end{document}